\documentclass[a4paper,11pt]{article}
\usepackage[inline]{enumitem}
\usepackage[multidot]{grffile} 
\usepackage{dcolumn}
\usepackage{bm}
\usepackage{amsmath}
\usepackage{amsthm}
\usepackage{framed}
\usepackage{soul}
\usepackage{amsfonts}
\usepackage{bbold}
\usepackage{dsfont}
\usepackage{amssymb}
\usepackage{color}
\usepackage{latexsym}
\usepackage{stackrel}
\usepackage{slashed} 
\usepackage{empheq}
\usepackage{fancybox}
\usepackage{pstricks}
\usepackage{indentfirst}
\usepackage{mathrsfs}
\usepackage{tablefootnote}
\usepackage{longtable}
\usepackage{multirow}
\usepackage{epsfig,psfrag}
\usepackage{subfigure}
\usepackage{mathtools}
\usepackage{setspace} 
\usepackage[utf8]{inputenc} 
\usepackage[scientific-notation=true]{siunitx} 
\usepackage{verbatim}
\usepackage{comment}
\usepackage[many]{tcolorbox} 
\usepackage{JHEPpub}
\usepackage{JHEPorcid} 
\usepackage[T1]{fontenc} 
\graphicspath{fig/}
\newcommand{\uvec}{\boldsymbol}
\newcommand{\ud}{\text{d}}
\title{Radiative corrections in neutral-current (anti)neutrino elastic scattering at $\text{GeV}$ energies I: Nucleon targets}
\author[a]{Yi Chen\orcidlink{0000-0002-7444-1971}}
\emailAdd{physchen@mail.ustc.edu.cn}
\affiliation[a]{Department of Physics and Center for High Energy Physics, Tsinghua University, Beijing 100084, P. R. China}
\author[b]{Oleksandr Tomalak\orcidlink{0000-0002-4827-5842}}
\emailAdd{tomalak@itp.ac.cn}
\affiliation[b]{Institute of Theoretical Physics, Chinese Academy of Sciences, Beijing 100190, P. R. China}
\author[a, b]{Bing-Song Zou\orcidlink{0000-0002-3000-7540}}
\emailAdd{zoubs@mail.tsinghua.edu.cn}
\abstract{We introduce radiative corrections in neutral-current (anti)neutrino-nucleon elastic scattering at $\text{GeV}$ energies within the effective field theory framework. We factorize cross sections into soft and hard functions, clarify the (anti)neutrino flavor dependence at both amplitude and cross-section levels, and improve the quantum chromodynamics (QCD) contributions to low-energy neutral-current processes. The radiative corrections at the single-nucleon level reach a magnitude comparable to the contributions from strange quarks. We also compare our results with the experimental data from BNL E734 and MiniBooNE collaborations, finding excellent agreements with the experimental data.}


\begin{document} 
\maketitle

\section{Introduction}
\label{sec:intro}

Nucleons, protons and neutrons, are the building blocks of atomic nuclei and contribute to more than 99\% of the mass of visible matter in the Universe~\cite{Gao:2021sml}. Owing to the complicated non-perturbative quantum chromodynamics (QCD) interactions between quarks and gluons, nucleons and nuclei inherit particularly rich and intricate internal structures.

Information on the spatial distribution of the nucleon is encoded in Lorentz-invariant functions, known as form factors (FFs). For example, a nucleon has four electromagnetic FFs: $G_E^\gamma$ (electric), $G_M^\gamma$ (magnetic), $F_A^\gamma$ (anapole), and $F_D^\gamma$ (electric dipole), which are usually extracted from electron-proton (or muon-proton) and electron-deuteron (or muon-deuteron) elastic scattering data~\cite{Chambers:1956zz,Hofstadter:1956qs,A1:2010nsl,Mihovilovic:2019jiz,Xiong:2019umf,PRad:2020oor,Atac:2020hdq,Atac:2021wqj,A1:2022wzx,Strauch:2024imt}. They describe how electric charge, current, polarization, and magnetization are spatially distributed inside the nucleon~\cite{Hofstadter:1956qs,Yennie:1957rmp,Ernst:1960zza,Burkardt:2002hr,Miller:2007uy,Carlson:2007xd,Lorce:2020onh,Kim:2021kum,Li:2022hyf,Epelbaum:2022fjc,Panteleeva:2022khw,Chen:2022smg,Chen:2023dxp,Freese:2023jcp}. By analogy with electron-proton elastic scattering, (anti)neutrinos can also be used as incident particles to probe the internal axial-vector structure of a nucleon/nucleus in weak interactions through elastic~\cite{Lee:1976gxa,Cline:1976in,Cline:1976ba,Horstkotte:1981ne,Ahrens:1986xe,MiniBooNE:2010xqw,MiniBooNE:2013dds,COHERENT:2017ipa,Ren:2022qop,Ren:2022qut} or quasielastic~\cite{Mann:1973pr,Baker:1981su,Belikov:1983kg,Ahrens:1988rr,MiniBooNE:2010bsu,MINERvA:2013bcy,MicroBooNE:2020fxd,MINERvA:2023avz} (anti)neutrino scattering experiments.

Weak neutral-current (NC) interactions are mediated by $Z$ bosons and typically occur in elastic (anti)neutrino scattering experiments. Unlike the electromagnetic current, the generic weak NC of a spin-$\frac{1}{2}$ target involves six weak NC FFs~\cite{LlewellynSmith:1971uhs,Ohlsson:1998bk,Bernard:2001rs,Singh:2006xp,Chen:2024oxx,Chen:2024ksq,Chen:2025gmg}: $G_E^Z$ (electric), $G_M^Z$ (magnetic), $G_S^Z$ (induced scalar), $G_A^Z$ (axial), $G_P^Z$ (induced pseudoscalar), and $G_T^Z$ (induced pseudotensor). They describe the weak interaction with an incoming (anti)neutrino in an elastic scattering reaction, thereby encoding clean internal vector and axial-vector structure information~\cite{Chen:2024oxx,Chen:2024ksq,Chen:2025gmg}.

It is generally acknowledged that weak NC (anti)neutrino elastic scattering is a useful tool for studying the strange-quark content of the nucleon/nucleus~\cite{Kim:1980sa,Horstkotte:1981ne,Ahrens:1986xe,Kaplan:1988ku,Mckeown:1989ir,Garvey:1992cg,Garvey:1993sg,Musolf:1993tb,Pate:2003rk,Beise:2004py,Pate:2008va,Butkevich:2011fu,Sufian:2016pex,Maas:2017snj,Pate:2024acz,Alexandrou:2026ait}, e.g. the strange-quark contribution to the nucleon spin $\Delta s$~\cite{Jaffe:1989jz,Garvey:1992cg,Pate:2003rk,Pate:2008va,Pate:2024acz}. For such scattering reactions, high-precision measurements have been performed at Brookhaven~\cite{Horstkotte:1981ne,Ahrens:1986xe} and Fermilab~\cite{MiniBooNE:2010xqw,MiniBooNE:2013dds,Ren:2022qop,Ren:2022qut} national laboratories in the United States. The same nucleon strange-quark contribution became available recently with the rapid development of lattice QCD and computing techniques~\cite{Green:2015wqa,Djukanovic:2019jtp,Sufian:2016pex,Alexandrou:2018zdf,Alexandrou:2019olr,Alexandrou:2026ait,Barone:2026uyx}. To compare lattice-QCD predictions with measurements, the experimental data points must be unfolded with proper electroweak and quantum electrodynamics (QED) radiative corrections that might have a size comparable to that of the nucleon strange-quark contributions but were not presented or included in previous experimental analyses~\cite{Horstkotte:1981ne,Ahrens:1986xe,MiniBooNE:2010xqw,MiniBooNE:2013dds,Pate:2024acz}.

In this paper, we fill this gap by formulating and evaluating radiative corrections to NC (anti)neutrino elastic scattering at the single-nucleon level within the low-energy effective field theory approach\footnote{Radiative corrections to NC neutrino-induced reactions in the Standard Model were presented in the framework of current algebra in~\cite{Marciano:1980pb}.} and confronting our results with the experimental data~\cite{Entenberg:1979wc,Horstkotte:1981ne,Ahrens:1986xe,MiniBooNE:2010xqw,MiniBooNE:2013dds,Pate:2024acz}. We find that the nucleon strange-quark contributions can reach $(4-9)\%$ and $(3-15)\%$ at the single-proton and single-neutron levels, respectively, comparable to the size of radiative corrections. We also observe excellent agreement of our theoretical predictions for the flux-averaged (unpolarized) differential cross sections with the experimental data from both BNL E734~\cite{Horstkotte:1981ne,Ahrens:1986xe} and MiniBooNE~\cite{MiniBooNE:2010xqw,MiniBooNE:2013dds}, and find that the total nuclear effects in the carbon nucleus for elastic neutrino ($\nu_\ell N \to \nu_\ell N$) and antineutrino ($\bar\nu_\ell N \to \bar\nu_\ell N$) scattering on the $\mathrm{C} \mathrm{H}_2$ target are irrelevant to the MiniBooNE data, especially at large squared four-momentum transfers. Our result might be of interest for high-precision measurements and background simulations of NC cross sections in the MicroBooNE~\cite{MicroBooNE:2015bmn,MicroBooNE:2016pwy,MicroBooNE:2021tya,Ren:2022qop,Ren:2022qut}, NOvA~\cite{NOvA:2007rmc,NOvA:2019cyt,NOvA:2026zup}, DUNE~\cite{DUNE:2020ypp,DUNE:2021tad}, T2K~\cite{T2K:2011qtm,T2K:2019bcf}, and Hyper-K~\cite{Hyper-KamiokandeProto-:2015xww} experiments, as well as for a detailed description of complex astrophysical systems.

The paper is organized as follows. In section~\ref{sec:tree_level}, we present the general formalism for the neutral-current (anti)neutrino-nucleon elastic scattering, define invariant amplitudes, and relate them to nucleon FFs at tree level in both the low-energy effective field theory (cf. section~\ref{sec:leading_order_LEFT}) and the Standard Model (cf. section~\ref{sec:leading_order_SM}). In section~\ref{sec:closed_fermion_loops}, we present radiative corrections from closed fermion loops as shifts in the Wilson coefficients of the low-energy effective field theory and, equivalently, as shifts in the tree-level single-nucleon vector FFs. We first discuss perturbative fermion-loop corrections from heavy quarks and charged leptons, and their dependence on the (anti)neutrino flavor (cf. section~\ref{sec:closed_lepton_loops}), and then evaluate non-perturbative hadronic contributions from light quarks at low momentum transfers and connect them to perturbative results at larger energies (cf. section~\ref{sec:closed_hadron_loops}). We then elaborate in section~\ref{sec:nucleon_ffs_and_brems} on all the electromagnetic corrections specific to the nucleon target within the factorization framework. We describe QED radiative corrections with the soft function and rigorously define the Born FFs (cf. section~\ref{subsec:factorization}), compare the soft function with corresponding structure-independent QED radiative corrections in the standard treatment of elastic electron-proton scattering (cf. section~\ref{subsec:traditional}), and perform the resummation of large perturbative logarithms and over the logarithmically enhanced contributions from an arbitrary number of radiated photons (cf. section~\ref{subsec:resummation}). Our results and discussions are presented in section~\ref{sec:results} at the single-nucleon level, with a particular emphasis on the radiative corrections and nucleon strangeness. Finally, we present our conclusions and outlook in section~\ref{sec:summary}. In appendix~\ref{app:flavor_decomposition}, we report the quark-flavor form factors inside the proton used in this work. In appendix~\ref{app:flux_distributions}, we provide details of the neutrino and antineutrino fluxes used for the evaluation of flux-averaged differential cross sections. We express the quark-flavor form factors in terms of generalized parton distributions in appendix~\ref{app:GPDs}. Supplemental material and \href{https://github.com/tomalak7/NCelasticXsec}{github.com/tomalak7/NCelasticXsec} include a Mathematica notebook for the fast, accurate evaluation of neutral-current (anti)neutrino-nucleon elastic scattering cross sections, accounting for radiative corrections.

\section{Neutral-current (anti)neutrino-nucleon elastic scattering}
\label{sec:tree_level}

\begin{figure}[htb!]
	\centering
	{\includegraphics[angle=0,scale=0.5]{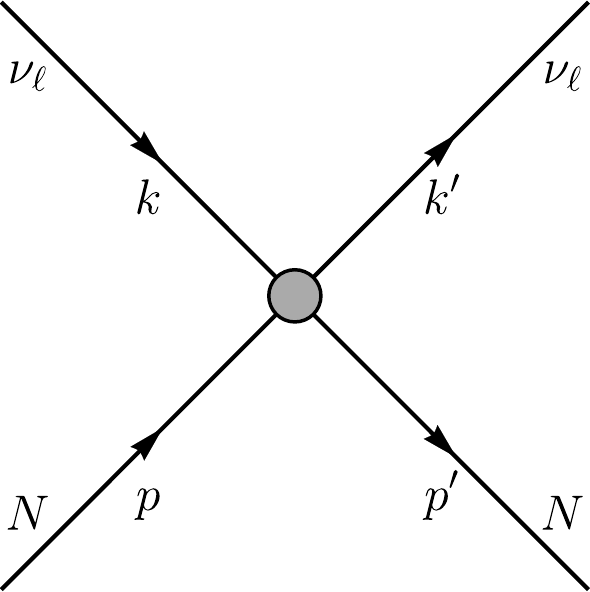}}
	\caption{The Feynman diagram for the low-energy contact interaction of weak NC neutrino-nucleon elastic scattering $\nu_\ell N \to \nu_\ell N$ is shown. The four-momentum transfer is $\Delta=p'-p$.}
	\label{fig:general_interaction_NC}
\end{figure}

Let us first consider the weak NC neutrino-nucleon elastic scattering reaction:
\begin{equation} \label{eq:vN-elastic-reaction}
	\nu_{\ell}(k,r) + N(p,s) \to \nu_{\ell}(k',r') + N(p',s'),
\end{equation}
where $\nu_\ell$ represents a neutrino of mass $m_{\nu_\ell}$ with $\ell \in \{ e, \mu, \tau \}$ being the flavor label of a charged lepton, $N \in \{p, n\}$ represents a nucleon of mass $M$, $k^2=k'^2=m_{\nu_\ell}^2$, and $p^2 = p'^2 = M^2$. The corresponding Feynman diagram for the low-energy contact interaction in the reaction (\ref{eq:vN-elastic-reaction}) is illustrated in figure~\ref{fig:general_interaction_NC}.

For simplicity, we set $m_{\nu_\ell}=0$ throughout this paper, since the neutrino-mass effect on the differential cross section~\cite{LlewellynSmith:1971uhs} is extremely small, i.e. $\left( m_{\nu_\ell}/M \right)^2 \lesssim 2.3\times 10^{-19}$ with $m_{\nu_\ell} \lesssim 0.45~\text{eV}$~\cite{KATRIN:2024cdt} and $m_{\nu_\ell} \lesssim 0.1-0.2~\text{eV}$~\cite{Elbers:2025vlz,Naredo-Tuero:2024sgf,Allali:2024aiv}, in comparison to the current precision of measurements in neutrino scattering experiments~\cite{Horstkotte:1981ne,Ahrens:1986xe,MiniBooNE:2010xqw,MiniBooNE:2013dds,Ren:2022qop,Ren:2022qut}. In this section, we will explore the tree-level calculations both in the low-energy effective field theory (LEFT) and the Standard Model (SM).

The scattering amplitude $\mathcal M$ for the NC neutrino-nucleon elastic scattering process~(\ref{eq:vN-elastic-reaction}) can be expressed in terms of four independent invariant amplitudes $g_E$, $g_M$, $f_A$, $f_T \equiv f_{A3}$ as~\cite{Borah:2024hvo}
\begin{equation}
	\begin{aligned}\label{eq:NC_amplitude_decomposition}
		\mathcal M
		= \frac{ G_F }{ \sqrt{2} } \ell_\mu H^{\mu},
	\end{aligned}
\end{equation}
with
\begin{equation}
	\begin{aligned}\label{eq:NC_leptonic-hadronic-currents}
		\ell_\mu
		&\equiv \bar u(k',r') \gamma_\mu (1-\gamma^5) u(k,r),\\
        H^{\mu}
		&\equiv \bar u(p',s') \left[ \gamma^\mu \left( {g}_M + {f}_A \gamma^5 \right) - \left( \frac{ g_M-g_E }{1+\tau} - 2 f_T \gamma^5 \right) \frac{{P}^\mu}{M} \right] u(p,s),
	\end{aligned}
\end{equation}
where $G_F$ is the Fermi coupling constant, $P \equiv (p^\prime+p)/2$ is the averaged nucleon four-momentum, $Q^2 \equiv -(p'-p)^2$ is the squared four-momentum transfer, $\tau \equiv Q^2/(4M^2)$, $u(k,r)$ ($u(k',r')$) and $u(p,s)$ ($u(p',s')$) represent the Dirac spinors of the neutrino and nucleon in the initial (final) state, respectively. Each of the amplitudes in (\ref{eq:NC_leptonic-hadronic-currents}) is a complex function of two kinematic invariants, in general. Within the accuracy of our analysis at GeV energies, we note that all these amplitudes $f_{A,T}$ and $g_{E,M}$ depend solely on $Q^2$. In this paper, we choose $Q^2$ and the crossing-symmetric variable $\nu \equiv (s-u)/4 = M E_\nu - Q^2/4 $ as the two independent variables, where $s \equiv (k+p)^2 $ and $u\equiv (k'-p)^2$ are the standard Mandelstam invariants and $E_\nu=k^0$ is the incident neutrino energy in the laboratory frame (or the rest frame of the initial-state nucleon).

For NC neutrino-nucleon elastic scattering (\ref{eq:vN-elastic-reaction}), the unpolarized differential cross section in the laboratory frame, after neglecting the neutrino mass, is given by~\cite{Chen:2024ksp}
\begin{equation}
	\begin{aligned}\label{eq:differential-crosec-lab}
		\frac{\ud \sigma }{\ud Q^2 } = \frac{\overline{|\mathcal M|^2}}{4\pi (s-u+Q^2)^2 } = \frac{\overline{|\mathcal M|^2}}{ 16\pi (s-M^2)^2 },
	\end{aligned}
\end{equation}
where $\overline{|\mathcal M|^2}$ is spin-averaged (over the initial states) and spin-summed (over the final states) scattering amplitude squared. For antineutrino-nucleon elastic scattering, one simply needs to replace $\ell_\mu$ in (\ref{eq:NC_leptonic-hadronic-currents}) with
\begin{equation}
	\begin{aligned}\label{eq:NC-leptonic-current-antileutrino}
		\ell_\mu
		&\equiv \bar v(k,r) \gamma_\mu (1-\gamma^5) v(k',r').
	\end{aligned}
\end{equation}
Using the invariant amplitude decomposition of (\ref{eq:NC_amplitude_decomposition}), the unpolarized differential cross section in weak NC (anti)neutrino-nucleon elastic scattering in the laboratory frame is obtained as~\cite{LlewellynSmith:1971uhs,Ahrens:1986xe,Chen:2024ksp,Chen:2026uhf}
\begin{equation} \label{eq:NC_cross_section_eff}
	\frac{\ud \sigma}{\ud Q^2} (E_\nu, Q^2) 
	= \frac{ G_F^2 }{ 2\pi } \frac{ M^2 }{ E_\nu^2 } \left[ \tau A(\nu,~Q^2) - \eta \frac{\nu}{M^2} B(\nu,~Q^2) + \frac{\nu^2}{M^4} \frac{C(\nu,~Q^2)}{ 1 + \tau } \right],
\end{equation}
with the structure-dependent quantities $A$, $B$, and $C$ given by
\begin{equation}
	\begin{aligned}
		A &= \tau | g_M |^2 - | g_E |^2 + (1+ \tau) | f_A |^2 - 4 \tau (1+ \tau) |f_T|^2, \\
		B &= 4 \tau \mathfrak{Re} \Big[f^*_A g_M \Big], \\
		C &= \tau |g_M|^2 + |g_E|^2 + (1 + \tau) |f_A|^2 + 4 \tau (1 + \tau) |f_T|^2,
	\end{aligned}
\end{equation}
where $\eta = 1$ ($\eta = -1$) corresponds to neutrino (antineutrino) scattering, respectively.

In the following sections~\ref{sec:leading_order_LEFT} and~\ref{sec:leading_order_SM}, we describe the NC (anti)neutrino-nucleon elastic scattering process at tree level in the LEFT and SM, respectively.

\subsection{Tree-level results in the low-energy effective field theory}
\label{sec:leading_order_LEFT}

To provide a microscopic description of the weak NC (anti)neutrino-nucleon scattering process,\footnote{Such a description for the charged-current (anti)neutrino-nucleon scattering is presented in~\cite{Tomalak:2020zlv}.} we start with the LEFT below the electroweak scale that includes only quarks, leptons, QED and QCD interactions, as well as dimension-6 four-fermion electroweak operators and (anti)neutrino-photon couplings. The effective Lagrangian density of this theory can be written as~\cite{Tomalak:2019ibg}
\begin{align} \label{eq:effective_Lagrangian_all}
		{\cal L}_{\rm eff} &= - \sum_{\ell,\ell^\prime} \left[ \bar{\nu}_\ell \gamma_\mu \mathrm{P}_\mathrm{L} \nu_\ell \right] \bar{\ell}^\prime \gamma^\mu \Big( c_\mathrm{L}^{\nu_\ell \ell^\prime} \mathrm{P}_\mathrm{L} + c_\mathrm{R}^{\nu_\ell \ell^\prime} \mathrm{P}_\mathrm{R} \Big) \ell^\prime - \sum_{\ell,q} \left[ \bar{\nu}_\ell \gamma_\mu \mathrm{P}_\mathrm{L} \nu_\ell \right] \bar{q} \gamma^\mu \Big( c_\mathrm{L}^{q} \mathrm{P}_\mathrm{L} + c_\mathrm{R}^{q} \mathrm{P}_\mathrm{R} \Big)q \nonumber \\
		&\quad - \frac{1}{e}\sum_\ell c^{\nu_\ell \gamma} \partial_\mu F^{\mu\nu} \left[ \bar{\nu}_\ell \gamma_\nu \mathrm{P}_\mathrm{L} \nu_\ell \right],
\end{align}
where $e=|e|$ is the electric charge of the positron, $\mathrm{P}_\mathrm{L,R}=(1 \mp \gamma^5)/2$ are the left- and right-handed projection operators, $q \in \{ u,d,s,c,b \}$ denotes the quark fields, $\nu_\ell \in \{ \nu_e, \nu_\mu, \nu_\tau \}$ denotes the (anti)neutrino fields, $F^{\mu\nu} \equiv \partial^\mu A^\nu - \partial^\nu A^\mu$ is the electromagnetic field strength tensor with $A^\mu$ being the photon field, $\{ c_\mathrm{L,R}^{\nu_\ell \ell^\prime},~c_\mathrm{L,R}^{q},~c^{\nu_\ell \gamma} \}$ are the corresponding effective couplings.\footnote{Determining the effective couplings, we have neglected corrections of order $\alpha m_f^2/m_W^2$ and $|Q^2|/m_W^2$~\cite{Tomalak:2019ibg}, with the electromagnetic fine structure constant $\alpha \equiv e^2/(4\pi) \approx 1/137$, the masses of fermion and $W$ boson $m_f$ and $m_W$, respectively, and the typical kinematic scale of the scattering $\sim Q$.} The standard kinetic, mass, and QED+QCD gauge coupling terms for (anti)neutrinos, charged leptons, and quarks (including the neutrino mixing matrix) are implied as part of this effective field theory. The sums in (\ref{eq:effective_Lagrangian_all}) run over all three lepton flavors and quark flavors of the particles present in the effective theory.

For applications to (anti)neutrino scattering, it is convenient to replace the (anti)neutrino-photon operator by an equivalent combination of four-fermion operators in the effective theory, obtained via the field redefinition:
\begin{equation}\label{eq:field_redef}
	A^\mu \to A^\mu + \frac{1}{e} \sum_\ell c^{\nu_\ell \gamma} \bar{\nu}_\ell \gamma^\mu \mathrm{P}_\mathrm{L} \nu_\ell.
\end{equation}
Under (\ref{eq:field_redef}), the Wilson coefficients in the effective Lagrangian (\ref{eq:effective_Lagrangian_all}) should be modified as~\cite{Tomalak:2019ibg}
\begin{equation}\label{eq:field_redef_couplings}
	c^{\nu_\ell \gamma} \to 0,\qquad 
	c_\mathrm{L,R}^{\nu_\ell \ell^\prime} \to c_\mathrm{L,R}^{\nu_\ell \ell^\prime} + c^{\nu_\ell \gamma},\qquad
	c_\mathrm{L,R}^{q} \to c_\mathrm{L,R}^{q} - Q_q c^{\nu_\ell \gamma},
\end{equation}
where $Q_q$ is the electric charge of the quark flavor $q$ in units of $e$.

For each quark flavor, the nucleon matrix elements of the vector $\hat j^\mu_q(x) \equiv \hat{\bar q}(x) \gamma^\mu \hat q(x)$ and axial-vector $\hat j^\mu_{5,q}(x) \equiv \hat{\bar q}(x) \gamma^\mu \gamma^5 \hat q(x)$ quark-current operators in the case of the spin-$\frac{1}{2}$ nucleon $N$ are in general parametrized as~\cite{LlewellynSmith:1971uhs,Ohlsson:1998bk,Bernard:2001rs,Singh:2006xp,Chen:2024ksq,Chen:2026uhf}
\begin{equation}
	\begin{aligned}\label{eq:quark-V-A-matrix-elements}
		\langle N(p',s')| \hat j^\mu_q(0) |N(p,s) \rangle
		&= \bar u(p',s') \left[ \gamma^\mu G_M^q + \frac{P^\mu (G_E^q - G_M^q)}{M(1+\tau)} + \frac{i\Delta^\mu }{2M} G_S^q \right] u(p,s),\\
		\langle N(p',s')|\hat j^\mu_{5,q}(0)|N(p,s) \rangle
		&= \bar u(p',s') \left[ \gamma^\mu G_A^q + \frac{\Delta^\mu}{2M} G_P^q - \frac{\sigma^{\mu\nu} \Delta_\nu }{2M} G_T^q \right]\gamma^5 u(p,s),\\
	\end{aligned}
\end{equation}
where $\Delta \equiv p'-p$ is the four-momentum transfer, $p^2=p'^2=M^2$, and for the sake of clarity, we have omitted the explicit $Q^2$ dependence of the quark-flavor FFs $G_{E,M,S,A,P,T}^q(Q^2)$. 

From the above effective Lagrangian (\ref{eq:effective_Lagrangian_all}), one can easily read off the effective weak NC operator $\hat j^\mu (x)$ as
\begin{equation}\label{eq:weak-NC-operator-EFT}
	\hat j^\mu (x) 
= \sum_q \Big[ c^q_V \, \hat j^\mu_q(x) - c^q_A \, \hat j^\mu_{5,q}(x) \Big] = \sum_q \hat{\bar q}(x) \gamma^\mu \Big[ c^q_V - c^q_A\, \gamma^5 \Big] \hat q(x),
\end{equation}
where $c^q_V \equiv (c_\mathrm{L}^{q} + c_\mathrm{R}^{q})/(2\sqrt{2} G_F)$ and $c^q_A \equiv (c_\mathrm{L}^{q} - c_\mathrm{R}^{q})/(2\sqrt{2} G_F)$ are effective vector- and axial-vector couplings of quarks to the (anti)neutrino current, respectively. One then naturally expects~\cite{Chen:2024ksq,Chen:2026uhf}
\begin{equation}
	\begin{aligned}\label{eq:weak-NC-matrix-elements-EFT}
		\langle N(p',s')|\hat j^\mu (0)|N(p,s) \rangle
		&= \bar u(p',s') \Gamma^\mu(P,\Delta) u(p,s),\\
	\end{aligned}
\end{equation}
with the vector $\Gamma^\mu_{V}$ and the axial-vector $\Gamma^\mu_{A}$ parts given by
\begin{equation}
	\begin{aligned}\label{eq:weak-NC-vertex-functions-EFT}
		\Gamma^\mu(P,\Delta)
		&= \Gamma^\mu_{V}(P,\Delta) - \Gamma^\mu_{A}(P,\Delta),\\
		\Gamma^\mu_{V}(P,\Delta)
		&\equiv \gamma^\mu G_M + \frac{P^\mu (G_E - G_M)}{M(1+\tau)} + \frac{i\Delta^\mu }{2M} G_S ,\\
		\Gamma^\mu_{A}(P,\Delta)
		&\equiv \gamma^\mu G_A + \frac{\Delta^\mu}{2M} G_P - \frac{\sigma^{\mu\nu} \Delta_\nu }{2M} G_T,
	\end{aligned}
\end{equation}
where expressions for the resulting FFs are given as sums over individual quark contributions:
\begin{equation}\label{eq:weak-NC-FFs-EFT}
	G_{E,M,S}(Q^2) 
	= \sum_q c^q_V \, G_{E,M,S}^q(Q^2),\qquad
	G_{A,P,T}(Q^2) 
	= \sum_q c^q_A \, G_{A,P,T}^q(Q^2).
\end{equation}

It is straightforward to make connections between these FFs and the amplitudes $g_{E,M}$ and $f_{A,T}$ in (\ref{eq:NC_amplitude_decomposition}). At tree level in the LEFT, we find the following relations
\begin{equation}
	\begin{aligned}
		&g_E = G_E,\qquad g_M = G_M,\qquad 
		f_A = -G_A,\qquad f_T = \frac{i}{2}G_T.
	\end{aligned}
\end{equation}
Following the general decomposition of the nucleon current in~(\ref{eq:weak-NC-vertex-functions-EFT}), two more invariant amplitudes can be introduced, $g_S = i G_S$ and $f_P = -G_P$. However, they do not contribute to any observable in the NC (anti)neutrino-nucleon elastic scattering when the (anti)neutrino mass is neglected.

\subsection{Tree-level results in the Standard Model}
\label{sec:leading_order_SM}

In the SM, the weak NC at leading order is described by the exchange of the $Z$ boson. The corresponding nucleon current at the quark level can be obtained by replacing the Wilson coefficients $c_V^q$ and $c_A^q$ in~(\ref{eq:weak-NC-operator-EFT}) with the tree-level couplings $g_V^q$ and $g_A^q$, respectively. The NC operator $\hat j^\mu_Z(x)$ in the SM is given by
\begin{equation}
	\begin{aligned}\label{eq:weak-NC-operator-SM}
		\hat j^\mu_Z(x)
		\equiv \sum_q \hat{\bar q}(x) \gamma^\mu \left( g_V^q - \gamma^5 g_A^q \right) \hat q(x) = \sum_q \left[ g_V^q\, \hat j^\mu_q(x) - g_A^q\, \hat j^\mu_{5,q}(x) \right],
	\end{aligned}
\end{equation}
where the vector $g_V^q$ and axial-vector $g_A^q$ couplings of quarks to the $Z$ boson can be expressed in terms of the Weinberg angle $\theta_W$ as~\cite{Peskin:1995ev}
\begin{equation}
	\begin{aligned}\label{eq:weak-NC-couplings-SM}
		g_V^{u,c,t}
		= \frac{1}{2} - \frac{4}{3}\sin^2\theta_W,\qquad
		g_V^{d,s,b}
		= -\frac{1}{2} + \frac{2}{3}\sin^2\theta_W,\qquad
		g_A^{u,c,t}
		= -g_A^{d,s,b} = \frac{1}{2}.
	\end{aligned}
\end{equation}
The resulting weak NC FFs of the spin-$\frac{1}{2}$ nucleon $N$ in the SM are given by
\begin{equation}
	\begin{aligned}\label{eq:weak-NC-FFs-SM}
		G_{E,M,S}^{Z,N}(Q^2) 
		= \sum_q g_V^q\, G_{E,M,S}^{q,N}(Q^2),\qquad
		G_{A,P,T}^{Z,N}(Q^2) = \sum_q g_A^q\, G_{A,P,T}^{q,N}(Q^2),
	\end{aligned}
\end{equation}
where $G_{E,M,S,A,P,T}^{q,N}(Q^2)= G_{E,M,S,A,P,T}^{q}(Q^2)$ are the FFs of quark flavor $q$ of the nucleon $N$.

Based on the $\text{SU}(3)$ flavor symmetry, the $u$ and $d$ quark-flavor contributions to the tree-level nucleon electromagnetic FFs $G_{E,M}^{\gamma,N} (Q^2) = \sum \limits_q Q_q\, G_{E,M}^{q,N}(Q^2)$, with the quark electric charge $Q_q$, are given by\footnote{In extracting form factors from the experimental data, we neglect isospin-breaking effects. Based on estimates for electromagnetic form factors from the constituent quark model and chiral perturbation theory, these effects are expected to be at the permille~\cite{Behrends:1960nf,Dmitrasinovic:1995jt,Miller:1997ya,Lewis:2006vve} and percent~\cite{Kubis:2006cy} level, respectively.}
\begin{equation}
	\begin{aligned}\label{GEM-q}
		G_{E,M}^{u} &\equiv G_{E,M}^{u,p} = 2 G_{E,M}^{\gamma, p} + G_{E,M}^{\gamma, n} + G_{E,M}^{s}, \\
        G_{E,M}^{d} &\equiv G_{E,M}^{d,p} = 2 G_{E,M}^{\gamma, n} + G_{E,M}^{\gamma, p} + G_{E,M}^{s},
	\end{aligned}
\end{equation}
where $G_{E,M}^s=G_{E,M}^{s,p}$ are the total strangeness contributions (i.e. sum of the $s$ and $\bar s$ quarks) to the nucleon electromagnetic FFs. In the $\text{SU}(3)$ flavor symmetric case, the weak NC electric and magnetic FFs of the nucleon in the SM are given by~\cite{Kim:1980sa,Ahrens:1986xe,Kaplan:1988ku,Mckeown:1989ir,Garvey:1992cg,Garvey:1993sg,Musolf:1993tb,Alberico:2001sd,Pate:2003rk,Beise:2004py,Pate:2008va,Sufian:2016pex,Maas:2017snj,Pate:2024acz,Alexandrou:2026ait,Chen:2026uhf}
\begin{equation}
	\begin{aligned}\label{GEM-Z}
		G_{E,M}^{Z,p (n)}
		= \tilde g_{V}^{p} G_{E,M}^{\gamma, p (n)} + \tilde g_V^{n} G_{E,M}^{\gamma,n (p)} + \tilde g_V^s G_{E,M}^{s},
	\end{aligned}
\end{equation}
with vector couplings of the proton $\tilde g_V^p$, the neutron $\tilde g_V^n$, and the strange quark $\tilde g_V^s$ given by
\begin{equation}
	\tilde g_{V}^{p} = \frac{1}{2} - 2 \sin^2\theta_W,\qquad
	\tilde g_V^{n} = -\frac{1}{2},\qquad \tilde g_V^s = -\frac{1}{2}.
\end{equation}

For the axial-vector FF of the nucleon, we employ the charged-current (CC) FF $G_A^{W}$ of the nucleon to fix the $u$-$d$ flavor difference $G_A^{W} = G^u_A - G^d_A$. To disentangle these two flavors, we take the fit for the flavor sum $G_A^{u+d} \equiv G_A^{u}+G_A^{d}$ from lattice-QCD studies~\cite{Alexandrou:2021wzv}, in the dipole functional form, cf. also~\cite{Barone:2026uyx}. Hence, in the $\text{SU}(3)$ flavor symmetric case, we have~\cite{Alberico:2001sd,Golan:2013jtj,Chen:2024ksp,Chen:2026uhf}
\begin{align}
	G^{Z, p (n)}_{A} = \tilde g^{p (n)}_A \left( G_A^u - G_A^d \right) + \tilde g_A^s G_A^s = \tilde g^{p (n)}_A G_A^{W} + \tilde g_A^s G_A^s,
\end{align}
with the axial-vector couplings to the proton $\tilde g_A^p$, the neutron $\tilde g_A^n$, and the strange quark $\tilde g_A^s$
\begin{equation}
	\tilde g_{A}^{p} = \frac{1}{2},\qquad \tilde g_A^{n} = -\frac{1}{2},\qquad \tilde g_A^s = -\frac{1}{2}.
\end{equation}
Measurements of parity-violating elastic scattering of polarized electrons on unpolarized protons and those of the NC (anti)neutrino-proton elastic scattering pave the way for the experimental determination of strangeness contributions to the electromagnetic $G_{E,M}^{s}$ and axial-vector $G_A^s$ FFs of the nucleon~\cite{Pate:2024acz}, respectively.

For numerical evaluations, we assume three quark flavors and exploit the flavor decomposition for the nucleon FFs. For the electromagnetic Born FFs of the nucleon $G_{E,M}^{\gamma,N}$, we use fits to the $ep$ and $en$ cross-section data, measurements of the neutron scattering length, and energy levels in muonic hydrogen~\cite{Borah:2020gte}, assuming that the experimental extraction is performed at the renormalization scale $\mu \approx M$. The nucleon axial-vector FF $G_A^{W}$ is taken from fits to the radiatively corrected data of the MINERvA experiment~\cite{MINERvA:2023avz,Tomalak:2026wsu}. For the flavor sum $G_A^u + G_A^d$ and the axial-vector contribution $G_A^s$ from the strange-flavor quarks, we take the lattice-QCD calculations from~\cite{Alexandrou:2021wzv,Alexandrou:2026ait}, respectively, improved with the known perturbative-QCD behaviour~\cite{Lepage:1980fj,Chernyak:1983ej}.

Similarly to (\ref{eq:weak-NC-matrix-elements-EFT}), the hadronic matrix elements of the weak NC operator $\hat j^\mu_Z(0)$ for the spin-$\frac{1}{2}$ nucleon are in general parametrized as~\cite{LlewellynSmith:1971uhs,Ohlsson:1998bk,Bernard:2001rs,Singh:2006xp,Chen:2024ksq,Chen:2025gmg,Chen:2026uhf}
\begin{equation}
	\begin{aligned}\label{elements-NCO-proton}
		\langle N(p',s')|\hat j^\mu_Z(0)|N(p,s)\rangle
		= \bar u(p',s') \Gamma_Z^\mu(P,\Delta) u(p,s),
	\end{aligned}
\end{equation}
with the decomposition of the vector $\Gamma_V^\mu(P,\Delta)$ and axial-vector $\Gamma_A^\mu(P,\Delta)$ terms given by
\begin{equation}
	\begin{aligned}\label{Vertex-Fun-Weak}
		\Gamma_Z^\mu(P,\Delta)
		&= \Gamma_V^\mu(P,\Delta) - \Gamma_A^\mu(P,\Delta),\\
		\Gamma_V^\mu(P,\Delta)
		&\equiv \gamma^\mu G_M^Z + \frac{P^\mu (G_E^Z - G_M^Z)}{M(1+\tau)} + \frac{i\Delta^\mu }{2M} G_S^Z,\\
		\Gamma^\mu_A(P,\Delta)
		&\equiv \left[ \gamma^\mu G_A^Z + \frac{\Delta^\mu}{2M} G_P^Z - \frac{\sigma^{\mu\nu} \Delta_\nu }{2M} G_T^Z \right]\gamma^5,
	\end{aligned}
\end{equation}
where $G_X^Z \equiv G_X^Z(Q^2)$ for $X=E,M,S,A,P,T$ are the weak NC FFs of the nucleon, called electric, magnetic, induced scalar, axial, induced pseudoscalar, and induced pseudotensor FFs, respectively. By virtue of hermiticity conditions $\Gamma_Z^{\mu,\dag}(P,\Delta)=\gamma^0 \Gamma_Z^\mu(P,-\Delta)\gamma^0$, one can explicitly show that all six FFs $G_X^Z(Q^2)$ are real-valued functions~\cite{Chen:2024ksq,Chen:2025gmg,Chen:2026uhf}. 

Note that for the weak CC interactions mediated by $W^\pm$ bosons between two different spin-$\tfrac{1}{2}$ nucleons of masses $M_1=\sqrt{p^2}$ and $M_2=\sqrt{p'^2}$, the generic weak CC vertex function $\Gamma^\mu_W$ assumes the same Lorentz-invariant decomposition as that of the weak NC in (\ref{Vertex-Fun-Weak}). It also contains six weak CC FFs $G_X^W(Q^2)$. For $\Gamma^\mu_Z \to \Gamma^\mu_W$, one needs to apply the replacements: $G_X^Z(Q^2) \to G_X^W(Q^2)$ and $M \to (M_1+M_2)/2$.

\begin{figure}[tb!]
	\centering
	{\includegraphics[angle=0,scale=0.525]{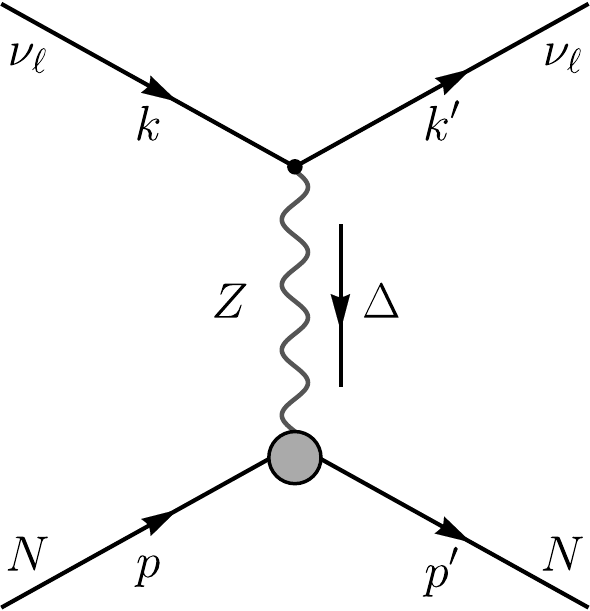}}
	\caption{The tree-level Feynman diagram for the $t$-channel weak NC elastic scattering (\ref{eq:vN-elastic-reaction}) in the first Born approximation (i.e. one virtual $Z$ boson exchange) is shown. The four-momentum transfer is $\Delta = p'-p = k-k'$.}
	\label{Fig_FeynmanDiagram_tree}
\end{figure}

Now, let us work out the tree-level scattering amplitude for (\ref{eq:vN-elastic-reaction}) in the SM, where the corresponding Feynman diagram is illustrated in figure~\ref{Fig_FeynmanDiagram_tree}. Using the unitary gauge for the propagator, the tree-level scattering amplitude is given by
\begin{equation}
	\begin{aligned}\label{eq:NC-amplitude-tree-SM}
		\mathcal M_\text{SM}
		&= \frac{G_F}{ \sqrt{2} } \left( \frac{ m_Z^2 }{ m_Z^2 + Q^2} \right) \cdot \ell^\nu \left( g_{\mu\nu} - \frac{\Delta_\mu \Delta_\nu}{m_Z^2} \right) H_\text{SM}^{\mu},
	\end{aligned}
\end{equation}
with
\begin{equation}
	\begin{aligned}\label{eq:Hadronic-current-SM}
		H_\text{SM}^{\mu} \equiv \bar u(p',s') \Gamma_Z^\mu(P,\Delta) u(p,s),
	\end{aligned}
\end{equation}
where $m_Z$ is the mass of the $Z$ boson, and $\ell_\mu$ is defined in (\ref{eq:NC_leptonic-hadronic-currents}). Exploiting (\ref{eq:differential-crosec-lab}), the tree-level unpolarized differential cross section of NC (anti)neutrino-nucleon elastic scattering within the SM is given by~\cite{LlewellynSmith:1971uhs,Ahrens:1986xe,Chen:2024ksp,Chen:2025gmg,Chen:2026uhf}
\begin{equation}\label{NC-diffSgm-tree}
	\frac{\ud \sigma^{\pm}_\text{SM} }{\ud Q^2}(E_\nu,Q^2)
	= \frac{G_F^2 M^2 }{ 8\pi E_{\nu}^2} 
	\!\left(\frac{m_Z^2}{ m_Z^2+Q^2 } \right)^2
	\!\left[ \tilde A(Q^2) \pm \frac{s-u}{M^2} \tilde B(Q^2) + \frac{(s-u)^2 }{M^4} \frac{ \tilde C(Q^2) }{1+\tau} \right]\!,
\end{equation}
with
\begin{equation}
	\begin{aligned}
		\tilde A
		&\equiv 4\tau \left[ (1+\tau) (G_A^Z)^2 + \tau (G_M^Z)^2 - (G_E^Z)^2 - \tau(1+\tau) (G_T^Z)^2 \right] + \tilde A_{ m_{\nu_\ell} },\\
		\tilde B
		&\equiv 4\tau G_A^Z G_M^Z,\\
		\tilde C
		&\equiv \frac{1}{4} \left[ (1+\tau) (G_A^Z)^2 + \tau (G_M^Z)^2 + (G_E^Z)^2 + \tau(1+\tau) (G_T^Z)^2 \right],
	\end{aligned}
\end{equation}
where the $+$ ($-$) sign is for neutrino (antineutrino) scattering, and $E_\nu$ is the incident (anti)neutrino energy in the laboratory frame. We notice that although we explicitly include $G_S^Z$ and $G_P^Z$ in the full vertex function (\ref{elements-NCO-proton}), these two FFs do not contribute to the differential cross sections (\ref{NC-diffSgm-tree}) when $m_{\nu_\ell}=0$, since their total contribution $\tilde A_{ m_{\nu_\ell} }$ is given by~\cite{Chen:2026uhf}
\begin{equation}
	\begin{aligned}
		\tilde A_{ m_{\nu_\ell} }\!
        &\equiv \! \frac{2 m_{\nu_\ell}^2 }{ M^2 } \bigg[ \left( G_A^Z-\tau G_P^Z \right)^2 + \tau(1+\tau) (G_S^Z)^2 - (G_E^Z)^2 - \tau (1+\tau) (G_T^Z)^2 \bigg]\\
        &+ \frac{16 m_{\nu_{\ell} }^2 \tau }{ m_Z^2 }\left( 1 + \frac{2M^2 \tau }{m_Z^2} \right) \left[ \left( G_A^Z-\tau G_P^Z \right)^2 + \tau(1+\tau) ( G_S^Z)^2 \right],\\
	\end{aligned}
\end{equation}
which is highly suppressed by the factors $(m_{\nu_\ell}/M)^2 \lesssim 2.3 \times 10^{-19}$ and $(m_{\nu_\ell}/m_Z)^2 \lesssim 2.4 \times 10^{-23}$~\cite{KATRIN:2024cdt}.

\section{Closed fermion loops}
\label{sec:closed_fermion_loops}

In this section, we include radiative corrections from the closed fermion (i.e. quark and lepton) loops that couple via a photon to the nucleon; see the Feynman diagram in the LEFT in figure~\ref{one_loop_ET}. Such contributions can be conveniently included by replacing the tree-level values of the vector Wilson coefficients in expressions for the nucleon FFs (\ref{eq:weak-NC-FFs-EFT}) as $c^i_{V} \to \tilde{c}^{i,\nu_\ell}_{V}(Q^2;\mu)$~\cite{Marciano:1980pb,Tomalak:2020zfh}, with $\tilde{c}^{i,\nu_\ell}_{V}(Q^2;\mu)$ given by
\begin{equation}
    \begin{aligned}\label{eq:radiative_correction}
        \tilde{c}^{i,\nu_\ell}_V 
        &= c^i_V + \frac{\alpha}{\pi} Q_i \left[ \delta^{\nu_\ell} + \delta^{\mathrm{QCD}} \right],\\
        \delta^{\nu_\ell} 
        &= c_{V}^{\nu_{\ell} e} \mathrm{\Pi} \left( Q^2,m_e; \mu \right) + c_{V}^{\nu_{\ell} \mu} \mathrm{\Pi} \left( Q^2,m_\mu; \mu \right)+ c_{V}^{\nu_{\ell} \tau} \mathrm{\Pi} \left( Q^2,m_\tau; \mu \right),\\
        \delta^{\mathrm{QCD}}
        &= \hat{\Pi}_{3 \gamma}^{(3)}(Q^2; \mu) - 2 \sin^2 \theta_\mathrm{W} \hat{\Pi}_{\gamma \gamma}^{(3)}(Q^2; \mu) - N_c Q_c c_{V}^{c} \mathrm{\Pi} \left( Q^2, m_c; \mu \right),
    \end{aligned}
\end{equation}
where $Q_c=2/3$ is the electric charge of the charm quark, $N_c = 3$ is the number of colors in QCD, $\hat{\Pi}_{3 \gamma}^{(3)}$ ($\hat{\Pi}_{\gamma \gamma}^{(3)}$) is the renormalized nonperturbative current-current charge-isospin (charge-charge) correlator in the theory with $n_f=3$ quark flavors,\footnote{$\hat{\Pi}_{3 \gamma}^{(3)}$ and $\hat{\Pi}_{\gamma \gamma}^{(3)}$ in this paper enter with a factor $2$ compared to~\cite{Tomalak:2025tls} and with an opposite sign compared to~\cite{Tomalak:2019ibg,Hill:2019xqk,Tomalak:2020zfh,Tomalak:2021lif}.} and $\Pi \left( Q^2, m_f; \mu \right)$ is the vacuum polarization function (\ref{QED-VP-function}) for the perturbative treatment of loops involving charged leptons or heavy quarks [i.e. the quark mass $m_q \gg \Lambda_{\rm QCD}\approx (0.2-0.3)~\text{GeV}$]. Starting from the $n_f=4$ flavor theory, we treat the charm quark as a heavy quark, while the up, down, and strange quarks as light quarks contributing to the nonperturbative correlation functions $\hat{\Pi}_{\gamma \gamma}^{(3)}$ and $\hat{\Pi}_{3 \gamma}^{(3)}$. We discuss and present these inputs in the following subsections. Being $Q^2$-dependent, $\tilde{c}^{i,\nu_\ell}_V$ are not proper Wilson coefficients and also depend on the (anti)neutrino flavor.

The (anti)neutrino-quark vector coupling constants are modified with corrections proportional to the quark electric charges, resulting in shifts of the tree-level vector FFs $G_{E,M,S}$ in~(\ref{eq:weak-NC-FFs-EFT}) with corrections proportional to the electromagnetic FFs $G^{\gamma}_{E,M,S}$:
\begin{equation}\label{eq:weak-NC-FFs-EFT_closed_loop}
	G_{E,M,S}(Q^2) \to G_{E,M,S} \left( Q^2; \mu \right) + \frac{\alpha}{\pi} \left[ \delta^{\nu_\ell} \left( Q^2; \mu \right) + \delta^{\mathrm{QCD}} \left( Q^2; \mu \right) \right] G^{\gamma}_{E,M,S}(Q^2),
\end{equation}
where the scale dependence of $\delta^{\nu_\ell}$ and $ \delta^\mathrm{QCD}$ is exactly cancelled by the corresponding scale dependence of the Wilson coefficients that enter $G^{\gamma}_{E,M,S}(Q^2)$. In contrast, tree-level axial-vector FFs $G_{A,P,T} \left( Q^2 \right)$ are not affected by closed fermion loops.

\subsection{Perturbative lepton and quark contributions}
\label{sec:closed_lepton_loops}

\begin{figure}[tb]
    \centering
    {\includegraphics[scale=0.50]{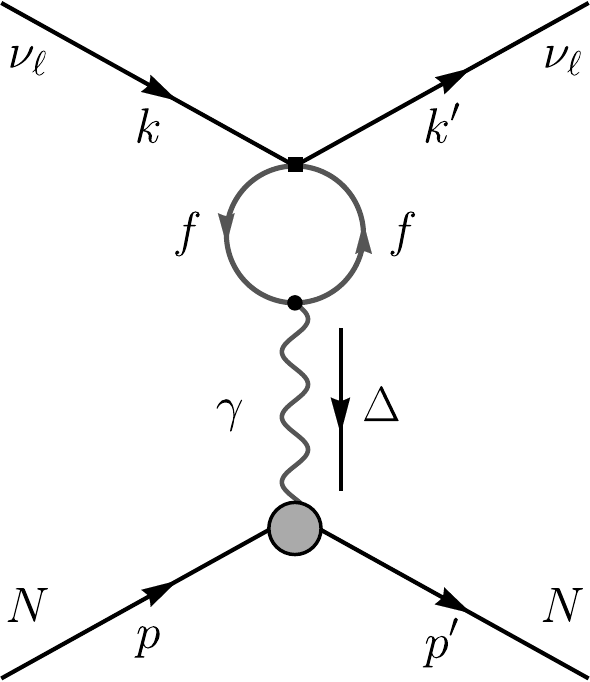}}
	\caption{Long-range dynamics in elastic neutrino-nucleon scattering in the LEFT. Loops with all interacting fields in the theory are summed up.}
    \label{one_loop_ET}
\end{figure}

\begin{figure}[tb!]
	\centering
	{\includegraphics[angle=0,scale=0.50]{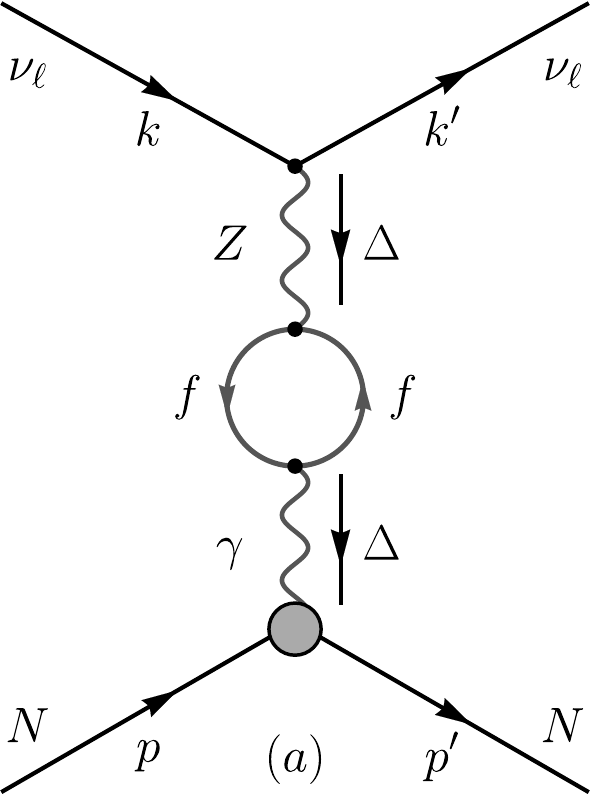}}
    \qquad\qquad\qquad
    {\includegraphics[angle=0,scale=0.50]{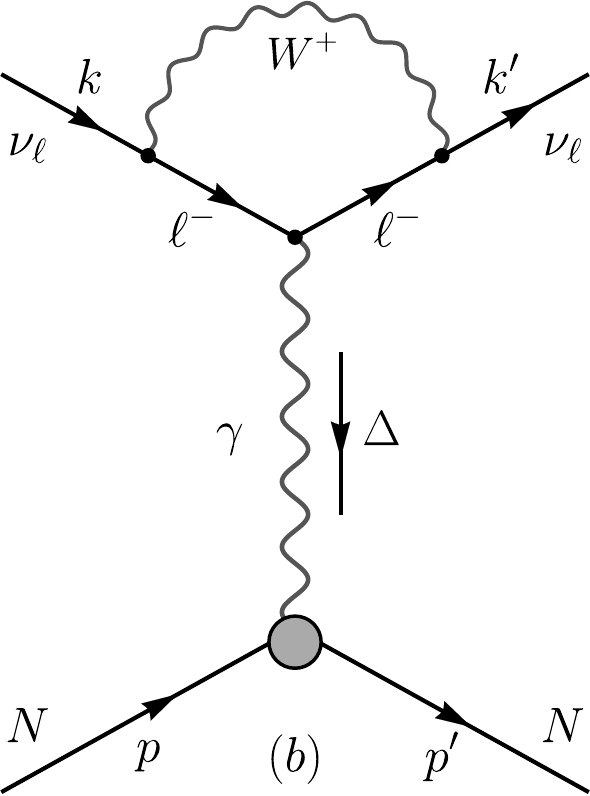}}
	\caption{The Feynman diagrams in the Standard Model that give rise to long-range dynamics in EFT are shown: ({\em a}) the $\gamma$-$Z$ mixing diagram  and ({\em b}) the penguin-type diagram.}
	\label{Fig_FeynmanDiagram_closeloop}
\end{figure}

In the LEFT formalism, the corrections with a closed fermion loop are represented diagrammatically in figure~\ref{one_loop_ET}. These corrections correspond to the effects of $\gamma$-$Z$ mixing and diagrams of penguin type in the Standard Model (see figure~\ref{Fig_FeynmanDiagram_closeloop}). Such contributions introduce the kinematic dependence of electroweak radiative corrections at low energies; see e.g.~\cite{Sarantakos:1982bp,Bardin:1983yb}.

\begin{figure}[tb!]
	\centering
 	{\includegraphics[angle=0,scale=0.55]{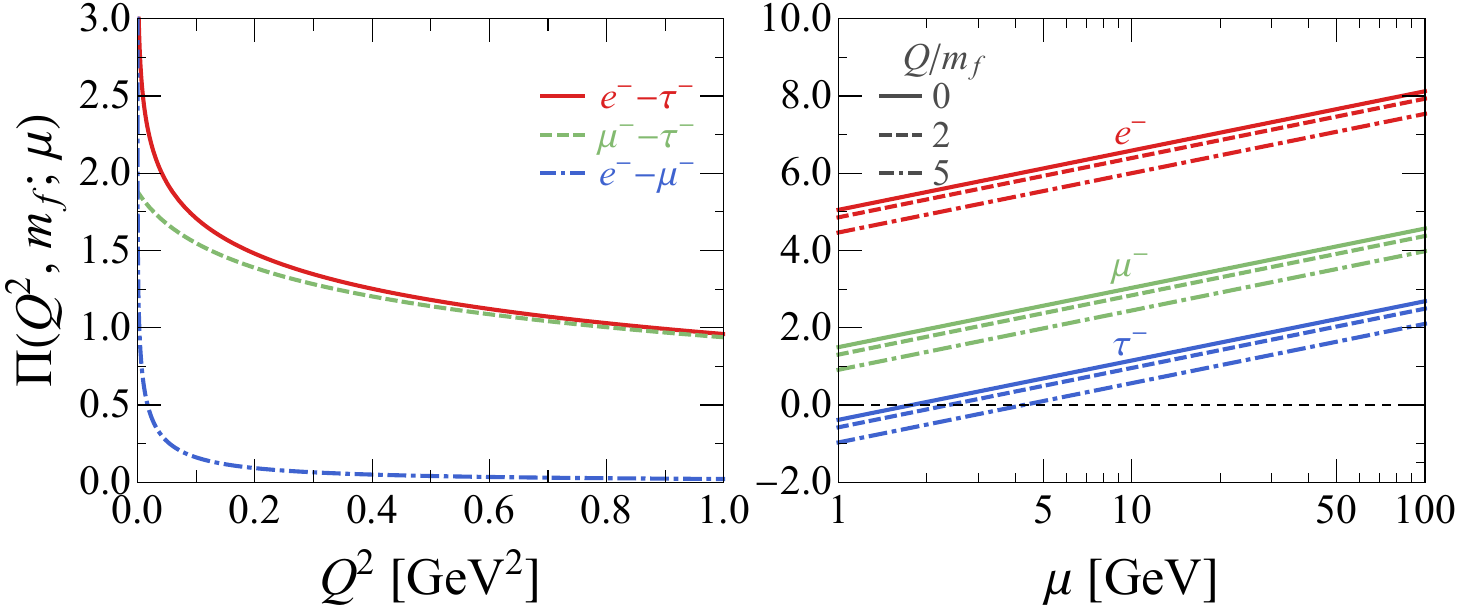}}
	\caption{The renormalization-scale independent charged lepton flavor differences of $\mathrm{\Pi}\left( Q^2,m_f;\mu \right)$ are shown as functions of $Q^2$ (left panel). $\mathrm{\Pi}\left( Q^2,m_f;\mu \right)$ in (\ref{QED-VP-function}) is shown as a function the renormalization scale $\mu$ for charged leptons at different values of $Q^2$ (right panel).}
	\label{Fig_PiDiff_leptons}
\end{figure}

At the $\overline{\mathrm{MS}}$ renormalization scale $\mu$, the QED vacuum polarization function at one-loop order is given by~\cite{Pauli:1936zz,Feynman:1949zx,Tsai:1960zz,Peskin:1995ev,Vanderhaeghen:2000ws,Heller:2019dyv}
\begin{equation}\label{QED-VP-function}
	\mathrm{\Pi}\left( Q^2,m_f;\mu \right) = \frac{1}{3} \bigg\{ \ln \frac{\mu^2}{m_f^2 } - \left[ (v^2 - \frac{8}{3} ) + \frac{v}{2} (3-v^2) \ln \frac{v+1}{v-1} \right] \bigg\}, 
\end{equation}
where $m_f$ is the mass of the fermion, $Q^2 = -\Delta^2$, and $v \equiv \sqrt{1 + 4m_f^2/Q^2 } > 1$. In particular, we find $\mathrm{\Pi}\left(0,m_f;\mu \right)=\frac{1}{3} \ln(\mu^2/m_f^2)$. The difference in $\mathrm{\Pi}\left( Q^2,m_f;\mu \right)$ for two particles of different masses does not depend on the scale $\mu$. In figure~\ref{Fig_PiDiff_leptons}, we show the $Q^2$ dependence of the $\mu$-independent flavor differences of $\mathrm{\Pi}\left( Q^2,m_f;\mu \right)$ between two charged leptons (left panel) and the $\mu$-dependence of $\mathrm{\Pi}\left( Q^2,m_f;\mu \right)$ at different $Q^2$ values for individual charged leptons (right panel). The difference in $\mathrm{\Pi}\left( Q^2,m_f;\mu \right)$ between the electron and tau is larger than that between the muon and tau, and the three curves will eventually get close to each other at large $Q^2$ above the scale of charged lepton masses, since $m_e<m_\mu<m_\tau$ and $\mathrm{\Pi}\left( Q^2,m_f;\mu \right)-\mathrm{\Pi}\left( Q^2,m_{f'};\mu \right) \approx 2(m_{f'}^2-m_f^2)/Q^2 + \mathcal O(1/Q^{4}) $ at large $Q^2$.

\subsection{Non-perturbative light-quark contributions}
\label{sec:closed_hadron_loops}

To evaluate non-perturbative hadronic contributions, we take the charge-charge correlation function at $Q^2=0$, i.e. $\hat{\Pi}_{\gamma \gamma}^{(3)} \left( 0; \mu = 2~\mathrm{GeV} \right) = 3.597(21)$, from the data-driven analysis of~\cite{Erler:1998sy,Erler:2004in,Erler:2017knj} and determine its $Q^2$ dependence in the theory with three quark flavors using the {\em alphaQED2025} package~\cite{Jegerlehner:2011mw}. At low momentum transfers (e.g. $Q^2 \lesssim m^2_\rho$), the charge-isospin correlation function can be estimated from the precise relation of $\text{SU}(2)$ chiral perturbation theory (ChPT): $\hat{\Pi}_{3 \gamma}^{(3)} = \hat{\Pi}_{\gamma \gamma}^{(3)}$~\cite{Tomalak:2025tls}. As the momentum transfer increases, contributions from the strange quarks become important. Hence, for three (and even more) quark flavors, $\hat{\Pi}_{3 \gamma}^{(3)}$ and $\hat{\Pi}_{\gamma \gamma}^{(3)}$ are, in general, different.

\begin{figure}[tb!]
	\centering
 	{\includegraphics[angle=0,scale=0.52]{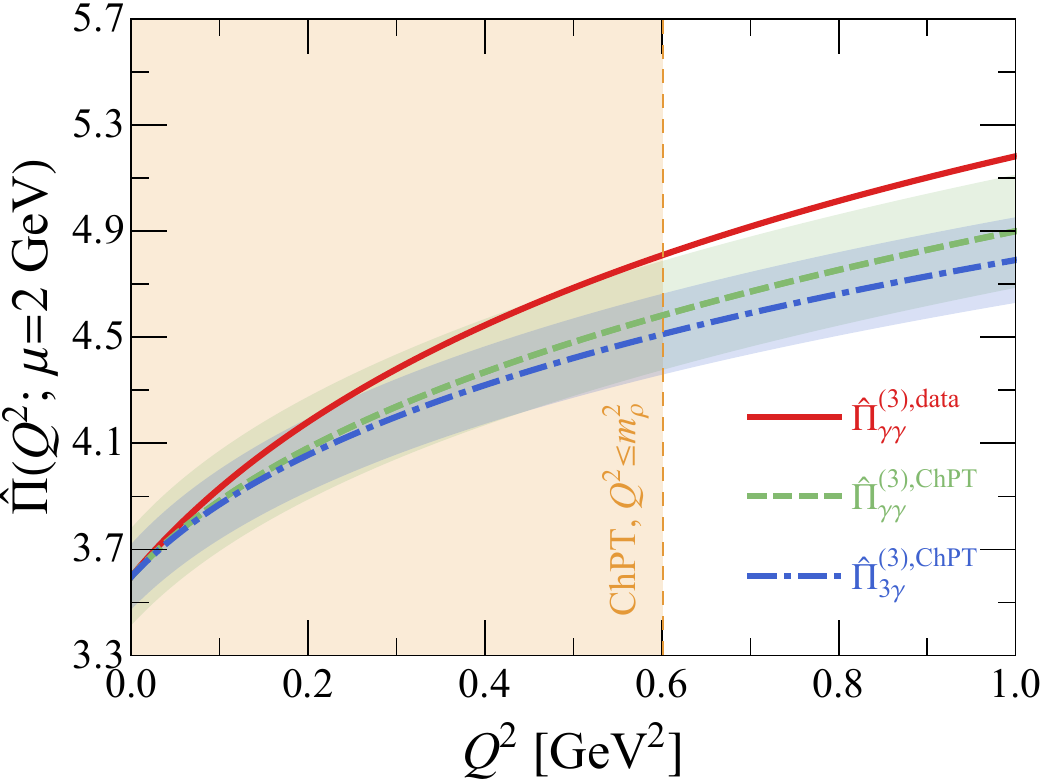}}
	\caption{The data-driven hadronic vacuum polarization function $\hat{\Pi}_{\gamma \gamma}^{(3),\text{data}}(Q^2)$ at the renormalization scale $\mu = 2~\mathrm{GeV}$ is compared with the corresponding one-loop $\text{SU}(3)$ ChPT results $\hat{\Pi}_{\gamma \gamma}^{(3),\text{ChPT}}(Q^2)$ and $\hat{\Pi}_{3 \gamma}^{(3),\text{ChPT}}(Q^2)$. The theoretical uncertainties of $\hat{\Pi}_{\gamma \gamma}^{(3),\text{ChPT}}$ and $\hat{\Pi}_{3 \gamma}^{(3),\text{ChPT}}$ are indicated by the shaded bands. The $Q^2$ validity range of ChPT is also shown, with the upper bound specified by the mass $m_\rho$ of the $\rho(770)$ meson.}
	\label{fig:Pi_light_quarks}
\end{figure}

In figure~\ref{fig:Pi_light_quarks}, we compare the data-driven result of $\hat{\Pi}_{3 \gamma}^{(3),\text{data}} =\hat{\Pi}_{\gamma \gamma}^{(3),\text{data}}$, which is based on the $\text{SU}(2)$ ChPT ansatz $\hat{\Pi}_{3 \gamma}^{(3)} = \hat{\Pi}_{\gamma \gamma}^{(3)}$, with the one-loop $\text{SU}(3)$ ChPT results of $\hat{\Pi}_{\gamma \gamma}^{(3),\text{ChPT}}$ and $\hat{\Pi}_{3 \gamma}^{(3),\text{ChPT}}$~\cite{Tomalak:2025tls}. By comparing $\hat{\Pi}_{\gamma \gamma}^{(3),\text{ChPT}}$ and $\hat{\Pi}_{3 \gamma}^{(3),\text{ChPT}}$, we find that the $\text{SU}(2)$ ChPT ansatz works reasonably well, within uncertainty estimates, in the $\text{SU}(3)$ case, even outside the validity range of ChPT (i.e. $ m^2_\rho \lesssim Q^2 \lesssim 1.0~\mathrm{GeV}^2$). In addition, we observe that the data-driven result $\hat{\Pi}_{\gamma \gamma}^{(3),\text{data}}$~\cite{Jegerlehner:2011mw} is also within the uncertainty of the one-loop $\text{SU}(3)$ ChPT calculation $\hat{\Pi}_{\gamma \gamma}^{(3),\text{ChPT}}$ for $Q^2 \leq m_\rho^2$. Based on the low-$Q^2$ $\text{SU}(2)$ ChPT relation, we take the data-driven $\hat{\Pi}_{\gamma \gamma}^{(3),\text{data}}$ as the central value of $\hat{\Pi}_{3 \gamma}^{(3)}$, and estimate its relative uncertainty as twice the relative difference between $\hat{\Pi}_{\gamma \gamma}^{(3),\text{ChPT}}$ and $\hat{\Pi}_{3 \gamma}^{(3),\text{ChPT}}$ obtained in one-loop $\text{SU}(3)$ ChPT. For conservative error estimates of $\hat{\Pi}_{3 \gamma}^{(3)}$, we choose its relative error as the theoretical error of $\hat{\Pi}_{\gamma \gamma}^{(3),\text{ChPT}}$. At higher momentum transfers, we consider the contribution from each light quark perturbatively substituting the $\overline{\text{MS}}$ masses to describe the dependence on $Q^2$ and include $\mathcal O(\alpha \alpha_s)$ corrections according to~\cite{Djouadi:1987gn,Djouadi:1987di,Kniehl:1989yc,Fanchiotti:1992tu,Chetyrkin:1996ela,Chetyrkin:1996cf,Chetyrkin:1997un,Erler:1998sy,Tomalak:2019ibg,ParticleDataGroup:2024cfk}, where $\alpha_s \equiv g_s^2/(4\pi)$ is the running strong coupling constant. We match the dependence on the squared momentum transfer of the non-perturbative and perturbative functions $\delta^\mathrm{QCD}$ at $Q_0^2 = 1~\mathrm{GeV}^2$, and study the variation of $Q_0^2$ for $m_\rho^2 \le Q_0^2 \le Q_\mathrm{max}^2 = 2~\mathrm{GeV}^2$ to conservatively estimate the uncertainty $\delta \left[ \delta^{\mathrm{QCD}} \right]$ as 
\begin{align} \label{eq:deltaQCDerror}
	&\left( \frac{\delta^{\mathrm{QCD}} \left( Q^2, Q^2_0 = Q_\mathrm{max}^2 \Theta \left( Q^2 - Q_\mathrm{max}^2 \right) + Q^2 \Theta \left( Q_\mathrm{max}^2 - Q^2 \right) \right) - \delta^{\mathrm{QCD}} \left( Q^2, Q^2_0 = m^2_\rho \right)}{2}\right)^2 \nonumber \\
	&\times \Theta \left( Q^2 - m^2_\rho \right) + \left[ \left( \delta \hat{\Pi}_{\gamma \gamma}^{(3)}(0) \right)^2 + \left( \delta \hat{\Pi}_{\gamma \gamma}^{(3),\text{ChPT}} \left(m^2_\rho \Theta \left( Q^2 - m^2_\rho \right) + Q^2 \Theta \left( m^2_\rho - Q^2\right)\right) \right)^2 \right] \nonumber \\
	&\times \left( 1 - \sin^2 \theta_W \right)^2 = \left( \delta \left[ \delta^{\mathrm{QCD}} \left( Q^2 \right) \right] \right)^2.
\end{align}
To account for large electroweak logarithms, we substitute $\sin^2 \theta_W \to \frac{c_A^u-c_V^u}{4Q_u} + \frac{c_A^d -c_V^d}{4Q_d} \approx 0.23281$ in $\delta^{\mathrm{QCD}}$ and neglect the uncertainty of $\mathcal{O} \left( \alpha \right)$ in this determination, which results in a much smaller uncertainty than our conservative estimate of hadronic errors above.

\begin{figure}[tb!]
	\centering
 	{\includegraphics[angle=0,scale=0.52]{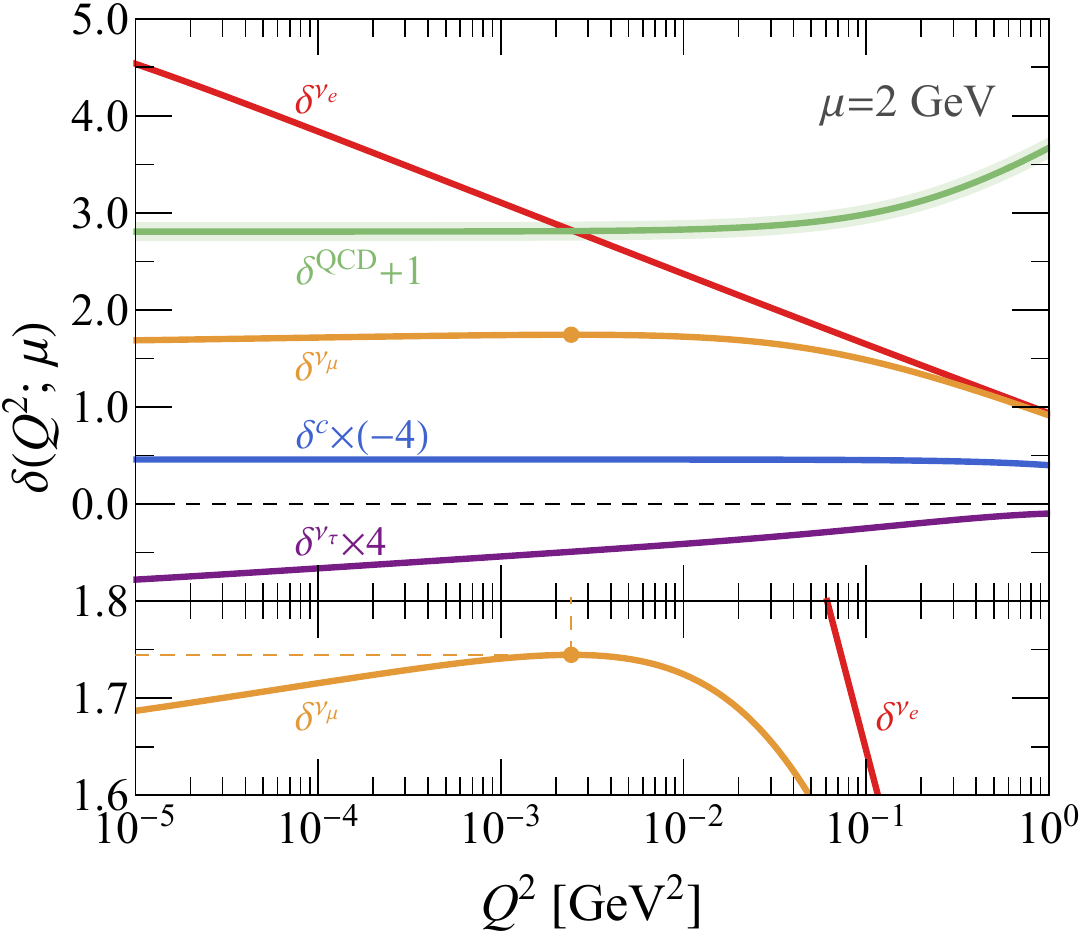}}
	\caption{Contributions from radiative corrections $\delta^{\nu_e}(Q^2;\mu)$, $\delta^{\nu_\mu}(Q^2;\mu)$, $\delta^{\nu_\tau}(Q^2;\mu)$, $\delta^{\text{QCD}}(Q^2;\mu)$, and $\delta^c(Q^2;\mu)$ in (\ref{eq:radiative_correction}) are shown as functions of $Q^2$ at the renormalization scale $\mu=2~\text{GeV}$. The dominant uncertainty from light-quark loops is also presented as the error band.}
	\label{Fig_deltaFuns}
\end{figure}

\begin{figure}[tb!]
	\centering
 	{\includegraphics[angle=0,scale=0.55]{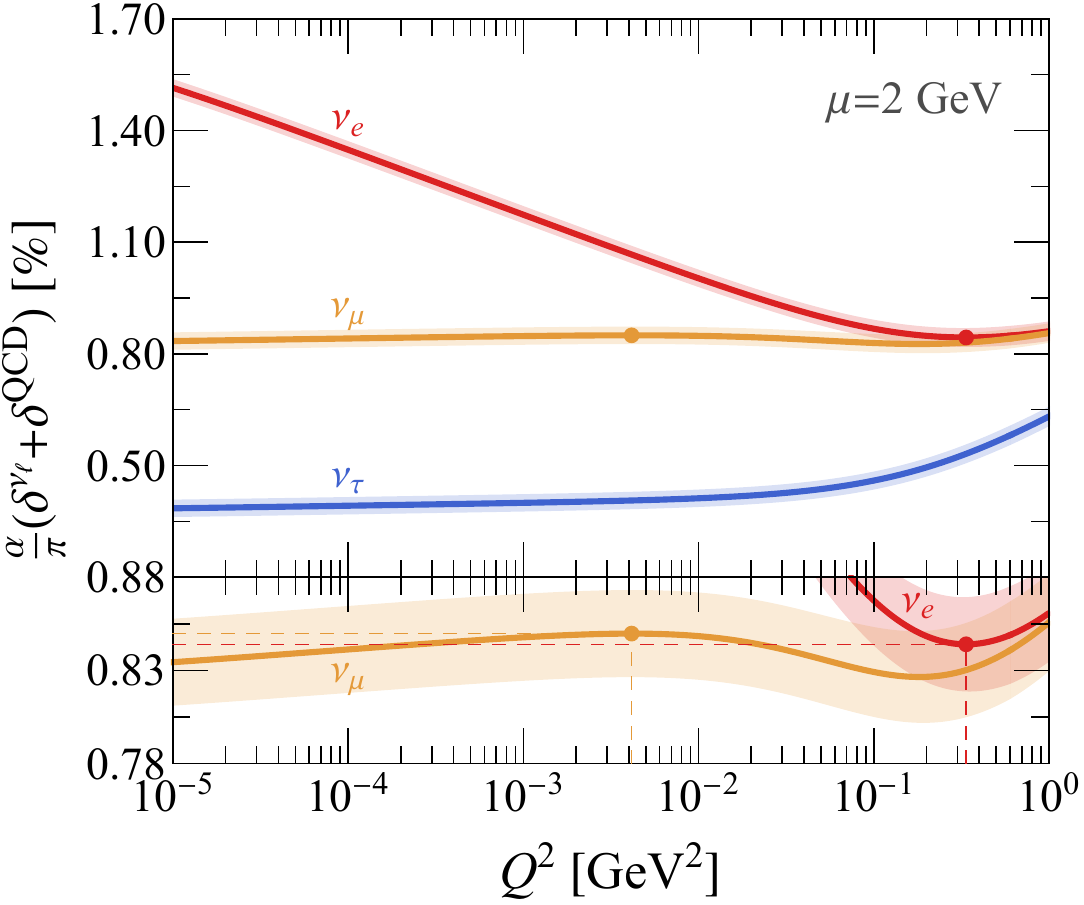}\quad\,}
	\caption{The total closed fermion-loop contributions to $\frac{\alpha}{\pi} \left[ \delta^{\nu_\ell}\left( Q^2; \mu \right) + \delta^{\mathrm{QCD}}\left( Q^2; \mu \right) \right]$ from (\ref{eq:radiative_correction}) at the renormalization scale $\mu = 2~\mathrm{GeV}$, with uncertainties, are shown as functions of $Q^2$ for the electron-, muon-, and tau-flavor (anti)neutrino beams.}
	\label{fig:closed_loop_contributions}
\end{figure}

To get an idea of the relative size of various radiative corrections, we present in figure~\ref{Fig_deltaFuns} the individual contributions to $ \delta^{\nu_\ell}\left( Q^2; \mu \right) + \delta^{\mathrm{QCD}}\left( Q^2; \mu \right)$ of (\ref{eq:radiative_correction}) at the renormalization scale $\mu = 2~\mathrm{GeV}$ from the charm quark, light quarks, and the resulting leptonic corrections for the electron, muon, and tau (anti)neutrino beams. For numerical estimates, we take the Wilson coefficients of four-fermion interactions at the scale $\mu = 2~\mathrm{GeV}$ from the table~3 of~\cite{Hill:2019xqk} and the table~I of~\cite{Tomalak:2019ibg}, cf. also~\cite{Erler:2004cx,Erler:2013xha,Erler:2017knj}. At small momentum transfers, the closed fermion-loop correction from charged leptons dominates with an enhancement of the electron contribution by the electron mass for the electron flavor (anti)neutrinos. At larger momentum transfers, the contribution from light quarks is the most important. In figure~\ref{fig:closed_loop_contributions}, we compare the resulting closed fermion-loop contributions to $\frac{\alpha}{\pi} \left[ \delta^{\nu_\ell}\left( Q^2; \mu \right) + \delta^{\mathrm{QCD}}\left( Q^2; \mu \right) \right]$ from (\ref{eq:radiative_correction}) at the renormalization scale $\mu = 2~\mathrm{GeV}$ for the electron, muon, and tau (anti)neutrino beams. The radiative correction is the largest for the electron (anti)neutrinos, smaller for the muon (anti)neutrinos, and the smallest for the tau (anti)neutrinos.

\section{Nucleon form factors and bremsstrahlung}
\label{sec:nucleon_ffs_and_brems}

QED interactions with the nucleon in NC (anti)neutrino-nucleon elastic scattering are significantly different in the neutron and proton cases. The radiation of a real photon and the vertex correction for the neutron are significantly suppressed compared to that for the proton (which has a nonzero electric charge $Z=1$). In this section, we will formulate the QED radiative corrections on the nucleon side in the factorization framework, including the resummation of large perturbative logarithms, and compare them with the traditional approaches to radiative corrections in elastic electron-proton scattering.

In this paper, we consider two experimental observables. First, we define an exclusive cross section including bremsstrahlung with total undetected energy in the radiation below a detector-specific cutoff $\Delta E$. Second, we study an inclusive observable that includes photons of arbitrary energy and direction. For the sufficiently inclusive second observable that is  free from soft and collinear enhancements~\cite{Bloch:1937pw,Nakanishi:1958ur,Kinoshita:1962ur,Lee:1964is}, we do not consider radiative corrections on the proton line, since these contributions are typically small.

\subsection{Factorization of QED radiative corrections}
\label{subsec:factorization}

Taking into account QED interactions, the proton FFs in (\ref{eq:weak-NC-matrix-elements-EFT}) and~(\ref{elements-NCO-proton}) become IR divergent. These FFs can be factorized into a product of the infrared-finite Born FFs $G_i^{\mathrm{Born}, \left( p \right)} \left( Q^2; \mu \right)$ and $F_i^{\mathrm{Born}, \left( p \right)} \left( Q^2; \mu \right)$, and the universal soft FF $F^{\left( p \right)}_\text{soft} \left( Q^2, \mu \right)$ that describes virtual interactions with soft photons and is independent of the nucleon structure,
\begin{equation}
	\begin{aligned}\label{eq:QED_virtual_correction}
		G^{\left( p \right)}_i \left( Q^2 \right) 
        &= G_i^{\mathrm{Born}, \left( p \right)} \left( Q^2; \mu \right) F^{\left( p \right)}_\text{soft} \left( Q^2; \mu \right),\\
		F^{\left( p \right)}_i \left( Q^2 \right) 
        &= F_i^{\mathrm{Born}, \left( p \right)} \left( Q^2; \mu \right) F^{\left( p \right)}_\text{soft} \left( Q^2; \mu \right).
	\end{aligned}
\end{equation}
Since the soft virtual QED correction and corresponding QED field renormalization factors for electrically charged particles are UV divergent, we emphasize the resulting scale dependence of the soft and Born FFs.

Cross sections with radiation of soft real photons of energy $E_\gamma$ below the detection threshold (or cutoff) $\Delta E$ are measured together with the elastic scattering process. These contributions cancel the infrared divergence of virtual corrections, e.g. the dependence on the fictitious photon mass $\lambda_\gamma$ for the photon-mass regularization. Such soft bremsstrahlung is conveniently included in the soft contribution at the cross-section level, where the unpolarized differential cross section $\mathrm{d} \sigma$ is factorized into a hard $H^{\left( p \right)}$ and a soft $S^{\left( p \right)}$ functions:
\begin{equation}
	\begin{aligned}\label{eq:factorization}
		\mathrm{d} \sigma^{\left( p \right)} \sim H^{\left( p \right)} \left( \frac{Q^2}{M^2}; \frac{\mu}{M} \right) S^{\left( p \right)} \left( \frac{p^0}{M}, \frac{p'^0}{M}, \frac{Q^2}{M^2}; \kappa \right),
	\end{aligned}
\end{equation}
with $\kappa \equiv \mu/\Delta E$ and $\mu$ being the renormalization scale. Hence, the soft function contributes to the invariant amplitudes in the NC (anti)neutrino-nucleon scattering process as
\begin{equation}
	\begin{aligned}\label{eq:QED_soft_correction}
		g^{\left( p \right) }_i \left( \frac{p^0}{M}, \frac{p'^0}{M}, \frac{Q^2}{M^2}, \frac{\Delta E}{M} \right) 
        &= G_i^{\mathrm{Born}, \left( p \right)} \left( Q^2; \mu \right) \sqrt{S^{\left(p \right)} \left( \frac{p^0}{M}, \frac{p'^0}{M}, \frac{Q^2}{M^2}; \kappa \right)},\\
		f^{\left( p \right) }_i \left( \frac{p^0}{M}, \frac{p'^0}{M}, \frac{Q^2}{M^2}, \frac{\Delta E}{M} \right) 
        &= F_i^{\mathrm{Born}, \left( p \right)} \left( Q^2; \mu \right) \sqrt{S^{\left(p \right)} \left( \frac{p^0}{M}, \frac{p'^0}{M}, \frac{Q^2}{M^2}; \kappa \right)}.
	\end{aligned}
\end{equation}

For a heavy proton, the soft FF at the one-loop order is given by~\cite{Hill:2016gdf}
\begin{equation}
	\begin{aligned}\label{eq:one_loop_soft_form_factor}
		 F^{\left( p \right) }_\text{soft} \left( Q^2; \mu \right) = 1 - \frac{ \alpha}{2 \pi} \left[ \frac{w}{\sqrt{w^2 - 1}} \ln \left( w + \sqrt{w^2 - 1} \right) - 1 \right] \ln \frac{\mu^2}{\lambda^2_\gamma} + \mathcal{O} \left( \frac{\alpha^2}{\pi^2} \right),
	\end{aligned}
\end{equation}
with $w \equiv 1 + \frac{Q^2}{2M^2}$. At $\mathcal{O} \left( \alpha \right)$, the soft function $S^{\left(p \right)} \left( \frac{p^0}{M}, \frac{p'^0}{M}, \frac{Q^2}{M^2}; \kappa \right)$ can be expressed as
\begin{equation}
	\begin{aligned}\label{eq:soft_function_at_NLO}
		 S^{\left(p \right)} \left( \frac{p^0}{M}, \frac{p'^0}{M}, \frac{Q^2}{M^2}; \kappa \right) 
         &= 1 - \frac{\alpha}{\pi} \left[ \frac{w \ln w_+}{\sqrt{w^2 - 1}} - 1 \right] \ln \frac{\kappa^2}{4}+ \frac{\alpha}{\pi} G \left(w,\frac{p^0}{M},\frac{\left(p^\prime \right)^0}{M}\right) + \mathcal{O} \left( \frac{\alpha^2}{\pi^2} \right),
	\end{aligned}
\end{equation}
where the function $G$ is given by~\cite{tHooft:1978jhc,Hill:2016gdf,Tomalak:2022xup}
\begin{equation}
	\begin{aligned}
        G \left( w, x, y \right)
        &= \frac{x }{ \sqrt{ x^2-1} } \ln{x_+} + \frac{y }{\sqrt{ y^2 -1}} \ln{y_+} + \frac{w}{ \sqrt{w^2-1}} \bigg[ \ln^2 x_+ -\ln^2 y_+ \\
        &\quad + {\rm Li}_2\left( 1 - \frac{ x_+ }{ \sqrt{w^2-1}} ( w_+ x - y ) \right) + {\rm Li}_2\left( 1 - \frac{ x_- }{ \sqrt{w^2-1}} ( w_+ x - y ) \right)\\
        &\quad - {\rm Li}_2\left( 1 - \frac{ y_+ }{ \sqrt{w^2-1}} ( x - w_- y ) \right) - {\rm Li}_2\left( 1 - \frac{ y_- }{ \sqrt{w^2-1}} ( x - w_- y ) \right) \bigg],
	\end{aligned}
\end{equation}
with $a_\pm \equiv a \pm \sqrt{a^2-1}$. For the initial proton at rest, the soft function can be conveniently expressed in terms of the recoil proton speed $\beta$ in the laboratory frame as
\begin{equation}
	\begin{aligned}\label{eq:soft_function_at_NLO_nucleon_at_rest}
        S^{(p)} \left( \beta; \kappa \right) 
        &= 1 + \frac{\alpha}{\pi} \bigg[ 1 +2 \left(1 - \frac{1}{2\beta} \ln \xi_\beta \right) \ln \frac{\kappa}{2} + \frac{1}{ \beta}\left( \mathrm{Li}_2 \frac{1}{\xi_\beta} - \frac{\pi^2}{6} \right)\\
        &+ \frac{1}{2 \beta} \left( 1 + \ln \frac{\sqrt{1 - \beta^2} \left( 1+ \beta \right)}{4 \beta^2} \right)\ln \xi_\beta \bigg],
	\end{aligned}
\end{equation}
with $\xi_\beta \equiv (1+\beta)/(1-\beta)$. In particular, when $\beta \to 0$, we have $S^{(p)} \left( 0; \kappa \right) = 1$.

\subsection{QED vertex correction and soft bremsstrahlung}
\label{subsec:traditional}

It is instructive to compare the above soft function at different renormalization scales $\mu$ to the structure-independent QED radiative correction $[1+ Z^2\delta_2^{\left( 0 \right)} ]$, which comes from the proton vertex combined with the one-photon soft bremsstrahlung on the proton line in electron-proton elastic scattering~\cite{Vanderhaeghen:2000ws,Maximon:2000hm}, with $\delta_2^{\left( 0 \right)}$ given by
\begin{equation}
	\begin{aligned}\label{eq:delta2_from_papers}
		\delta_2^{(0)}(\beta, \tilde \kappa) 
        &= \frac{\alpha }{\pi} \bigg\{ 2\left( 1 - \frac{1}{2\beta} \ln \xi_\beta \right) \ln \frac{\tilde \kappa}{2} + \frac{1}{2\beta} \ln \xi_\beta \left[ 1 - \ln\left(2 \zeta_\beta \right) - \frac{1}{4} \ln \xi_\beta \right]\\
		&+ \frac{1}{\beta} \left[ \frac{\pi^2 }{6} + 2\text{Li}_2\left( -\sqrt{ \frac{1}{\xi_\beta} } \right) - \text{Li}_2\left( \frac{2\beta}{1+\beta} \right) \right] \bigg\} = \delta_2^{(0)}\left( \beta, \Delta E \right),
	\end{aligned}
\end{equation}
where $\zeta_\beta \equiv 1 + 1/\sqrt{1-\beta^2}$, $\tilde \kappa \equiv M /\Delta E_s $, $\Delta E_s = \Delta E/f_\text{recoil}$~\cite{Vanderhaeghen:2000ws,Maximon:2000hm}, and $f_\text{recoil}=E_\nu'/E_\nu=1-\frac{Q^2}{2ME_\nu}$ is the recoil factor in the laboratory frame~\cite{Chen:2023dxp}. This correction vanishes in forward scattering, e.g. $\delta_2^{(0)}(0, \Delta E) = 0$ for $\beta \to 0$.

\begin{figure}[tb!]
	\centering
 	{\includegraphics[angle=0,scale=0.57]{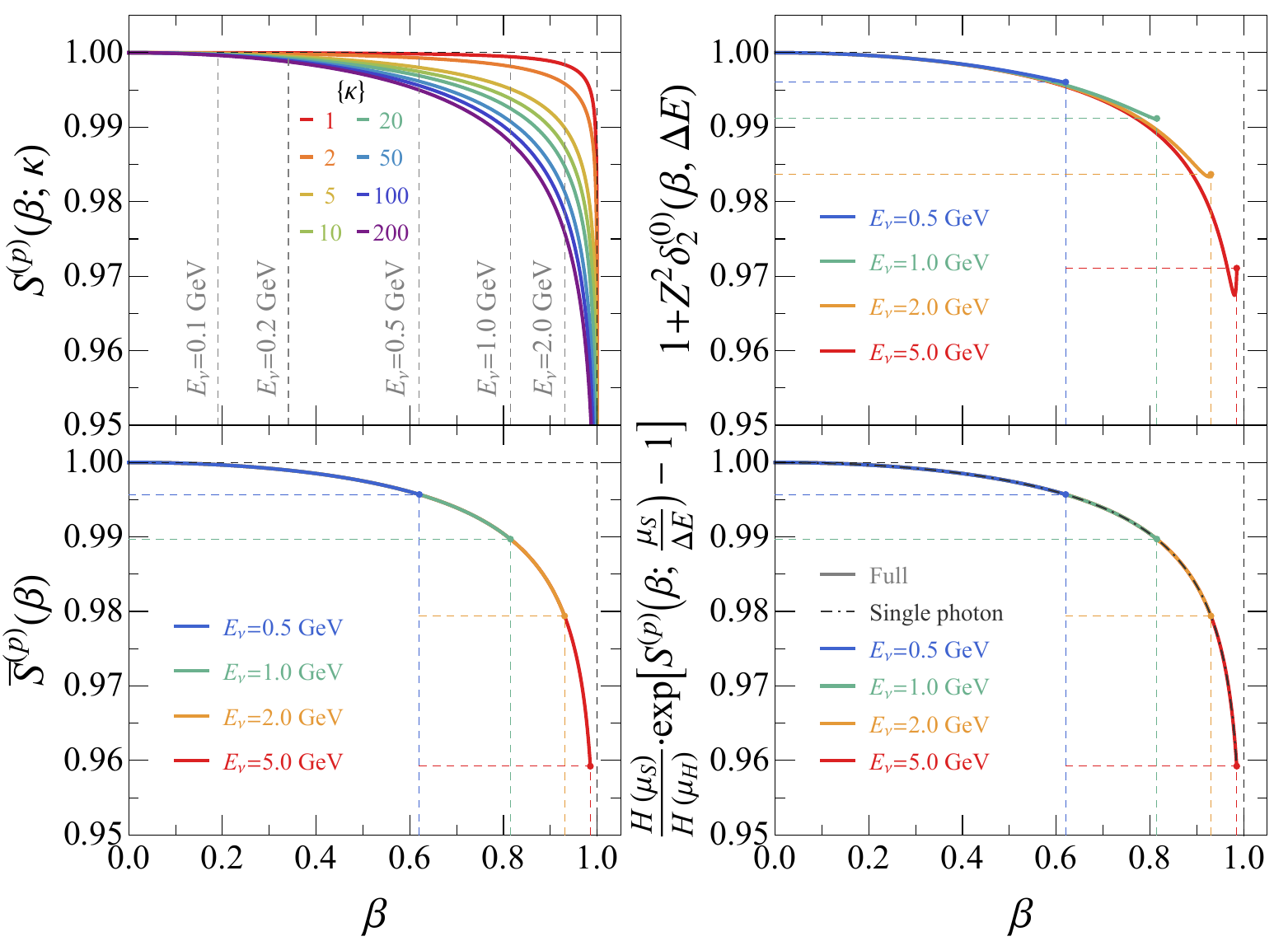}}
	\caption{The soft function $S^{(p)}(\beta;\kappa)$ from (\ref{eq:soft_function_at_NLO_nucleon_at_rest}), $ 1 +  Z^2 \delta_2^{(0)}\left( \beta, \Delta E \right)$ from (\ref{eq:delta2_from_papers}), $\overline{S}^{(p)}(\beta)$ from (\ref{eq:soft_function_at_NLO_nucleon_at_rest}), and $\frac{H \left( \mu_S \right)}{H \left( \mu_H \right)} e^{S^{(p)} \left( \beta;\frac{\mu_S}{\Delta E} \right) - 1}$ for the proton are presented on the upper left, upper right, lower left, and lower right panels, respectively, as functions of the recoil speed $\beta$ at different (anti)neutrino energies. The curves are shown at different values of $\kappa \in\{1,2,5,10,20,50,100,200\}$ and $\mu_S = \Delta E = 5~\text{MeV}$ in the laboratory frame. The vertical dashed lines indicate the maximal values of $\beta$ at fixed $E_\nu$.}
	\label{Fig_SpFun7}
\end{figure}

In figure~\ref{Fig_SpFun7}, we show the $\beta$-dependence of the soft function $S^{(p)}(\beta;\kappa)$ at different values of $\kappa$ in the laboratory frame. We indicate the maximal values of the proton recoil speed $\beta$ at fixed $E_\nu$ by the vertical dashed lines.\footnote{For $\nu_\ell N \to \nu_\ell N$ at fixed $E_\nu$ in the laboratory frame, $Q^2$ and $\beta$ satisfy $0 \leq Q^2 \leq \frac{ 4M E_\nu^2 }{M + 2E_\nu } $ and $0\leq \beta \leq \beta_\text{max}$, with $\beta_\text{max}=\sqrt{1-\left(1+\frac{2E_\nu^2}{M^2+2M E_\nu}\right)^{-2} }$.} We also present the $\beta$-dependence of $\big[ 1+ Z^2 \delta_2^{(0)}\left( \beta, \Delta E \right) \big]$ at different incident (anti)neutrino energies $E_{\nu}$ in the laboratory frame, where we take the photon energy detection threshold $\Delta E$ as $\Delta E = 5~\text{MeV}$. We find a very good approximate relation between $S^{\left(p \right)}$ and $[1 + Z^2 \delta_2^{\left( 0 \right)}]$, namely,
\begin{equation}
	\begin{aligned}\label{eq:soft_function_and_delta2}
		 S^{\left(p \right)} \left( \beta; \lambda \tilde \kappa \right) \approx 1 + Z^2 \delta_2^{\left( 0 \right)}\left( \beta, \tilde \kappa \right),
	\end{aligned}
\end{equation}
with $\frac{5}{4} \leq \lambda \leq \frac{3}{2}$, depending on the kinematic range to cover.\footnote{For high-energy scattering (i.e. $\beta_\text{max} \approx 1$), $\lambda \approx 10/7$ is a good choice over the whole kinematic range.} The correction for the vector Wilson coefficients in the soft function at $\mu=1~\text{GeV}$ and $E_\nu=2~\text{GeV}$ can be as large as $2\%$.

\subsection{Resummation}
\label{subsec:resummation}

To account for the bremsstrahlung of multiple photons with the sum of their energies below the cutoff $\Delta E$, the expression for the radiation of a single photon can be exponentiated~\cite{Yennie:1961ad}, i.e. $1 + Z^2 \delta_2^{(0)}\left( \beta, \Delta E \right)\to e^{ Z^2 \delta_2^{(0)}\left( \beta,\Delta E \right)}$, within the framework of~\cite{Vanderhaeghen:2000ws,Maximon:2000hm}.

In the factorization framework, we can also resum large perturbative logarithms. First, we express the hard function at the hard scale $\mu_H = M$ in terms of Born FFs. The renormalization group equation for the hard function at the first two orders in $\alpha$ is explicitly given by~\cite{Korchemsky:1987wg,Kilian:1992tj,Hill:2016gdf}
\begin{equation}\label{RGE-hShH}
    \frac{\mathrm{d}H \left( \mu \right)}{\mathrm{d} \ln \mu^2} = \frac{\alpha}{\pi} \left( \frac{w \ln w_+}{\sqrt{w^2 - 1}} - 1 \right) \left( \gamma_0 + \gamma_1 \frac{\alpha}{\pi} \right) H \left( \mu \right)\,,
\end{equation}
where $\gamma_0 = 1$ and $\gamma_1 = - \frac{5 n_f }{9}$ are anomalous dimensions, with the number of effective degrees of freedom in the effective field theory under consideration $n_f$, equivalent to the number of dynamical charged leptons. We resum large perturbative logarithms by evolving the hard function from the hard renormalization scale $\mu_H = M$ to the soft scale $\mu_S = \Delta E$. Subsequently, we evaluate the differential cross section $\mathrm{d} \sigma$ at the low energy scale $\mu_S$ as
\begin{equation}
    \mathrm{d} \sigma = H \left( \mu_S \right) S \left( \mu_S \right) = \frac{H \left( \mu_S \right)}{H \left( \mu_H \right)} H^\mathrm{exp} \left( \mu_H \right) S \left( \mu_S \right),
\end{equation}
where $H^\mathrm{exp} \left( \mu_H \right)$ represents the hard function expressed in terms of the experimentally-measured or theoretically-calculated FFs. The ratio $\frac{H \left( \mu_S \right)}{H \left( \mu_H \right)}$ can be determined analytically order by order in the electromagnetic coupling constant. As a result, the resummed factorizable correction can be expressed as
\begin{equation}\label{resummed-factorization-S}
    \overline{S}^{\left( p \right)} \left( \beta \right) = \frac{H \left( \mu_S \right)}{H \left( \mu_H \right)} S^{\left( p \right)} \left( \beta; \frac{\mu_S}{\Delta E} \right),
\end{equation}
with $\mu_H = M$ and $\mu_S = \Delta E$. To account for radiation of multiple photons with the sum of their energies below the cutoff $\Delta E$, the expression for the radiation of a single photon can be exponentiated~\cite{Yennie:1961ad}, i.e. $\overline{S}^{\left( p \right)} \to \frac{H \left( \mu_S \right)}{H \left( \mu_H \right)} e^{S^{(p)} \left( \beta;\frac{\mu_S}{\Delta E} \right) - 1} $. We find numerically that $\overline S^{(p)}(\beta)(\beta,\frac{\mu_S}{\Delta E})$ and $\frac{H(\mu_S)}{H (\mu_H)} e^{S^{(p)}(\beta,\frac{\mu_S}{\Delta E}) - 1}$ almost coincide with each other over a large kinematic range of $\beta$, see figure~\ref{Fig_SpFun7}, proving that we control large logarithms by evaluating the differential cross section at the low energy scale $\mu_S$.

For the neutron, the soft FF does not depend on the renormalization scale $\mu$ and is given by $F^{(n)}_S \left( Q^2; \mu \right) \equiv 1$ up to all orders. Likewise, the all-order expression for the soft function of the neutron in the weak NC (anti)neutrino-neutron elastic scattering process is $S^{\left(n \right)} \left( \frac{p^0}{M}, \frac{p'^0}{M}, \frac{Q^2}{M^2}; \kappa \right) \equiv 1$.

\section{Results and discussions}
\label{sec:results}

In this section, we present our results for weak NC (anti)neutrino-nucleon elastic scattering at $\text{GeV}$ energies and confront them with experimental data~\cite{Entenberg:1979wc,Horstkotte:1981ne,Ahrens:1986xe,MiniBooNE:2010xqw,MiniBooNE:2013dds,Pate:2024acz}.

\begin{figure}[th!]
	\centering
	{\includegraphics[angle=0,scale=0.50]{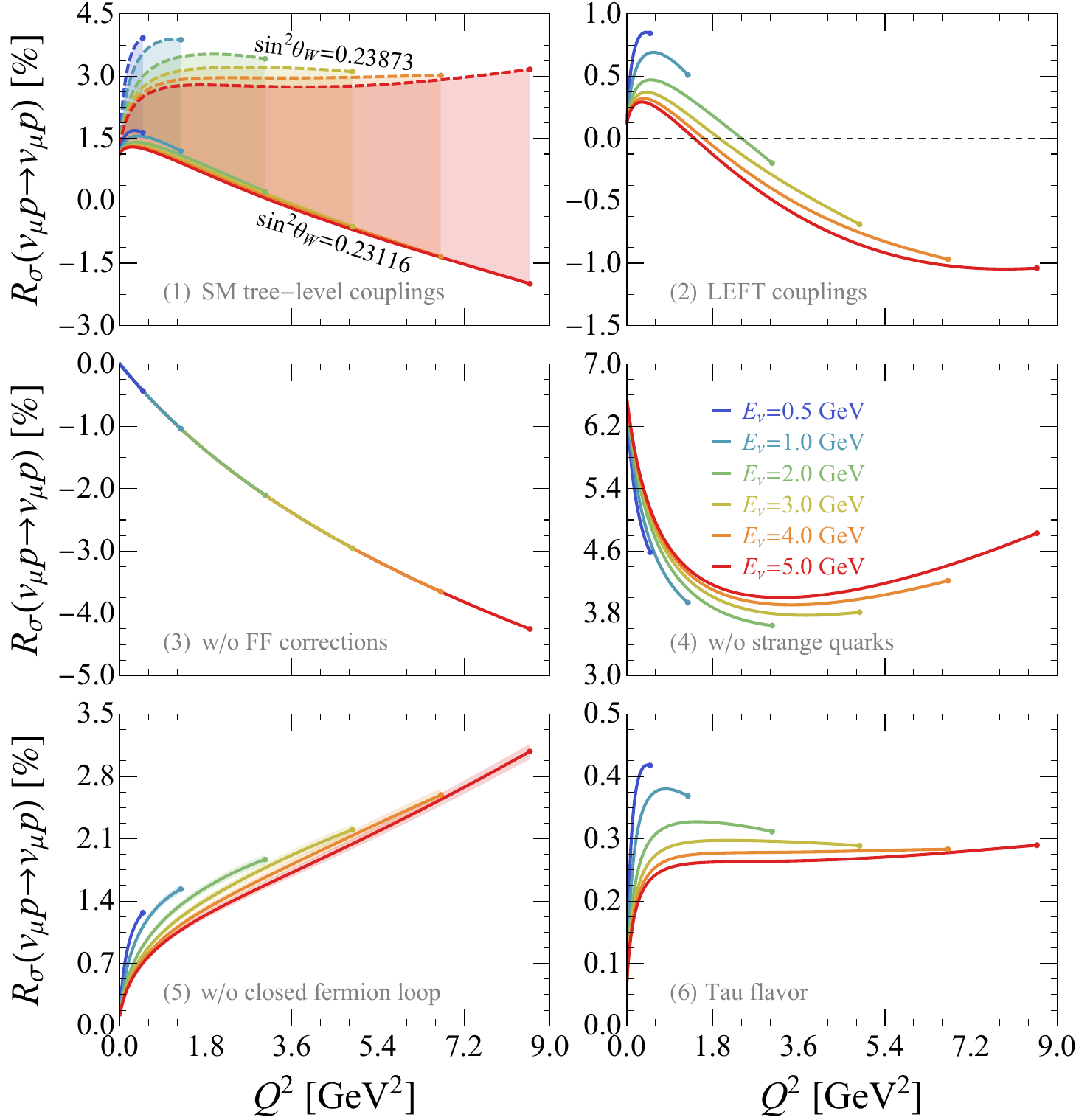}}
	\caption{The relative ratios $R_\sigma \equiv 1-\frac{ \ud \sigma^{(i)}/\ud Q^2 }{ \ud \sigma^{(\text{tot})}/\ud Q^2 } $, see~(\ref{eq:ratio_R}) for details, of differential cross sections for $\nu_\mu p \to \nu_\mu p$ elastic scattering in the laboratory frame are shown as functions of $Q^2$ at (anti)neutrino beam energies $E_\nu \in \{ 0.5,~1,~2,~3,~4,~5\}~\text{GeV}$ for the six cases under investigation [see the text below~(\ref{eq:ratio_R}) for detailed explanations]: (1) SM tree-level couplings, with $\sin^2 \theta_W = 0.23873$ (dashed curves) and $\sin^2 \theta_W = 0.23116$ (solid curves); (2) LEFT couplings; (3) w/o FF corrections; (4) w/o strange quarks; (5) w/o closed fermion loop, where the uncertainty of $\delta_\mathrm{QCD}$ is shown; (6) Tau flavor.}
	\label{Fig_SgmRatiovP}
\end{figure}

For convenience, we first define the variable $R_\sigma$ as
\begin{equation}
    \begin{aligned}\label{eq:ratio_R}
		R_\sigma \equiv 1-\frac{ \ud \sigma^{(i)}/\ud Q^2 }{ \ud \sigma^{(\text{tot})}/\ud Q^2 } = 1-\frac{ \ud \sigma^{(i)}/\ud \cos\theta }{ \ud \sigma^{(\text{tot})}/\ud \cos\theta},
	\end{aligned}
\end{equation}
where $\ud \sigma^{(\text{tot})}/\ud Q^2$ and $\ud \sigma^{(\text{tot})}/\ud \cos\theta$ denote the differential cross section for the muon flavor (anti)neutrino scattering. The cross section $\sigma^{(\text{tot})}$ includes all corrections that are considered in this paper: large electroweak contributions in LEFT couplings, QED radiative corrections on the nucleon side, contributions from the strange-quark FFs, and contributions due to the closed fermion loops. In contrast, $\ud \sigma^{(i)}/\ud Q^2$ and $\ud \sigma^{(i)}/\ud \cos\theta$ denote the incomplete (for $i=1,\dots,5$) and tau flavor (anti)neutrino scattering (for $i=6$) differential cross sections, namely
\begin{figure}[tb!]
	\centering
	{\includegraphics[angle=0,scale=0.50]{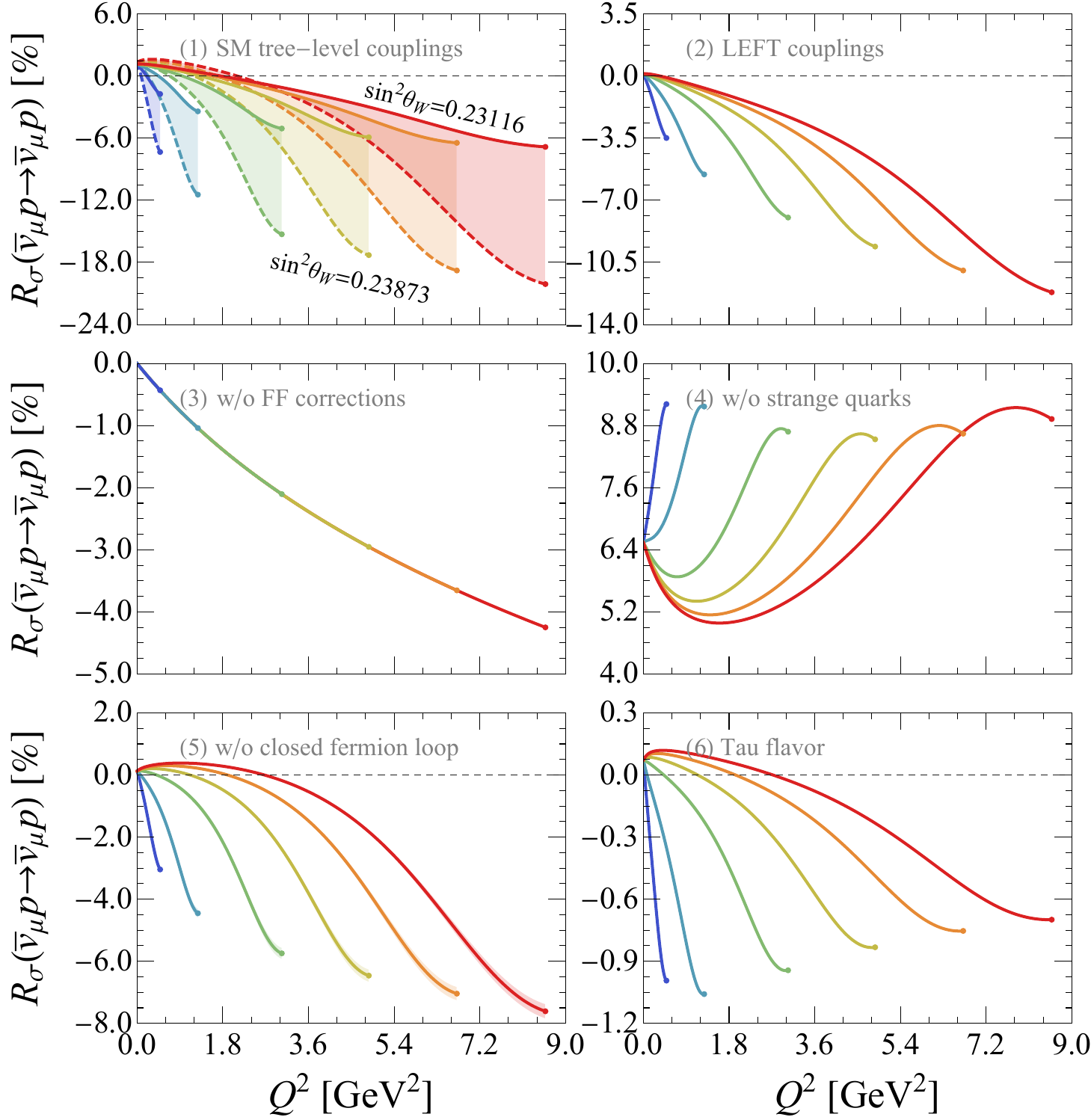}}
	\caption{Same as figure~\ref{Fig_SgmRatiovP} but for the antineutrino scattering.}
	\label{Fig_SgmRatiovP2}
\end{figure}
\begin{enumerate}
    \item the tree-level result with SM couplings as input, with $\sin^2 \theta_W = 0.23873$~\cite{ParticleDataGroup:2024cfk} and $\sin^2 \theta_W = 0.23116$~\cite{Pate:2024acz} for the band boundaries, labeled by ``(1) SM tree-level couplings'';\\
    \item the tree-level result with LEFT couplings as input (which include large electroweak contributions), labeled by ``(2) LEFT couplings'';\\
    \item the full result without (w/o) solely the radiative contributions to the Born FFs, labeled by ``(3) w/o FF corrections'';\\
    \item the full result but without (w/o) solely the contributions from the strange-flavor quarks to nucleon FFs, labeled by ``(4) w/o strange quarks'';\\
    \item the full result but without (w/o) solely the contributions due to closed fermion loops, with an error band due to the uncertainty of light-quark correction $\delta^\mathrm{QCD}$ in section~\ref{sec:closed_hadron_loops}, labeled by ``(5) w/o closed fermion loop'';\\
    \item For comparison with the muon flavor (anti)neutrino scattering, we also show the full result for the tau flavor (anti)neutrino scattering,\footnote{We have numerically checked that the corresponding results of $R_\sigma$ for the electron flavor (anti)neutrino beam are an order of magnitude smaller than the results for the tau flavor (anti)neutrino beam.} i.e. $\nu_\tau N \to \nu_\tau N$ or $\bar\nu_\tau N \to \bar\nu_\tau N$, labeled by ``(6) Tau flavor''.
\end{enumerate}

\begin{figure}[t!]
	\centering
	{\includegraphics[angle=0,scale=0.5]{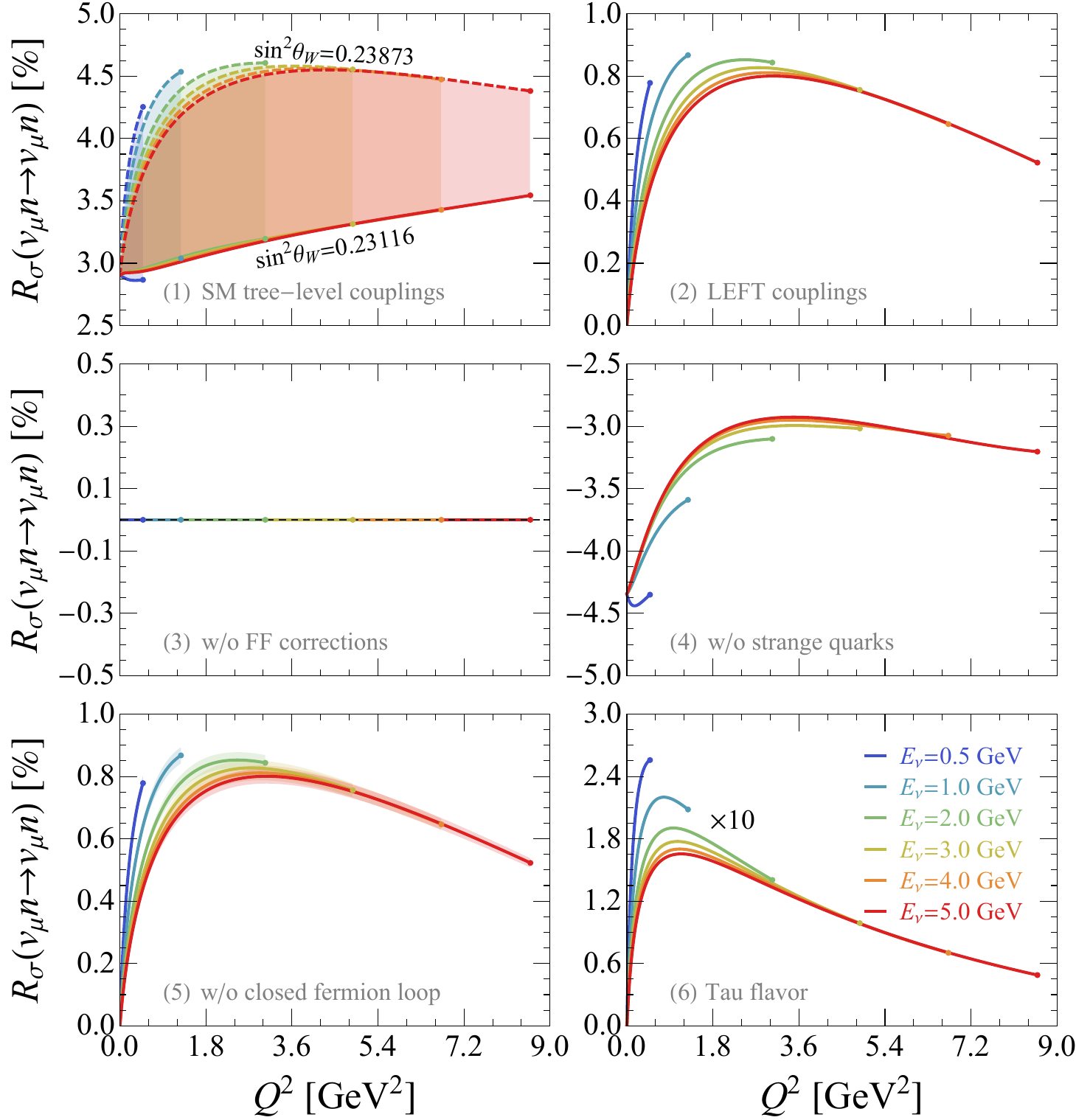}}
	\caption{Same as figure~\ref{Fig_SgmRatiovP} but for the scattering off the neutron.}
	\label{Fig_SgmRatiovN}
\end{figure}

\begin{figure}[ht!]
	\centering
	{\includegraphics[angle=0,scale=0.5]{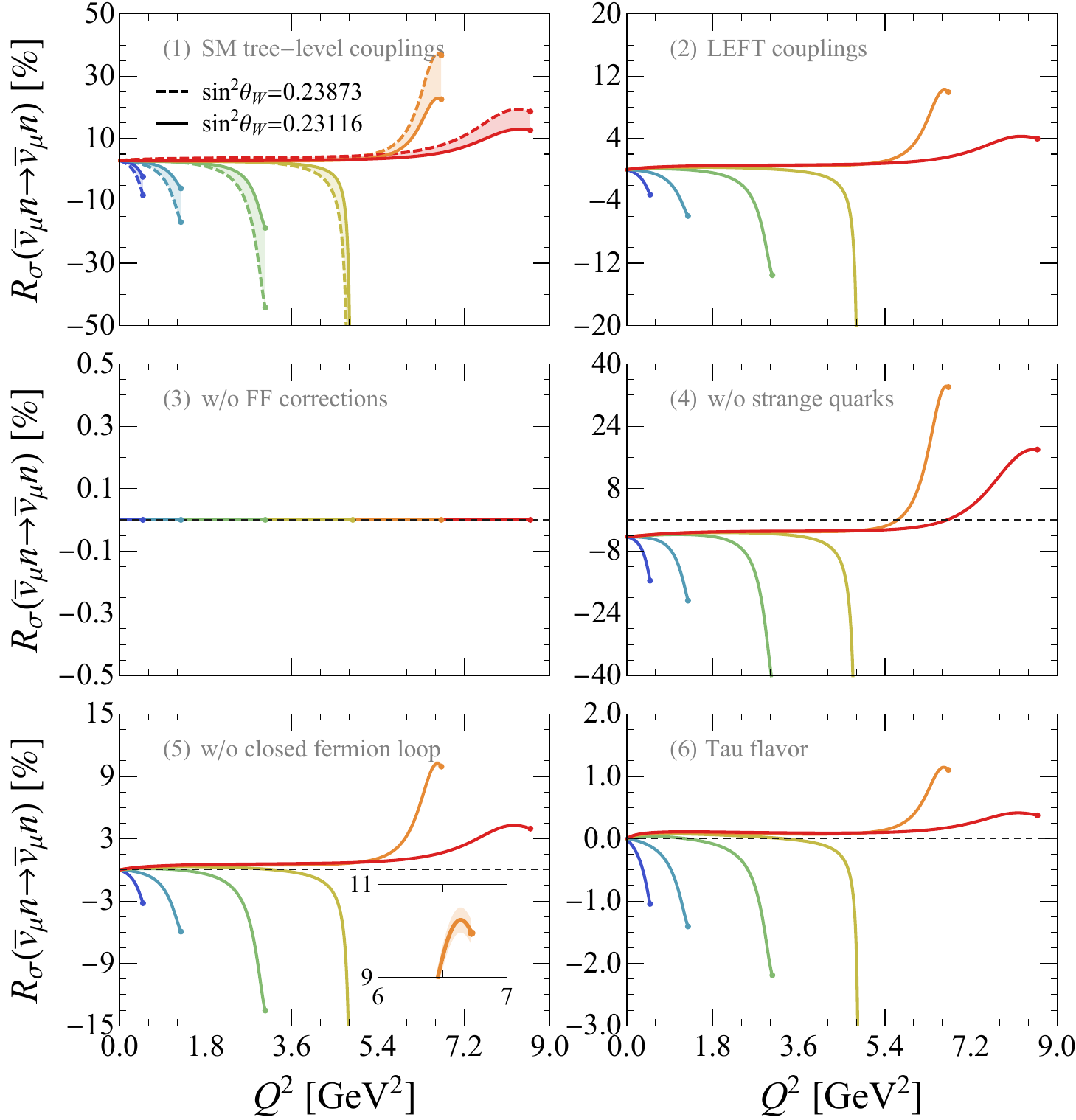}}
	\caption{Same as figure~\ref{Fig_SgmRatiovP2} but for the scattering off the neutron.}
	\label{Fig_SgmRatiovN2}
\end{figure}

First, we wish to quantify the relative size of the radiative effects. We present the corrections to the differential cross sections at the single-nucleon level $R_\sigma$ in figures~\ref{Fig_SgmRatiovP}-\ref{Fig_SgmRatiovN2} at different (anti)neutrino beam energies. We study different ratios $R_\sigma$, see (\ref{eq:ratio_R}), of differential cross sections for $\nu_\mu p \to \nu_\mu p$ ($\bar{\nu}_\mu p \to \bar{\nu}_\mu p$) elastic scattering in figure~\ref{Fig_SgmRatiovP} (figure~\ref{Fig_SgmRatiovP2}) at $E_\nu \in \{0.5,~1,~2,~3,~4,~5\}~\text{GeV}$ as functions of $Q^2$ for a photon energy detection threshold $\Delta E = 5~\mathrm{MeV}$. Similarly, we provide the same study for the neutron target in (anti)neutrino-neutron elastic scattering in figure~\ref{Fig_SgmRatiovN} (figure~\ref{Fig_SgmRatiovN2}).

We find that except for the case ``(3) w/o FF corrections'', $R_\sigma$ strongly depends on the incident (anti)neutrino energy $E_\nu$ and the squared four-momentum transfer $Q^2$. The strange-quark FF contributions to unpolarized NC (anti)neutrino-nucleon elastic scattering cross sections at $\mathrm{GeV}$ energies can reach $(4-9)\%$ and $(3-15)\%$ for the proton and neutron targets, respectively, with larger contributions for the antineutrino scattering. The variation in the unpolarized cross section as a function of the chosen value of $\sin^2 \theta_W$ is comparable to the contribution from the strange-quark FFs,\footnote{This finding, however, contradicts the assumption of small radiative corrections in~\cite{Marciano:1980pb,Ahrens:1986xe}.} while the inclusion of closed fermion-loop contributions from section~\ref{sec:closed_fermion_loops} removes this ambiguity and allows the definitive determination of the strange-quark FF at both low and high momentum transfers having a small residual uncertainty from the light-quark correction $\delta^\mathrm{QCD}$. Were the radiative corrections omitted from the experimental analysis, the extracted strange‑quark contribution to the nucleon spin $G_A^s \left( Q^2 = 0 \right) = \Delta s$ would shift from $-0.044\pm0.008$~\cite{Alexandrou:2021wzv} to $-0.053(-0.052)$ for scattering off the proton, and from $-0.044\pm0.008$~\cite{Alexandrou:2021wzv} to $-0.014(-0.014)$ for scattering off the neutron, taking $\sin^2\theta_W=0.23873$ ($\sin^2\theta_W=0.23116$).

However, $R_\sigma$ in the high-$Q^2$ region is rather sensitive to the precise $Q^2$ dependence of the nucleon electromagnetic FFs, motivating future experimental measurements of nucleon electromagnetic FFs in the high-$Q^2$ regions at the JLab 22~GeV upgrade~\cite{Accardi:2023chb} and electron-ion colliders~\cite{Accardi:2012qut,Burkert:2022hjz,Anderle:2021wcy}.

The cross-section difference between the tau- and muon-flavor neutrino (antineutrino) beams can reach around $0.5\%$ ($1.0\%$). In contrast, the cross-section difference between the electron- and muon-flavor neutrino (antineutrino) beams is negligibly small at $Q^2 > m_\mu^2$, in agreement with~\cite{Marciano:1980pb}, and therefore is not shown in figures~\ref{Fig_SgmRatiovP}-\ref{Fig_SgmRatiovN2}.

\begin{figure}[tb!]
	\centering
	{\includegraphics[angle=0,scale=0.50]{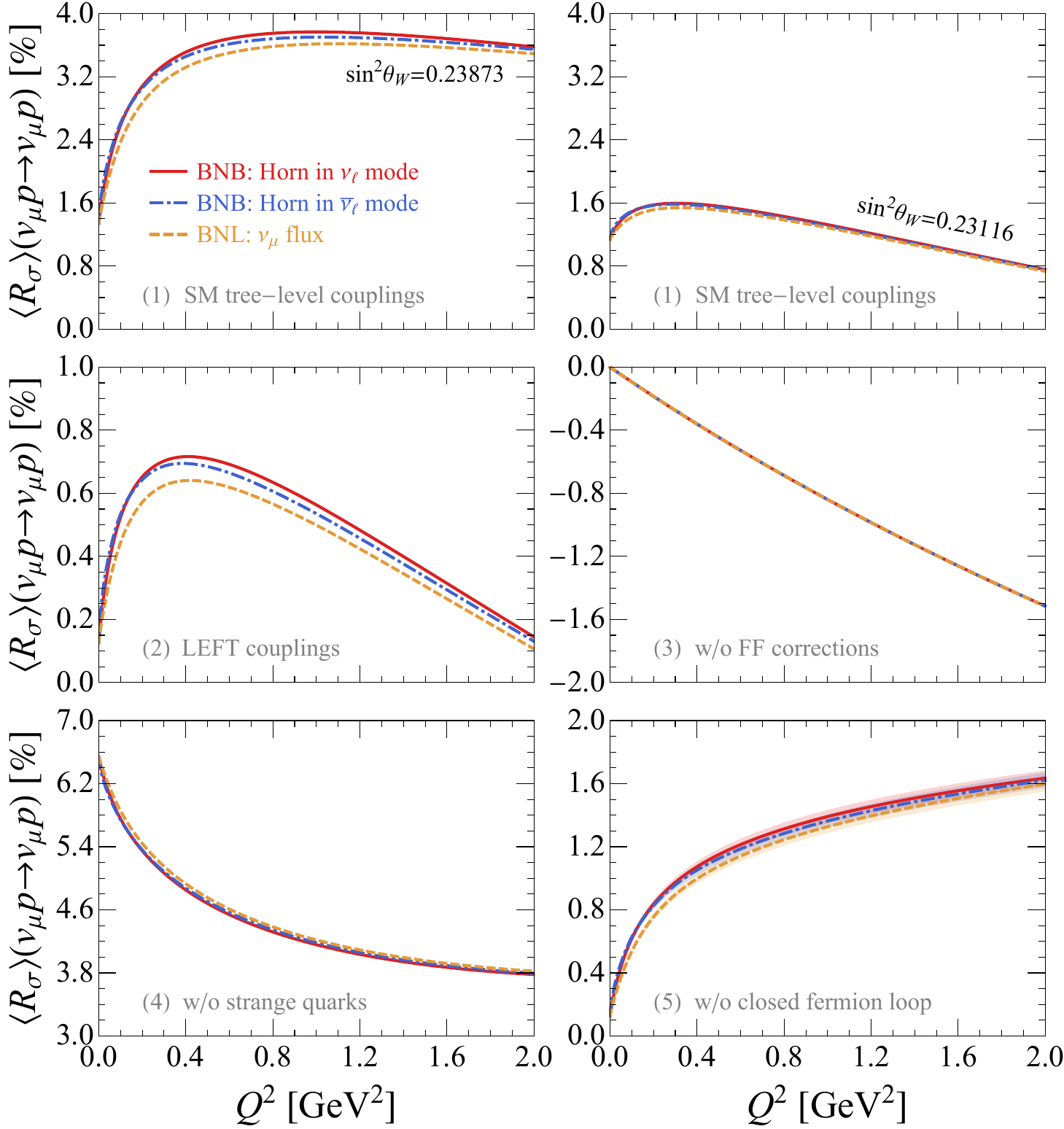}}
	\caption{Similar to figure~\ref{Fig_SgmRatiovP} but after averaging over the MiniBooNE $\nu_\mu$ flux at BNB with magnetic focusing horn in neutrino (solid red curves) and antineutrino (dash-dotted blue curves) modes~\cite{MiniBooNE:2008hfu}, as well as over the $\nu_\mu$ flux (dashed orange curves) at BNL E734~\cite{Ahrens:1986xe}.}
	\label{Fig_SgmRatioFluxedvP_BNBBNL}
\end{figure}

From figures~\ref{Fig_SgmRatiovP}-\ref{Fig_SgmRatiovN2}, we observe that both $R_\sigma$ and $\ud \sigma/\ud Q^2$ depend strongly on $E_\nu$, which shows the great necessity of properly taking into account the $E_\nu$-dependent (anti)neutrino fluxes (see appendix~\ref{app:flux_distributions}). According to~\cite{Ahrens:1986xe,Alberico:1998qw,Alberico:2001sd,Chen:2026uhf}, the flux-averaged differential cross sections are traditionally given by\footnote{According to the two-body kinematics (see also footnote 6) in $\nu_\ell N \to \nu_\ell N$, $Q^2= \frac{2M E_\nu^2 }{2E_\nu + M} (1-\cos\theta_\text{cm})$ for $\theta_\text{cm} \in [0, \pi]$. The (anti)neutrino energy $E_\nu$ is expressed in terms of the final-state kinematics as $E_\nu = \frac{Q^2 + \sqrt{ 2 M^2 Q^2 ( 1-\cos\theta_\text{cm} ) + (Q^2)^2 } }{ 2M (1-\cos\theta_\text{cm}) }$, which satisfies $ \big[ Q^2 + \sqrt{Q^2 (4M^2+ Q^2) } \big]/4M \leq E_\nu \leq \infty$ at fixed $Q^2$. This leads to a constrained lower bound $E_\text{L}'(Q^2)$, see (\ref{eq:normalized-Flux}), for the integral over $E_\nu$ with the $Q^2$-dependent integrand.}
\begin{equation}\label{eq:DiffSgm-fluxAve}
	\Big\langle \frac{\ud \sigma }{\ud Q^2} \Big\rangle
	= \frac{1}{\mathcal N_\Phi } \int_{ E_\text{L}' }^{E_\text{U}} \ud E_\nu\, \frac{\ud \sigma }{\ud Q^2}(E_\nu,Q^2)\, \Phi(E_\nu) = \int_{ E_\text{L}' }^{E_\text{U}} \ud E_\nu\, \frac{\ud \sigma }{\ud Q^2}(E_\nu,Q^2)\, \tilde \Phi(E_\nu),
\end{equation}
with
\begin{equation}\label{eq:normalized-Flux}
    \begin{aligned}
	\mathcal N_\Phi 
    &\equiv \int_{E_\text{L}}^{E_\text{U}} \ud E_\nu\, \Phi(E_\nu),\qquad 
	\int_{E_\text{L}}^{E_\text{U}} \ud E_\nu\, \tilde\Phi(E_\nu) = 1,\\
    E_\text{L}' 
    &= E_\text{L}'(Q^2) = \max \left\{ E_\text{L},~ \Big[ Q^2 + \sqrt{Q^2 (4M^2+ Q^2) } \Big]/4M \right\} \leq E_\text{U},
    \end{aligned}
\end{equation}
where $\tilde\Phi(E_\nu) = \Phi(E_\nu)/\mathcal N_\Phi$ is the normalized flux distribution, and $E_\text{L}$ ($E_\text{U}$) is the selected lower (upper) bound of $E_\nu$. Extending~(\ref{eq:ratio_R}) to flux-averaged cross sections, we define the flux-averaged ratio $\langle R_\sigma \rangle$ as
\begin{equation}
    \begin{aligned}\label{eq:ratio_R_ave}
		\langle R_\sigma \rangle \equiv 1-\frac{\langle \ud \sigma^{(i)}/\ud Q^2 \rangle}{ \langle \ud \sigma^{(\text{tot})}/\ud Q^2 \rangle } = 1-\frac{ \langle \ud \sigma^{(i)}/\ud \cos\theta\rangle }{ \langle \ud \sigma^{(\text{tot})}/\ud \cos\theta \rangle },
	\end{aligned}
\end{equation}
where $\ud \sigma^{(i)}/\ud Q^2$ and $\ud \sigma^{(\text{tot})}/\ud Q^2$ are specified in (\ref{eq:ratio_R}).

\begin{figure}[tb!]
	\centering
	{\includegraphics[angle=0,scale=0.50]{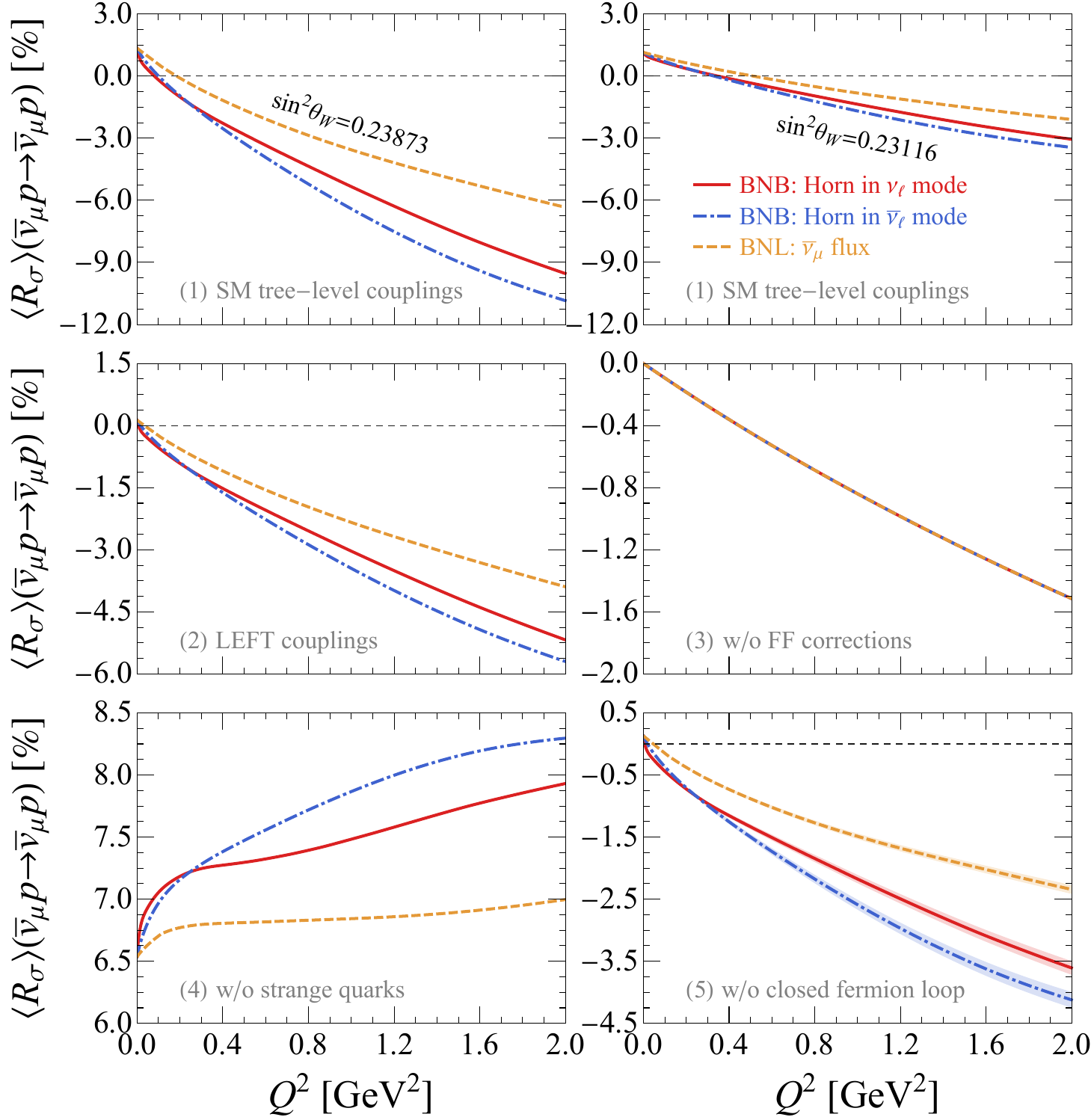}}
	\caption{Similar to figure~\ref{Fig_SgmRatiovP2} but after averaging over the MiniBooNE $\bar\nu_\mu$ flux at BNB with magnetic focusing horn in neutrino (solid red curves) and antineutrino (dash-dotted blue curves) modes~\cite{MiniBooNE:2008hfu}, as well as over the $\bar\nu_\mu$ flux (dashed orange curves) at BNL E734~\cite{Ahrens:1986xe}.}
	\label{Fig_SgmRatioFluxedvP2_BNBBNNL}
\end{figure}

\begin{figure}[tb!]
	\centering
	{\includegraphics[angle=0,scale=0.50]{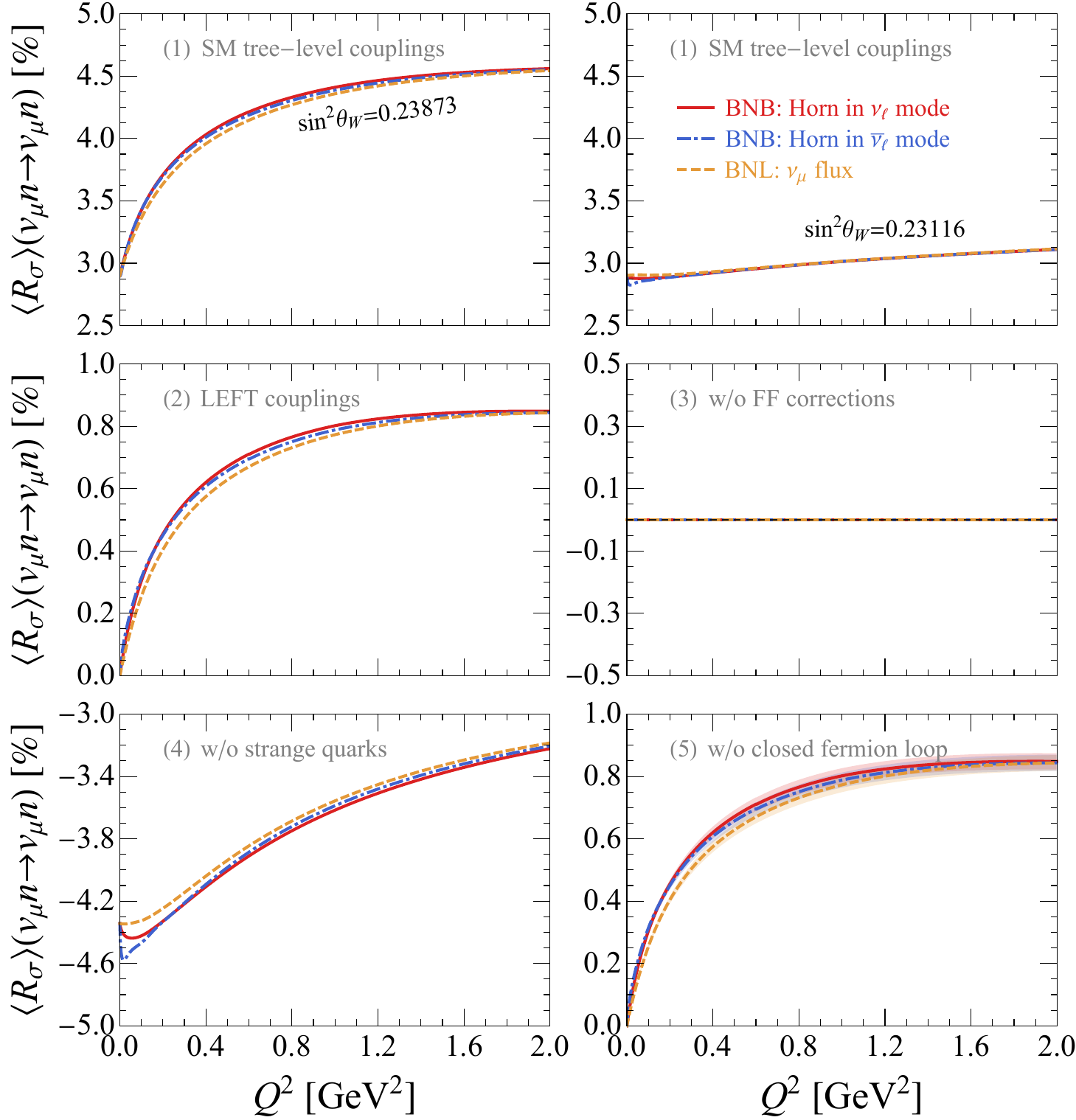}}
	\caption{Same as figure~\ref{Fig_SgmRatioFluxedvP_BNBBNL} but for the scattering off the neutron.}
	\label{Fig_SgmRatioFluxedvN_BNBBNL}
\end{figure}

\begin{figure}[tb!]
	\centering
	{\includegraphics[angle=0,scale=0.50]{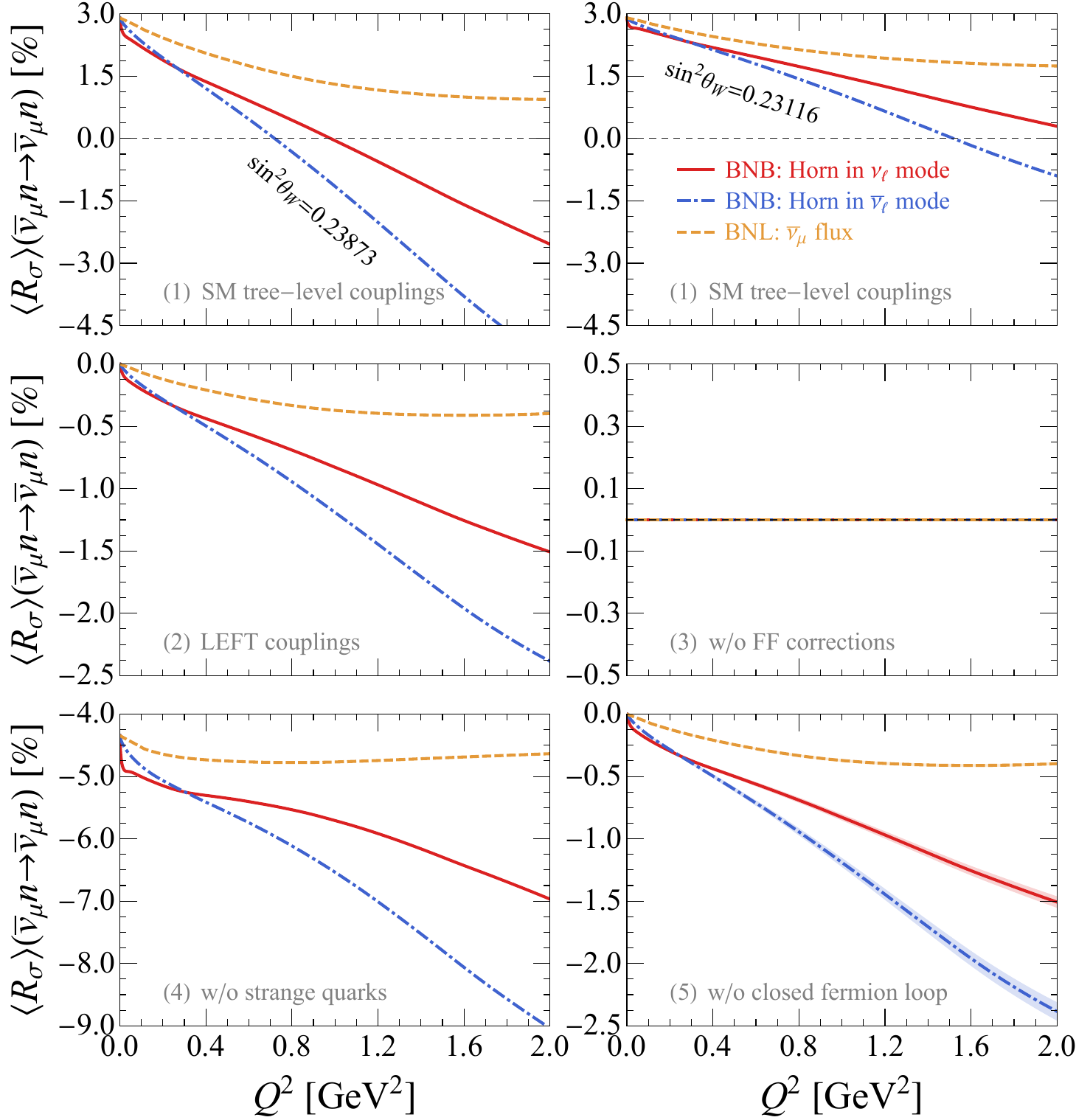}}
	\caption{Same as figure~\ref{Fig_SgmRatioFluxedvP2_BNBBNNL} but for the scattering off the neutron.}
	\label{Fig_SgmRatioFluxedvN2_BNBBNL}
\end{figure}

Subsequently, we average ${\ud \sigma}/{\ud Q^2}$ over the (anti)neutrino flux of the Boosted Neutrino Beam (BNB) at Fermilab~\cite{MiniBooNE:2008hfu} with limits $E_\text{L}=0.025~\text{GeV}$ and $E_\text{U}=5.0~\text{GeV}$ and present the (anti)neutrino-nucleon elastic scattering cross sections as functions of $Q^2$ in figures~\ref{Fig_SgmRatioFluxedvP_BNBBNL}-\ref{Fig_SgmRatioFluxedvN2_BNBBNL}. We show our results only at low momentum transfers $Q^2 \le 2~\mathrm{GeV}^2$ in order to avoid potentially large contributions from high-energy flux tails. Similarly, we average ${\ud \sigma}/{\ud Q^2}$ over the BNL E734 fluxes (see figure~\ref{fig:BNLflux}) with limits $E_\text{L}=0.2~\text{GeV}$ and $E_\text{U}=5.0~\text{GeV}$~\cite{Ahrens:1986xe}, and show our final results in figures~\ref{Fig_SgmRatioFluxedvP_BNBBNL}-\ref{Fig_SgmRatioFluxedvN2_BNBBNL}. We provide details of the (anti)neutrino fluxes in the appendix~\ref{app:flux_distributions}. 

Despite the $E_\nu$-dependence of the $\nu_\mu$ fluxes in the three scenarios above being distinctly different, the results of $\langle R_\sigma \rangle$ in the $\nu_\mu N$ elastic scattering for the BNB $\nu_\mu$ fluxes (in neutrino and antineutrino modes) and the BNL $\nu_\mu$ flux are very close to each other for all six cases under investigation. For the $\bar \nu_\mu N$ elastic scattering in figure~\ref{Fig_SgmRatioFluxedvP2_BNBBNNL} and figure~\ref{Fig_SgmRatioFluxedvN2_BNBBNL}, however, we observe that the results of $\langle R_\sigma \rangle$ for the BNB $\bar\nu_\mu$ fluxes (in neutrino and antineutrino modes) and the BNL $\bar\nu_\mu$ flux differ quite a lot, especially at larger $Q^2$ for all six cases [except for the case ``(3) w/o FF corrections'']. Although the $E_\nu$-dependences of the $\bar\nu_\mu$ fluxes in the three scenarios are distinctly different, roughly at the same level as that for the $\nu_\mu$ fluxes (see figures~\ref{fig:BNLflux} and \ref{fig:BNBflux} in appendix~\ref{app:flux_distributions}). This indicates that $\langle R_\sigma \rangle$ in (muon-flavor) antineutrino-nucleon elastic scattering is more sensitive to the corresponding flux distributions than that in (muon-flavor) neutrino-nucleon elastic scattering.

Moreover, we also observe an interesting feature in figures~\ref{Fig_SgmRatioFluxedvP_BNBBNL}-\ref{Fig_SgmRatioFluxedvN2_BNBBNL} that the ratios for the magnetic focusing horn in neutrino (solid red curves) and antineutrino (dash-dotted blue curves) modes at $Q^2=0$ are basically the same. According to (\ref{eq:NC_cross_section_eff}) and (\ref{NC-diffSgm-tree}), the corresponding differential cross sections at $Q^2=0$ do not depend on the (anti)neutrino energy and are exactly the same for both neutrino and antineutrino scattering. The cross section is given by~\cite{Horstkotte:1981ne}
\begin{equation}
	\begin{aligned}
    \frac{ \ud \sigma }{\ud Q^2}(E_\nu,0) 
    &= \frac{ G_F^2 }{ 2\pi } \left\{ [g_A(Q^2 = 0)]^2 + [g_E(Q^2 = 0)]^2 \right\},
	\end{aligned}
\end{equation}
which is evidently $E_\nu$-independent. Therefore, the corresponding flux-averaged differential cross sections and hence the flux-averaged ratios at $Q^2=0$ for the magnetic focusing horn in neutrino and antineutrino modes, with a similar flavor decomposition of the flux, are basically the same.

\begin{figure}[tb!]
	\centering
	{\includegraphics[angle=0,scale=0.57]{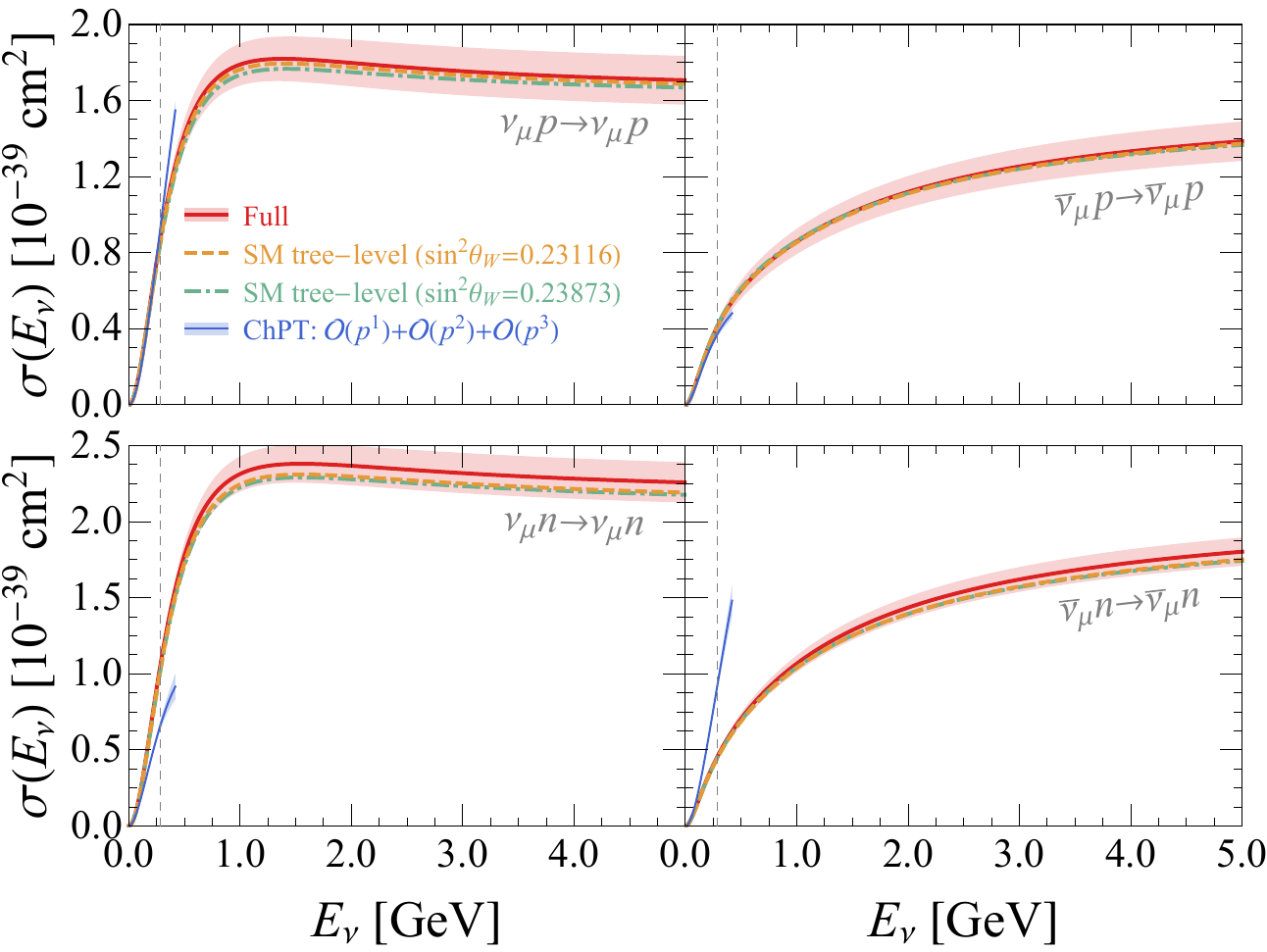}}
	\caption{Total cross sections (solid red curves with an error band), including radiative corrections, of $\nu_\mu p \to \nu_\mu p$ (upper left panel), $\overline{\nu}_\mu p \to \overline{\nu}_\mu p$ (upper right panel), $\nu_\mu n \to \nu_\mu n$ (lower left panel), and $\overline{\nu}_\mu n \to \overline{\nu}_\mu n$ (lower right panel) are shown as functions of the (anti)neutrino energy $E_\nu$ and compared to the tree-level SM results with $\sin^2\theta_W=0.23116$ (dashed orange curves) and $\sin^2\theta_W=0.23873$ (dash-dotted green curves), and the $\mathrm{SU}(2)$ ChPT calculation (thin solid blue curves) from~\cite{Chen:2024kbh}, with the dashed grey vertical lines indicating the upper bound ($E_\nu^\text{max} \approx 0.283~\text{GeV}$) of the validity range of results in~\cite{Chen:2024kbh}.}
	\label{fig:TotSgmVPVN}
\end{figure}

The total cross section $\sigma$ is given by
\begin{align}\label{eq:tot-sigma-def}
    \sigma(E_\nu) = \int_0^{\frac{ 4M E_\nu^2 }{M + 2E_\nu }} \ud Q^2\, \frac{\ud \sigma }{\ud Q^2} (E_\nu,Q^2),
\end{align}
which depends on the incident (anti)neutrino energy $E_\nu$. In figure~\ref{fig:TotSgmVPVN}, we show our predictions that include radiative corrections for the total cross sections of $\nu_\mu p \to \nu_\mu p$, $\overline{\nu}_\mu p \to \overline{\nu}_\mu p$, $\nu_\mu n \to \nu_\mu n$, and $\overline{\nu}_\mu n \to \overline{\nu}_\mu n$ as functions of the (anti)neutrino energy $E_\nu$.\footnote{Note that $1~\text{fb}=10^{-15}~\text{barn}=10^{-39}~\text{cm}^2$~\cite{ParticleDataGroup:2024cfk}.} To illustrate the resulting size of radiative corrections, we show the central value of tree-level SM results using $\sin^2\theta_W=0.23116$ and $\sin^2\theta_W=0.23873$. At the level of total cross section, the relative size of radiative corrections is larger for the scattering off the neutron target and is determined mainly by large electroweak logarithms. We also compare our results with the $\mathrm{SU}(2)$ ChPT calculation from~\cite{Chen:2024kbh}, cf. figure~8 in~\cite{Chen:2024kbh}, with the dashed grey vertical lines indicating the upper bound ($E_\nu^\text{max} \approx 0.283~\text{GeV}$) of the validity range of this calculation.\footnote{For the $\mathrm{SU}(2)$ ChPT results in~\cite{Chen:2024kbh}, the upper bound of the validity range of $Q^2$ is $ Q_\text{max}^2 = 0.2~\text{GeV}^2$, which leads to the validity range of $E_\nu$: $0\leq E_\nu \leq E_\nu^\text{max} = M\big[ \tau_\text{max} + \sqrt{\tau_\text{max}(1+\tau_\text{max}) } \big] \approx 0.283~\text{GeV}$ with $\tau_\text{max} \equiv Q_\text{max}^2/(4M^2) $.} We observe noticeable differences between the $\mathrm{SU}(2)$ ChPT results in~\cite{Chen:2024kbh} and our predictions for the neutron target, especially at larger $E_\nu$. However, we obtain perfect agreement after interchanging the neutrino with the antineutrino predictions for the neutron target in~\cite{Chen:2024kbh}.

\begin{figure}[t!]
	\centering
	{\includegraphics[angle=0,scale=0.57]{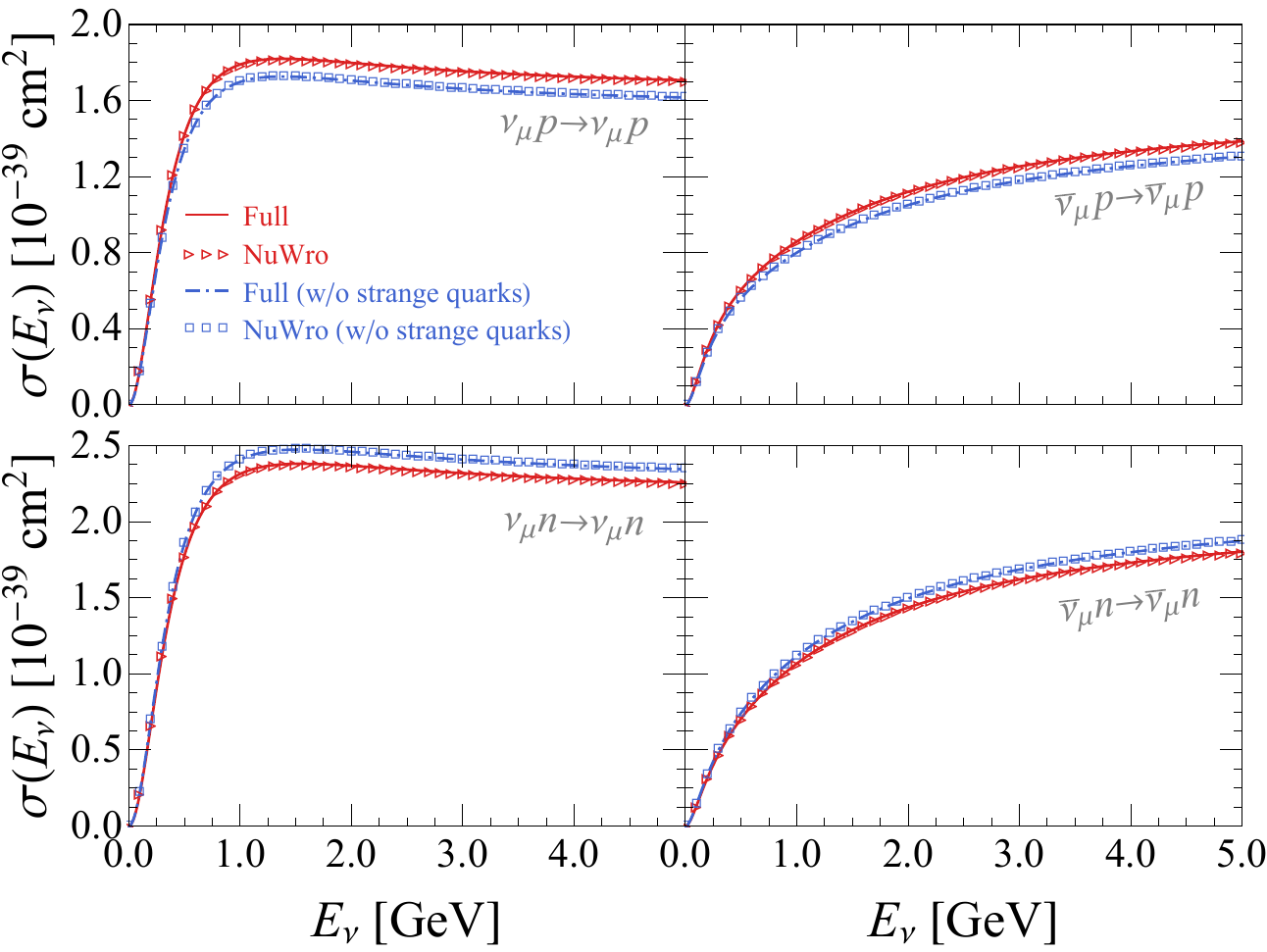}}
	\caption{Total cross sections, including radiative corrections, with (solid red curves) and without (dash-dotted blue curves) nucleon strange-quark contributions are shown as functions of the (anti)neutrino energy $E_\nu$, in comparison with the corresponding NuWro~\cite{Juszczak:2005zs,Juszczak:2005wk,Sobczyk:2008zz,Juszczak:2009qa,Golan:2013jtj} simulations, including radiative corrections (red-triangle/blue-square markers) and the quark-flavor FFs from figure~\ref{fig:quarkFFs}.}
	\label{fig:TotSgmVPVN2}
\end{figure}

We have implemented our results into the NuWro neutrino event generator framework~\cite{Juszczak:2005zs,Juszczak:2005wk,Sobczyk:2008zz,Juszczak:2009qa,Golan:2013jtj}. We test this implementation in figure~\ref{fig:TotSgmVPVN2}. First, we find a perfect agreement between the central values of our full results in figure~\ref{fig:TotSgmVPVN}, with (solid red curves) and without (dash-dotted blue curves) nucleon strange-quark contributions, and the corresponding NuWro simulations including radiative corrections, shown by red-triangle and blue-square markers, respectively. We find that the total nucleon strange-quark contribution is dominated by $G_A^s$, rather than by $G_E^s$ and $G_M^s$. This emphasizes the importance of $G_A^s(Q^2)$ in probing the strangeness content of the nucleon in NC (anti)neutrino-nucleon elastic scattering~\cite{Ahrens:1986xe,Garvey:1993sg,Pate:2003rk,Pate:2008va,Pate:2024acz}.

Following (\ref{eq:DiffSgm-fluxAve}), the flux-averaged total cross section $\langle \sigma \rangle$ is given by
\begin{align}\label{eq:tot-sigma-def-fluxAve}
    \langle \sigma \rangle = \int_0^{\infty} \ud Q^2\, \Big\langle \frac{\ud \sigma }{\ud Q^2} \Big\rangle = \int_{ E_\text{L}' }^{E_\text{U}} \ud E_\nu\, \int_0^{\frac{ 4M E_\nu^2 }{M + 2E_\nu }} \ud Q^2 \frac{\ud \sigma }{\ud Q^2}(E_\nu,Q^2)\, \tilde \Phi(E_\nu),
\end{align}
which is $E_\nu$-independent.

\begin{figure}[tb!]
	\centering
	{\includegraphics[angle=0,scale=0.52]{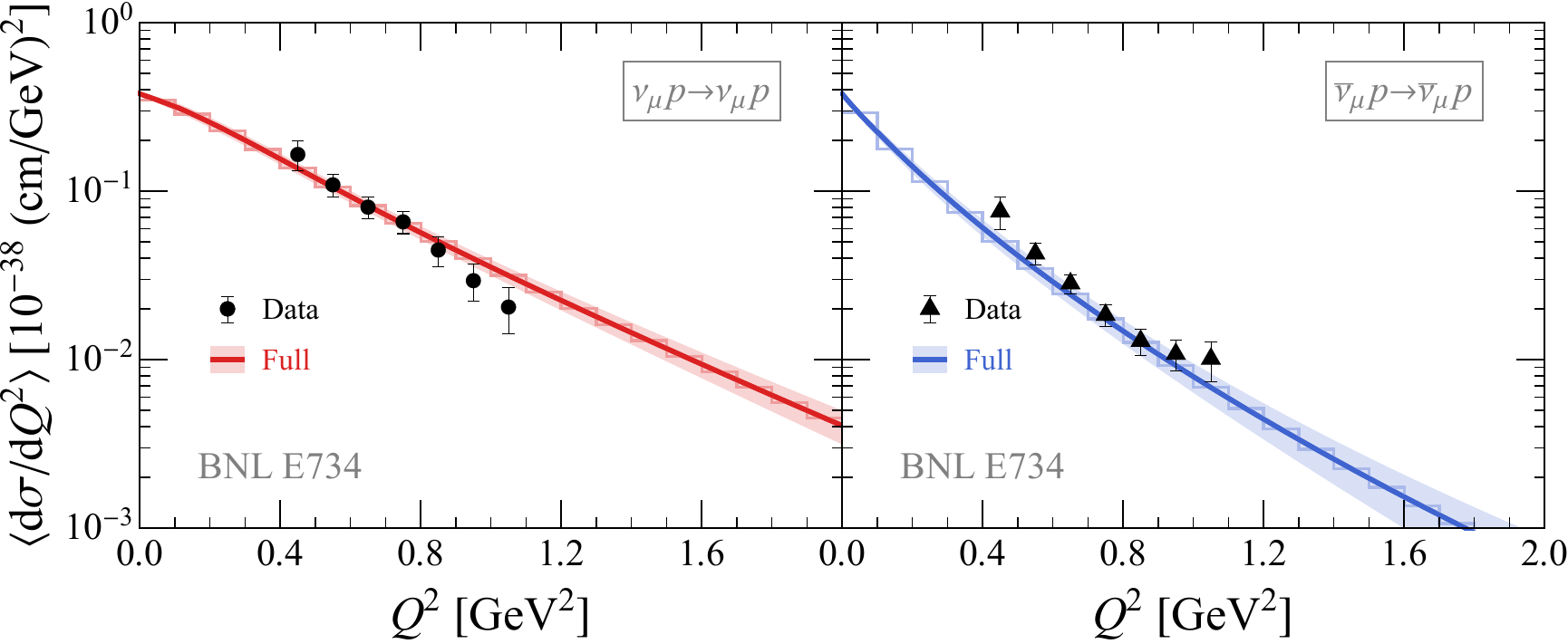}}
	\caption{Predictions for the flux-averaged differential cross sections of $\nu_\mu p \to \nu_\mu p$ (solid red curve) and $\overline{\nu}_\mu p \to \overline{\nu}_\mu p$ (solid blue curve) are compared with the experimental data (black markers) from the BNL E734 measurements~\cite{Ahrens:1986xe,Pate:2003rk,Pate:2024acz}. To visualize the possible bin-width effect in $Q^2$, we also show the step curves (thin light-red/light-blue curves) for all the $Q^2$ bins with a fixed width of $0.1~\text{GeV}^2$~\cite{Ahrens:1986xe} using the central value of our theoretical predictions.}
	\label{fig:BNL_DiffSgm_FluxedP}
\end{figure}

In figure~\ref{fig:BNL_DiffSgm_FluxedP}, we compare our predictions with the experimental data from the BNL E734 measurement~\cite{Ahrens:1986xe} (cf. figure 35 in this reference) using the BNL E734 flux distributions (see figure~\ref{fig:BNLflux}). Our results agree with the experimental data, within uncertainties propagated from the nucleon FFs (see figure~\ref{fig:quarkFFs}). Uncertainties from radiative corrections are much smaller than the errors associated with the nucleon structure and therefore can be neglected. Moreover, the nucleon-structure errors are much larger, by more than a factor $13$ and $23$ for $\nu_\mu p \to \nu_\mu p$ and $\overline{\nu}_\mu p \to \overline{\nu}_\mu p$, respectively, than the difference in the results for the three choices of radiative corrections on the proton line: $\Delta E = 5~\mathrm{MeV}$, $\Delta E = 20~\mathrm{MeV}$, and without FF corrections. Therefore, we choose the $\Delta E = 5~\mathrm{MeV}$ case as the default correction. This choice is also consistent with other figures in this paper. To visualize the possible bin-width effect in $Q^2$, we also show the step curves for all the $Q^2$ bins with a fixed width of $0.1~\text{GeV}^2$~\cite{Ahrens:1986xe} using the central value of our theoretical predictions.

\begin{figure}[tb!]
	\centering
	{\includegraphics[angle=0,scale=0.52]{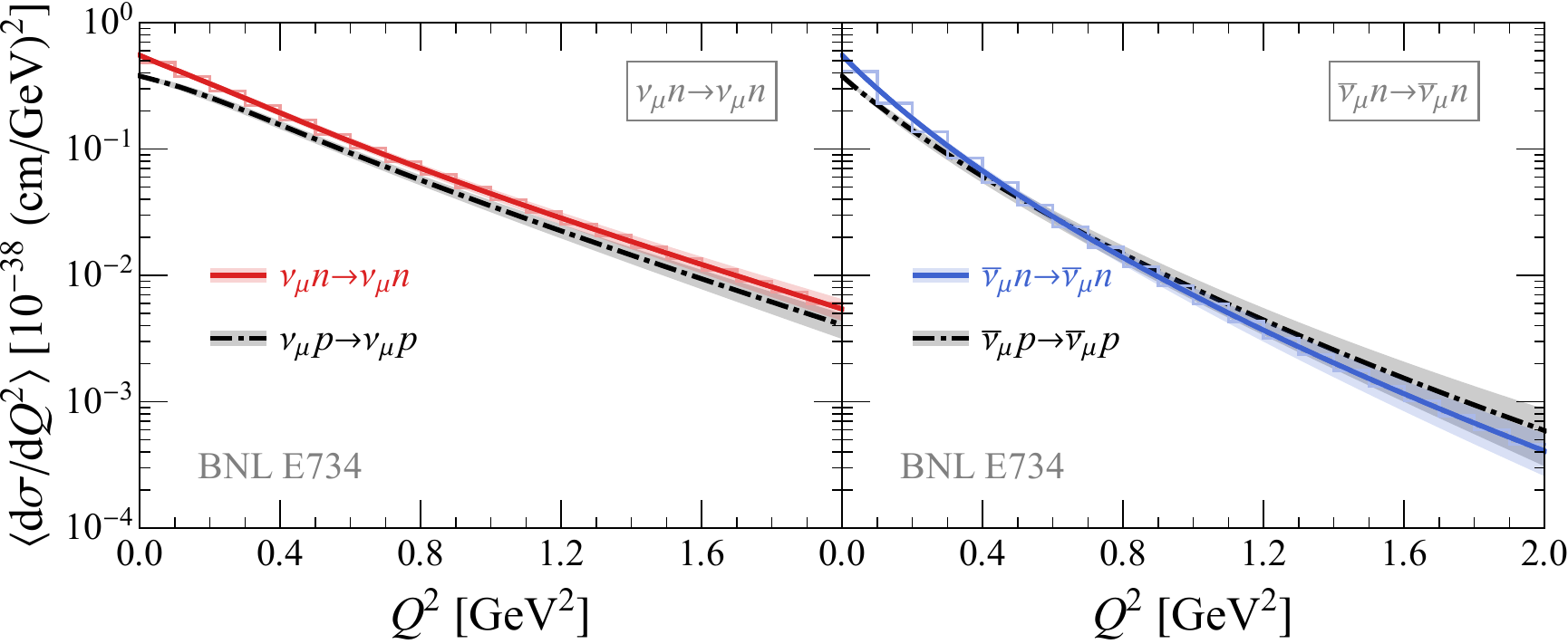}}
	\caption{Similar to figure~\ref{fig:BNL_DiffSgm_FluxedP} but for the predictions of the flux-averaged differential cross sections of $\nu_\mu n \to \nu_\mu n$ (solid red curve) and $\overline{\nu}_\mu n \to \overline{\nu}_\mu n$ (solid blue curve) including step curves for the bin-width effect. To visualize the difference between the neutron and proton targets, we also show our theoretical predictions for the proton target (dash-dotted black curves) in figure~\ref{fig:BNL_DiffSgm_FluxedP} for a comparison.}
	\label{fig:BNL_DiffSgm_FluxedN}
\end{figure}

We also present our predictions of the flux-averaged differential cross sections of $\nu_\mu n \to \nu_\mu n$ and $\overline{\nu}_\mu n \to \overline{\nu}_\mu n$ in figure~\ref{fig:BNL_DiffSgm_FluxedN} for the neutron target using the BNL E734 flux distributions (see figure~\ref{fig:BNLflux}). Since there is no such experimental data for the neutron target, we show our theoretical predictions for the proton target in figure~\ref{fig:BNL_DiffSgm_FluxedP} for a comparison. We find that the $\nu_\mu n \to \nu_\mu n $ cross sections are clearly larger than the $\nu_{\mu} p \to \nu_{\mu} p$ cross sections for $Q^2 \lesssim 2~\mathrm{GeV}^2$. In addition, we observe that the $\overline{\nu}_\mu n \to \overline{\nu}_\mu n$ cross sections are larger than the $\overline{\nu}_\mu p \to \overline{\nu}_\mu p$ cross sections for $Q^2 \lesssim 0.6~\mathrm{GeV}^2$, while they basically agree with each other within the uncertainty bands for $0.6~\text{GeV} \lesssim  Q^2 \leq 2~\text{GeV}^2$.

In addition to the calculations of the flux-averaged differential cross sections, we also predict the total cross sections. In what follows, we present our theoretical results for the NC to CC cross-section ratios $R_\nu^\text{th}$ and $R_{\bar\nu}^\text{th}$ in the $Q^2$ interval $0.5~\mathrm{GeV}^2 < Q^2 < 1.0~\mathrm{GeV}^2$, and for the antineutrino- to neutrino-nucleon NC cross-section ratio $R_{\overline{\nu}N/\nu N}^\text{th}$ in the $Q^2$ interval $0.4~\mathrm{GeV}^2 < Q^2 < 0.9~\mathrm{GeV}^2$ for three choices (i.e. $\Delta E = 5~\mathrm{MeV}$, $\Delta E = 20~\mathrm{MeV}$, and w/o FF corrections) of settings for the radiative corrections on the proton line:
\begin{align}\label{eq:ratios-totSgm}
    R^\mathrm{th}_\nu 
    &= \frac{\sigma \left( \nu_\mu p \to \nu_\mu p \right)}{\sigma \left( \nu_\mu n \to \mu^- p \right)} = 0.143\pm 0.004, \\
    R^\mathrm{th}_{\overline{\nu}} 
    &= \frac{\sigma \left( \overline{\nu}_\mu p \to \overline{\nu}_\mu p \right)}{\sigma \left( \overline{\nu}_\mu p \to \mu^+ n \right)} = 0.239\pm 0.022, \\
    {R}^\mathrm{th}_{\overline{\nu}N/\nu N}
    &= \frac{ \sigma(\bar\nu_\mu N \to \bar\nu_\mu N) }{ \sigma(\nu_\mu N \to \nu_\mu N) } = \begin{cases}
        0.318 \pm 0.016, &(N=p),\\
        0.262 \pm 0.008, & (N=n).
    \end{cases}
\end{align}
We notice that the three choices lead to the same result of ${R}^\mathrm{th}_{\overline{\nu} N/\nu N}$, up to the last significant digit. This justifies the validity of choosing $\Delta E=5~\text{MeV}$ as a default value and that the uncertainties of the radiative corrections are indeed small (at least they are basically independent of the three choices). We find that our theoretical results are in excellent agreement with the experimental data~\cite{Entenberg:1979wc,Ahrens:1985cm,Ahrens:1986xe}
\begin{align}
    R_\nu^\text{exp} &= \frac{\sigma \left( \nu_\mu p \to \nu_\mu p \right)}{\sigma \left( \nu_\mu n \to \mu^- p \right)} = 0.153\pm 0.007\left( \mathrm{stat.} \right)\pm 0.017\left( \mathrm{syst.} \right), \\
    R_{\overline{\nu}}^\text{exp} &= \frac{\sigma \left( \overline{\nu}_\mu p \to \overline{\nu}_\mu p \right)}{\sigma \left( \overline{\nu}_\mu p \to \mu^+ n \right)} = 0.218\pm 0.012\left( \mathrm{stat.} \right)\pm 0.023\left( \mathrm{syst.} \right), \\
    {R}^\mathrm{exp}_{\overline{\nu}p/\nu p}
    &= \frac{ \sigma(\bar\nu_\mu p \to \bar\nu_\mu p) }{ \sigma(\nu_\mu p \to \nu_\mu p) } = 0.53 \pm 0.17\left( \mathrm{stat.} \right),
\end{align}
despite the unfolding of nuclear effects in a simplified model~\cite{Entenberg:1979wc,Ahrens:1985cm,Ahrens:1986xe} that might result in large discrepancies and uncertainties~\cite{Mahn:2018mai,Giusti:2019cup,Dolan:2026nlr}. For our theoretical predictions above, radiative corrections in the CC reactions are taken from~\cite{Tomalak:2021hec,Tomalak:2022xup,Tomalak:2025vxa} as the fixed-order QED calculation in the inclusive measurement w.r.t. the kinematics of the photon, and nucleon FFs with uncertainties are taken from fits in~\cite{Borah:2020gte,Tomalak:2026wsu}, which incorporate state-of-the-art QED radiative corrections.

\begin{figure}[tb!]
	\centering
	{\includegraphics[angle=0,scale=0.55]{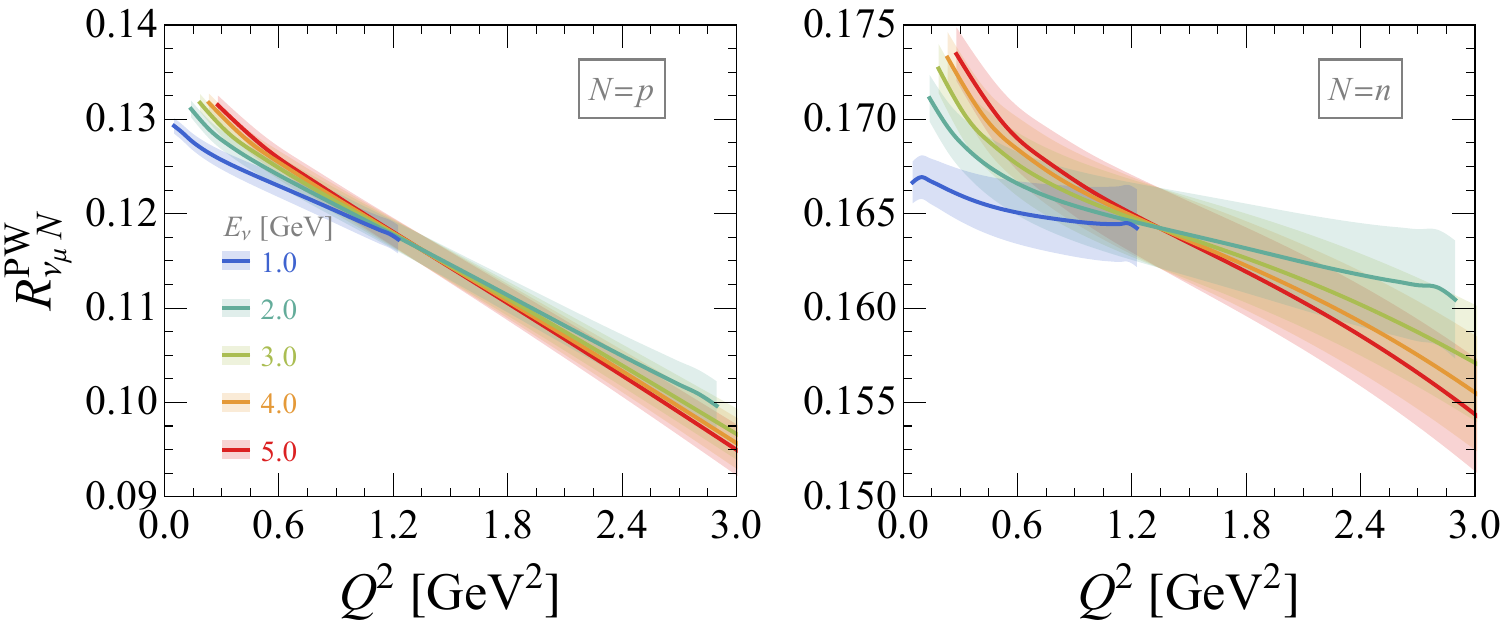}}
	\caption{$R_{\nu_\mu p}^ \text{PW}$ (left panel) and $R_{\nu_\mu n}^ \text{PW}$ (right panel) from (\ref{eq:deltaNC-to-deltaCCp}) and (\ref{eq:deltaNC-to-deltaCCn}) are shown as functions of $Q^2$ at (anti)neutrino beam energies $E_\nu \in \{1.0,~ 2.0,~3.0,~4.0,~5.0\}~\text{GeV}$.}
	\label{fig:PW_relation_vPvN}
\end{figure}

In the following, we investigate the well-known tree-level\footnote{Some corrections to the tree-level Paschos-Wolfenstein (PW) relation can be found in~\cite{Arbuzov:2004zr,Brodsky:2004qa,Gluck:2005xh,Moch:2007rq,Ding:2004dv,Wei:2007nb,Cloet:2009qs}.} Paschos-Wolfenstein relation for isoscalar targets~\cite{Paschos:1972kj,Cheng:1984vwu},
\begin{align}\label{eq:Paschos-Wolfenstein}
    R_{\nu_\ell}^ \text{PW}
    =\frac{ \sigma( \nu_\ell N \to \nu_\ell X) - \sigma(\bar\nu_\ell N \to \bar\nu_\ell X) }{\sigma(\nu_\ell N \to \ell X) - \sigma(\bar\nu_\ell N \to \bar{\ell} X) } = \frac{1-2\sin^2\theta_W}{2} =
		\begin{cases}
			0.2613,~ \sin^2\theta_W = 0.23873, \\
			0.2688,~ \sin^2\theta_W = 0.23116, \\
			0.2672,~ \sin^2\theta_W = 0.23281,
		\end{cases}
\end{align}
which is defined as the ratio of the NC to CC cross sections between the differences of neutrino- and antineutrino-nucleon inclusive deep-inelastic scattering~\cite{Moch:2007rq,Praet:2006kp}. We evaluate such a ratio for scattering on the proton $R_{\nu_\mu p}^ \text{PW}$ and neutron $R_{\nu_\mu n}^ \text{PW}$ targets~\cite{Bednaykov:1988vs},
\begin{align}
    R_{\nu_\mu p}^ \text{PW} 
    &= \frac{ \sigma(\nu_\mu p \to \nu_\mu p) - \sigma(\bar\nu_\mu p \to \bar\nu_\mu p) }{ \sigma(\nu_\mu n \to \mu^- p) - \sigma(\bar\nu_\mu p \to \mu^+ n) },\label{eq:deltaNC-to-deltaCCp} \\
    R_{\nu_\mu n}^ \text{PW} 
    &= \frac{ \sigma(\nu_\mu n \to \nu_\mu n) - \sigma(\bar\nu_\mu n \to \bar\nu_\mu n) }{ \sigma(\nu_\mu n \to \mu^- p) - \sigma(\bar\nu_\mu p \to \mu^+ n) },\label{eq:deltaNC-to-deltaCCn}
\end{align}
compare it with the relation that accounts for the quark mixing,
\begin{align}\label{eq:Paschos-Wolfenstein_Vud}
    R_{\nu_\ell}^ \text{PW} = \frac{1-2\sin^2\theta_W}{2 |V_{ud}|^2} =
		\begin{cases}
			0.2756,~ \sin^2\theta_W = 0.23873, \\
			0.2836,~ \sin^2\theta_W = 0.23116, \\
			0.2818,~ \sin^2\theta_W = 0.23281,
		\end{cases}
\end{align}
with the CKM matrix~\cite{Cabibbo:1963yz,Kobayashi:1973fv} element $|V_{ud}|=(0.97367 \pm 0.00032)$~\cite{ParticleDataGroup:2024cfk}, and investigate its dependence on $Q^2$ at different (anti)neutrino beam energies $E_\nu$ in figure~\ref{fig:PW_relation_vPvN}.

\begin{figure}[tb!]
	\centering
	{\includegraphics[angle=0,scale=0.50]{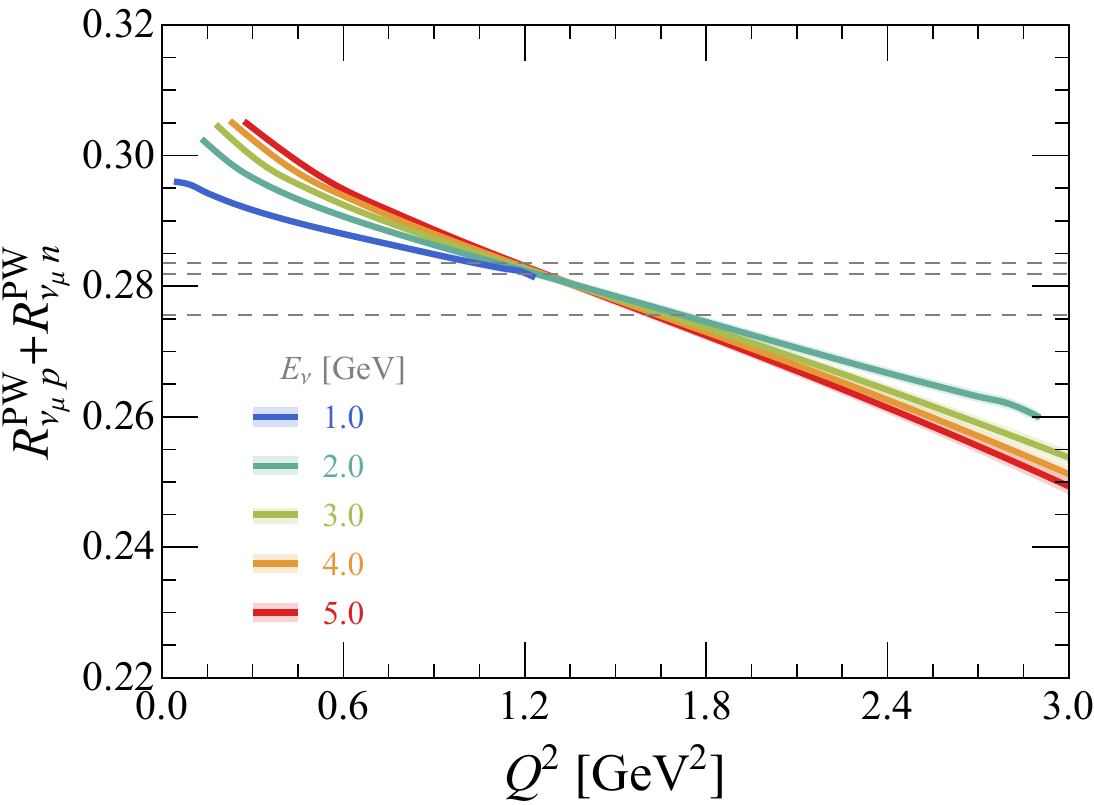}}
	\caption{$R_{\nu_\mu p}^ \text{PW}+R_{\nu_\mu n}^ \text{PW}$ from (\ref{eq:Paschos-Wolfenstein_approx}) is shown as a function of $Q^2$ at (anti)neutrino beam energies $E_\nu \in \{1.0,~ 2.0,~3.0,~4.0,~5.0\}~\text{GeV}$, where the horizontal dashed grey lines represent the tree-level results of $R_{\nu_\ell}^ \text{PW}$ in (\ref{eq:Paschos-Wolfenstein_Vud}).}
	\label{fig:PW_relation_vPmvN}
\end{figure}

Apparently, the PW relation does not hold at the single-nucleon level, as expected from the derivation within the quark model, since a single nucleon does not correspond to an isoscalar target (i.e. $A=2Z$). By fulfilling the isoscalar condition, we suggest the following approximate relation for isoscalar targets in terms of $R_{\nu_\ell p}^ \text{PW}$ and $R_{\nu_\ell n}^ \text{PW}$~\cite{Bednaykov:1988vs}
\begin{align}\label{eq:Paschos-Wolfenstein_approx}
	R_{\nu_\ell p}^\text{PW} + R_{\nu_\ell n}^\text{PW} \approx \frac{1-2\sin^2\theta_W}{2 |V_{ud}|^2}.
\end{align}
In figure~\ref{fig:PW_relation_vPmvN}, we present $R_{\nu_\mu p}^ \text{PW}+R_{\nu_\mu n}^ \text{PW}$ as a function of $Q^2$ at (anti)neutrino beam energies $E_\nu \in \{1.0,~ 2.0,~3.0,~4.0,~5.0\}~\text{GeV}$, in comparison with the tree-level results of $R_{\nu_\ell}^ \text{PW}$ in (\ref{eq:Paschos-Wolfenstein_Vud}) represented by the horizontal dashed grey lines. We observe small deviations of $R_{\nu_\ell p}^\text{PW} + R_{\nu_\ell n}^\text{PW}$, around $(5-10)\%$, from the tree-level expectations~(\ref{eq:Paschos-Wolfenstein_Vud}) at low momentum transfers (i.e. $Q^2 \lesssim 3~\text{GeV}^2$) and the $E_\nu$-dependence of $R_{\nu_\mu p}^ \text{PW}+R_{\nu_\mu n}^ \text{PW}$ is weak especially when $E_\nu$ is large. This indicates that the approximate relation (\ref{eq:Paschos-Wolfenstein_approx}) is roughly valid for low momentum transfers and high-energy (anti)neutrino beams. By comparing our results of $R_{\nu_\mu p}^\text{PW}+R_{\nu_\mu n}^ \text{PW}$ with that of $R_{\nu_\mu p}^\text{PW}$ and $R_{\nu_\mu n}^\text{PW}$ in figure~\ref{fig:PW_relation_vPvN}, we notice that the magnitude of relative change in $R_{\nu_\mu p}^\text{PW}+R_{\nu_\mu n}^\text{PW}$ lies in between $R_{\nu_\ell n}^\text{PW}$ and $R_{\nu_\ell p}^\text{PW}$, where $R_{\nu_\ell n}^\text{PW}$ depends least on the momentum transfer and the magnitude of relative change in $R_{\nu_\ell n}^ \text{PW}$ is the smallest.

The MiniBooNE experiment~\cite{MiniBooNE:2010xqw,MiniBooNE:2013dds}, using the BNB at Fermilab, has operated in two modes of the (anti)neutrino beam with opposite polarity of the magnetic focusing horn: neutrino mode and antineutrino mode~\cite{MiniBooNE:2008hfu}. In each mode, four species of (anti)neutrinos are involved, namely $\nu_\mu$, $\bar \nu_\mu$, $\nu_e$, and $\bar \nu_e$; see, e.g. figure~\ref{fig:BNBfluxRaw} in appendix~\ref{app:flux_distributions}. The MiniBooNE measurements~\cite{MiniBooNE:2010xqw} of the flux-averaged differential cross section of $\nu_\ell N \to \nu_\ell N$ for $N=p,n$ were performed in the neutrino mode, while the measurements~\cite{MiniBooNE:2013dds} of the flux-averaged differential cross section of $\bar\nu_\ell N \to \bar\nu_\ell N$ were performed in the antineutrino mode.

For each mode, the flux-averaged differential cross section on the $\mathrm{C} \mathrm{H}_2$ target at MiniBooNE~\cite{MiniBooNE:2010xqw,MiniBooNE:2013dds}, after incoherently adding the nucleon contributions without (w/o) applying the efficiency correction (EC), is given by
\begin{equation}
    \begin{aligned}\label{eq:MinoBooNE-woRC-DiffSgm}
        \Big\langle \frac{\ud \sigma}{\ud Q^2} \Big\rangle
        &=  \sum_{\chi} \alpha_\chi \left[ \varpi_p \Big\langle \frac{\ud \sigma}{\ud Q^2} \Big\rangle^{(\chi p )} + \varpi_n \Big\langle \frac{\ud \sigma}{\ud Q^2} \Big\rangle^{(\chi n)} \right],
	\end{aligned}
\end{equation}
where $\chi \in \{\nu_\mu, \bar\nu_\mu, \nu_e, \bar\nu_e\}$ denotes the (anti)neutrino flavor with its relative weight $\alpha_\chi $ defined as $\alpha_\chi \equiv \mathcal N_\Phi^\chi/ \sum_\chi \mathcal N_\Phi^\chi \in [0,1]$, the factors $\varpi_p=8/14$ and $\varpi_n=6/14$ account for the $8$ protons and $6$ neutrons in each $\mathrm{C} \mathrm{H}_2$ molecule, and $\langle \ud \sigma/ \ud Q^2 \rangle^{(\chi N)} $ denotes the flux-averaged differential cross section (\ref{eq:DiffSgm-fluxAve}) of $\chi N \to \chi N$. Evidently, we have normalization conditions $\sum_\chi \alpha_\chi =1$ and $\varpi_p+\varpi_n=1$.

\begin{figure}[tb!]
	\centering
	{\includegraphics[angle=0,scale=0.52]{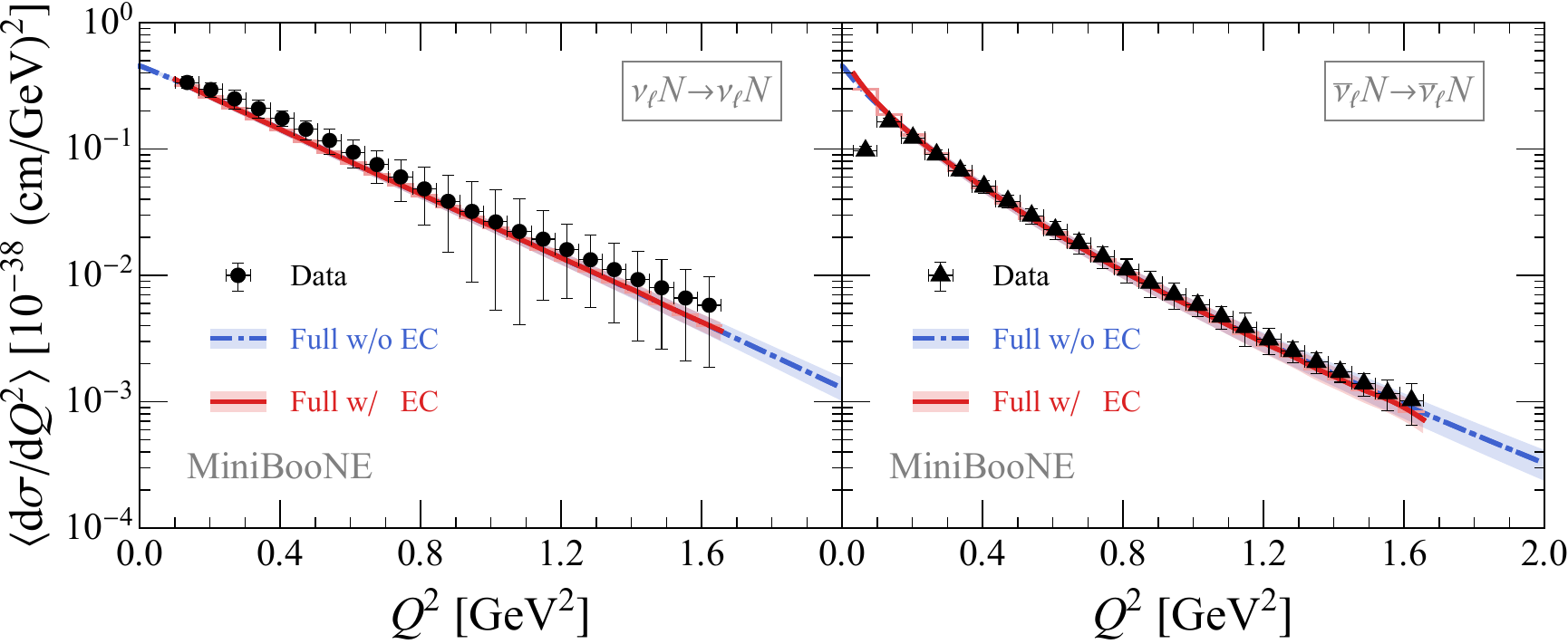}}
	\caption{Predictions for the flux-averaged differential cross sections of $\nu_\ell N \to \nu_\ell N$ (left panel) and $\bar\nu_\ell N \to \bar\nu_\ell N$ (right panel) on the $\mathrm{C} \mathrm{H}_2$ target are compared with the experimental data (black markers) from MiniBooNE~\cite{MiniBooNE:2010xqw,MiniBooNE:2013dds}, after incoherently adding the nucleon contributions with (solid red curves) and without (dash-dotted blue curves) applying the efficiency correction (EC). To visualize the possible bin-width effect in $Q^2$, we also show the thin light-red step curves for a comparison using exactly the same $Q^2$ bins and ECs as the experimental data~\cite{MiniBooNE:2010xqw,MiniBooNE:2013dds}.}
	\label{fig:MiniBooNE_DiffSgm}
\end{figure}

To directly compare with the experimental data from MiniBooNE~\cite{MiniBooNE:2010xqw,MiniBooNE:2013dds}, we consider different efficiency corrections (EC) for free protons, bound protons, and bound neutrons in the $\mathrm{C} \mathrm{H}_2$ target. In this case, the corresponding differential cross section for each beam mode is given by~\cite{MiniBooNE:2010xqw,MiniBooNE:2013dds,Sufian:2018qtw}
\begin{equation}
    \begin{aligned}\label{eq:MinoBooNE-wRC-DiffSgm}
        \Big\langle \frac{\ud \sigma}{\ud Q^2} \Big\rangle
        &=  \sum_{ \chi } \alpha_{ \chi }\left[ \varpi_{p,H}\, C_{p,H} \Big\langle \frac{\ud \sigma}{\ud Q^2} \Big\rangle^{(\chi p)} + \varpi_{p,C}\, C_{p,C} \Big\langle \frac{\ud \sigma}{\ud Q^2} \Big\rangle^{(\chi p )} + \varpi_{n,C}\, C_{n,C} \Big\langle \frac{\ud \sigma}{\ud Q^2} \Big\rangle^{(\chi n )} \right],
	\end{aligned}
\end{equation}
where $C_{p,H}(Q^2)$, $C_{p,C}(Q^2)$, and $C_{n,C}(Q^2)$ are the $Q^2$-dependent dimensionless ECs determined in the MiniBooNE measurements~\cite{MiniBooNE:2010xqw,MiniBooNE:2013dds}, which are different for the two modes even at the same $Q^2$ value (see, e.g. the figure~15 of~\cite{MiniBooNE:2010xqw} and the figure~10 of~\cite{MiniBooNE:2013dds}), and the factors $\{\varpi_{p,H},\varpi_{p,C}, \varpi_{n,C}\} =\{ 2/14,6/14,6/14 \}$~\cite{MiniBooNE:2010xqw,MiniBooNE:2013dds}. Evidently, we have $\varpi_{p,H}+\varpi_{p,C}=\varpi_p$ and $\varpi_{n,C}=\varpi_n$. Comparison of (\ref{eq:MinoBooNE-woRC-DiffSgm}) with (\ref{eq:MinoBooNE-wRC-DiffSgm}) can tell us how large the effect of ECs is and how it changes as a function of $Q^2$.

It should be mentioned that we do not consider any nuclear effects in (\ref{eq:MinoBooNE-woRC-DiffSgm}) and (\ref{eq:MinoBooNE-wRC-DiffSgm}). Therefore, if it turns out that our predictions based on (\ref{eq:MinoBooNE-wRC-DiffSgm}) have large (small) deviations from the true MiniBooNE data~\cite{MiniBooNE:2010xqw,MiniBooNE:2013dds}, it may signal that the nuclear effects involved in the carbon nucleus in the $\mathrm{C} \mathrm{H}_2$ target are large (small).

In figure~\ref{fig:MiniBooNE_DiffSgm}, we compare our theoretical predictions for the flux-averaged differential cross sections of $\nu_\ell N \to \nu_\ell N$ and $\bar\nu_\ell N \to \bar\nu_\ell N$ on the $\mathrm{C} \mathrm{H}_2$ target, based on~(\ref{eq:MinoBooNE-woRC-DiffSgm}) and (\ref{eq:MinoBooNE-wRC-DiffSgm}) at $\Delta E = 5~\mathrm{MeV}$, with the experimental data from MiniBooNE~\cite{MiniBooNE:2010xqw,MiniBooNE:2013dds} after incoherently adding the nucleon contributions with and without applying the EC. To visualize the possible bin-width effect in $Q^2$, we also show the thin light-red step curves using exactly the same $Q^2$ bins and ECs as the experimental data~\cite{MiniBooNE:2010xqw,MiniBooNE:2013dds}. Contrary to the conclusions of the review~\cite{Giusti:2019cup}, we find a good  agreement of our theoretical predictions with the MiniBooNE data~\cite{MiniBooNE:2010xqw,MiniBooNE:2013dds} for $Q^2 \gtrsim 0.1~\text{GeV}^2$. This indicates that either the nuclear effects are irrelevant (or at least small) for the flux-averaged differential cross sections or there is a large cancellation between various contributions when $Q^2 \gtrsim 0.1~\text{GeV}^2$. However, for $Q^2 \lesssim 0.1~\text{GeV}^2$, our theoretical prediction is inconsistent with the MiniBooNE data (see the first data point in the right panel), suggesting sizable nuclear effects at very low momentum transfers, where Pauli blocking may be one of the causes~\cite{Chen:2024kbh}.

It is worth mentioning that our formalism of radiative corrections at the single-nucleon level can be readily applied for background modelling and the search for new physics in the disappearance channel with short-baseline neutrino experiments~\cite{LSND:1996ubh,LSND:1997vun,Blanpied:1997zz,Harvey:2007ca,MiniBooNE:2007uho,MiniBooNE:2008yuf,Sparveris:2008jx,MiniBooNE:2009atu,Hill:2009ek,Hill:2010zy,Wang:2013wva,Garvey:2014exa,Wang:2014nat,Wang:2015ivq,Rosner:2015fwa,MiniBooNE:2018esg,Machado:2019oxb,T2K:2019odo,Ioannisian:2019kse,Giunti:2019sag,MiniBooNE:2020pnu,Chanfray:2021wie,MicroBooNE:2021zai,Goncalves:2022lmi,Butkevich:2022pzd,MicroBooNE:2025rsd,MicroBooNE:2025ovj,Ioannisian:2025bro}. Moreover, our improvements on radiative corrections~\cite{Tomalak:2020zfh,Tomalak:2025tls} in coherent elastic neutrino-nucleus scattering~\cite{Stodolsky:1966zz,Freedman:1973yd,Kopeliovich:1974mv,Sehgal:1985iu,Botella:1986wy,COHERENT:2017ipa,Tayloe:2017edz,COHERENT:2022nrm,COHERENT:2020iec,COHERENT:2024axu,RED-100:2024izi,Ackermann:2025obx} and incoherent neutral-current reactions enhance the sensitivity of Super-K, KamLAND, and JUNO to prompt supernova bursts~\cite{Super-Kamiokande:2007zsl,JUNO:2015zny,JUNO:2015sjr,KamLAND:2015dbn,Super-Kamiokande:2016kji,KamLAND:2024uia,JUNO:2021vlw} and the diffuse supernova background~\cite{Super-Kamiokande:2002hei,Super-Kamiokande:2013ufi,Super-Kamiokande:2021jaq,JUNO:2015zny,JUNO:2021vlw}. Updated QCD contributions extend the validity range and enhance both the accuracy and precision of radiative corrections for elastic (anti)neutrino–electron scattering~\cite{Tomalak:2019ibg,Tomalak:2025tls}, the Standard Candle for flux normalization in neutrino experiments~\cite{MINERvA:2015nqi,MINERvA:2019hhc,Marshall:2019vdy,MINERvA:2022vmb}.

\section{Summary}
\label{sec:summary}

In this paper, we formulate and evaluate the radiative corrections in weak neutral-current (anti)neutrino-nucleon elastic scattering at GeV energies within the effective field theory framework. We start our evaluation with four-fermion quark-level interactions in the low-energy effective field theory and determine the nucleon-level form factors in terms of effective couplings, which include large electroweak contributions. Subsequently, we define momentum-dependent QED radiative corrections from lepton and quark loops as modifications to these form factors. For the first time, we present the contribution from light quarks by connecting data-driven evaluation at low energies, guided by chiral-perturbation theory analysis, with perturbative-QCD dependence at large momentum transfers. We also include QED interactions on the nucleon line within the factorization framework and provide a rigorous formulation of QED radiative corrections in terms of process-independent Born form factors.

We find that nucleon strange-quark contributions can reach $(4-9)\%$ and $(3-15)\%$ at single-proton and single-neutron levels, respectively. The total contributions from QED and electroweak radiative corrections, dominated by large electroweak logarithms, are also at a few-percent level, comparable to the strange-quark contributions. A proper inclusion of effects of this size is of particular importance in phenomenological extractions of the strangeness content of nucleon/nuclei in terms of strange-quark form factors. The dominant uncertainty in our formalism at the single-nucleon level, assuming that the nucleon form factors are given, comes from light-quark contributions, but still remains below $0.1\%$. Consequently, this work paves the way for accurate determinations of the strangeness content of the nucleon/nucleus, as well as for high-precision studies of neutral-current (anti)neutrino scattering cross sections and analyses of (anti)neutrino disappearance rates.

For the (anti)neutrino flavor dependence, the cross-section difference between the tau- and muon-flavor neutrino (antineutrino) beams can reach around $0.5\%$ ($1.0\%$), while the cross-section difference between the electron- and muon-flavor neutrino (antineutrino) beams is negligible at momentum transfers above the muon mass scale.

We compare our theoretical predictions for the total NC (anti)neutrino-nucleon elastic scattering cross section, including radiative corrections and uncertainties from nucleon form factors, with the experimental data from BNL E734~\cite{Horstkotte:1981ne,Ahrens:1986xe} and MiniBooNE~\cite{MiniBooNE:2010xqw,MiniBooNE:2013dds}. We observe decent agreements within the uncertainty range for the $Q^2$ dependence of flux-averaged differential cross sections and various cross-section ratios, although with a noticeable difference for the first $\overline{\nu}_\ell N$ data point from MiniBooNE~\cite{MiniBooNE:2013dds} at $Q^2 \lesssim 0.1~\text{GeV}^2$. This indicates that the total nuclear effects in the carbon nucleus in the $\mathrm{CH}_2$ target are basically irrelevant to the MiniBooNE data when $Q^2 \gtrsim 0.1~\text{GeV}^2$, while at very small squared four-momentum transfers (i.e. $Q^2 \lesssim 0.1~\text{GeV}^2$) the involved nuclear effects, e.g. the Pauli blocking, may be sizable. We plan to investigate the interplay of radiative corrections and nuclear effects in future work.

\acknowledgments

The authors thank Han-Xue~Chen, Fan~Gao, Patrick~Green, Fred~Jegerlehner, Stephen F.~Pate, Jan~T.~Sobczyk, Adrian~Thompson, and De-Liang Yao for helpful communications, and Feng-Kun~Guo and Jia-Jun~Wu for valuable discussions. FeynCalc~\cite{Mertig:1990an,Shtabovenko:2016sxi}, LoopTools~\cite{Hahn:1998yk}, and Mathematica~\cite{Mathematica} were extremely useful in this work. This work is supported in part by the National Natural Science Foundation of China (NSFC) under Grants No.~12347105, No.~12547178, and No.~12447101; and by Tsinghua University. O.~T. is also supported by the Institute of Theoretical Physics, Chinese Academy of Sciences.

\appendix

\section{Proton quark-flavor form factors}
\label{app:flavor_decomposition}

\begin{figure}[tb!]
	\centering
	{\includegraphics[angle=0,scale=0.5]{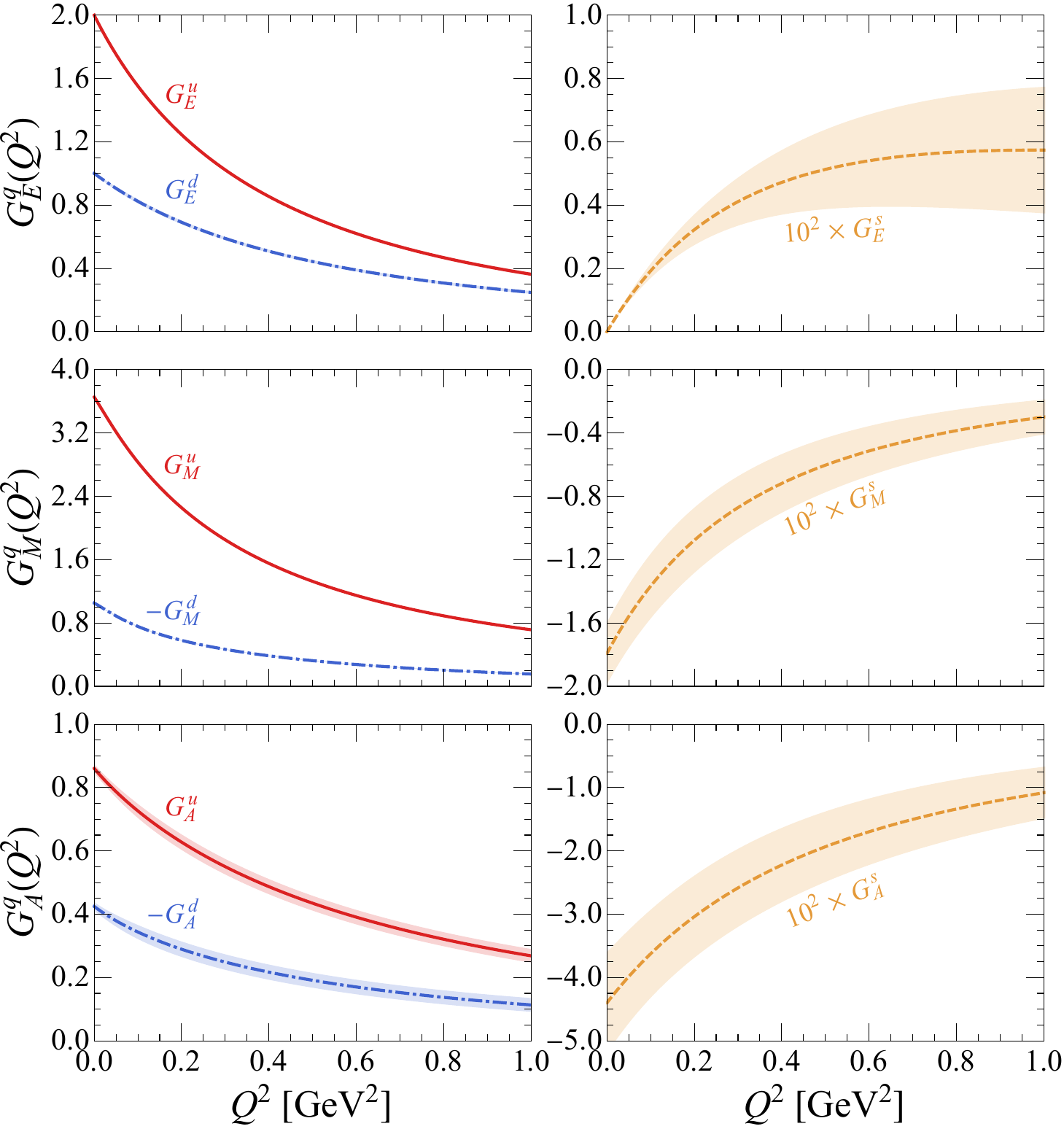}}
	\caption{Quark-flavor FFs $G_E^q(Q^2)$ (upper panel), $G_M^q(Q^2)$ (middle panel), and $G_A^q(Q^2)$ (lower panel) of the proton are shown as functions of $Q^2$ for $q=u,d,s$.}
	\label{fig:quarkFFs}
\end{figure}

In this appendix, we report numerical results for the quark-flavor FFs $G_{E,M,A}^q(Q^2)$ of the proton as functions of $Q^2$ for $q=u,d,s$, see figure~\ref{fig:quarkFFs}. These data-driven quark-flavor FFs are obtained with inputs from~\cite{Borah:2020gte,Alexandrou:2021wzv,MINERvA:2023avz,Tomalak:2026wsu,Alexandrou:2026ait}. We evaluate the quark-flavor FFs assuming the $\text{SU}(3)$ flavor symmetry, with detailed expressions and inputs described in section~\ref{sec:leading_order_SM}.

\section{Neutrino and antineutrino fluxes} 
\label{app:flux_distributions}

In this appendix, we provide numerical results for the flux distributions of neutrinos and antineutrinos used in this work. 

\begin{figure}[htb!]
	\centering
	{\includegraphics[angle=0,scale=0.5]{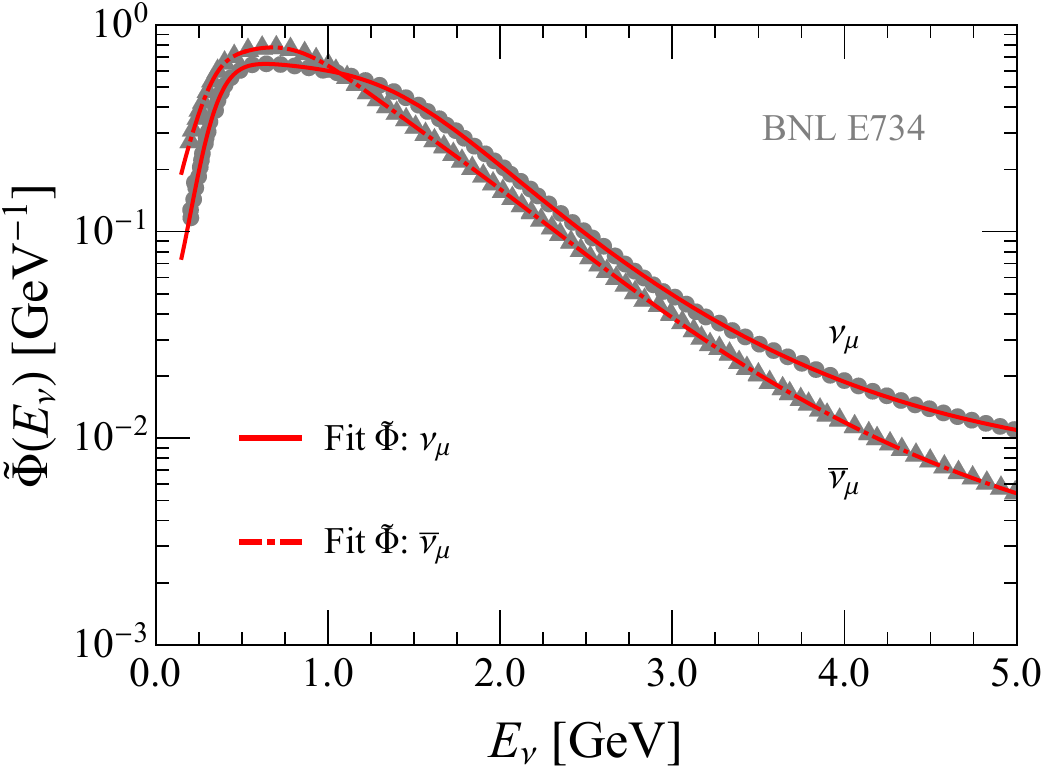}}
	\caption{The normalized flux distributions $\tilde\Phi(E_\nu)$ of the muon-flavor neutrino and antineutrino beams in the BNL E734 experiment~\cite{Ahrens:1986xe} are shown as functions of the (anti)neutrino energy $E_\nu$.}
	\label{fig:BNLflux}
\end{figure}

In figure~\ref{fig:BNLflux}, we present the normalized flux distributions $\tilde\Phi_\pm(E_\nu)$ of the muon flavor neutrino and antineutrino beams in the BNL E734 experiment~\cite{Ahrens:1986xe}, where the curves denote fits to the corresponding data points extracted from the solid curves in the figures~2 and 3 of~\cite{Ahrens:1986xe} with $E_\text{L}=0.2~\text{GeV}$ and $E_\text{U}=5.0~\text{GeV}$ [see e.g. (\ref{eq:normalized-Flux})].

\begin{figure}[tb!]
	\centering
	{\includegraphics[angle=0,scale=0.53]{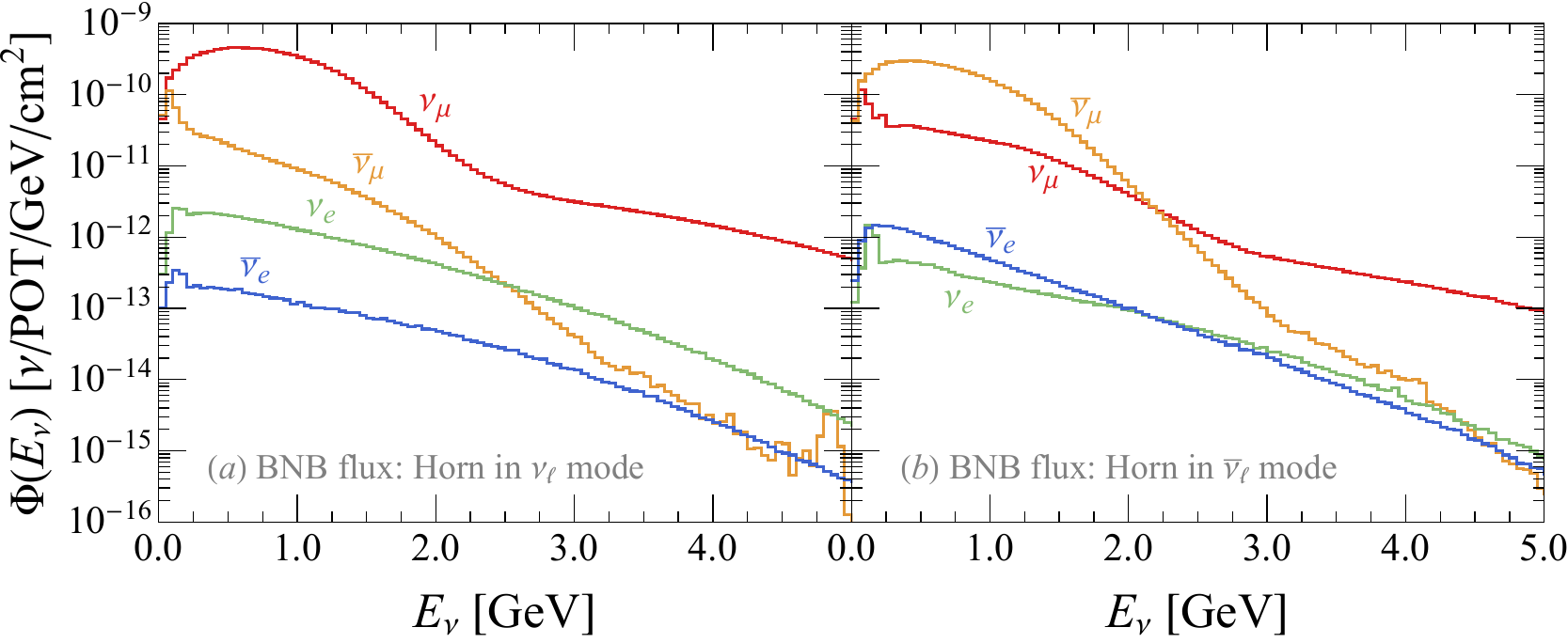}}
	\caption{The total flux distributions $\Phi(E_\nu)$ of the muon- and electron-flavor neutrino and antineutrino beams at the MiniBooNE detector in BNB (labeled by ``BNB flux'') with magnetic focusing horn in neutrino (left panel) and antineutrino (right panel) modes are shown as functions of the (anti)neutrino energy $E_\nu$, where curves are reproduced from the figures~27 and 28 of~\cite{MiniBooNE:2008hfu}.}
	\label{fig:BNBfluxRaw}
\end{figure}

\begin{figure}[tb!]
	\centering
	{\includegraphics[angle=0,scale=0.535]{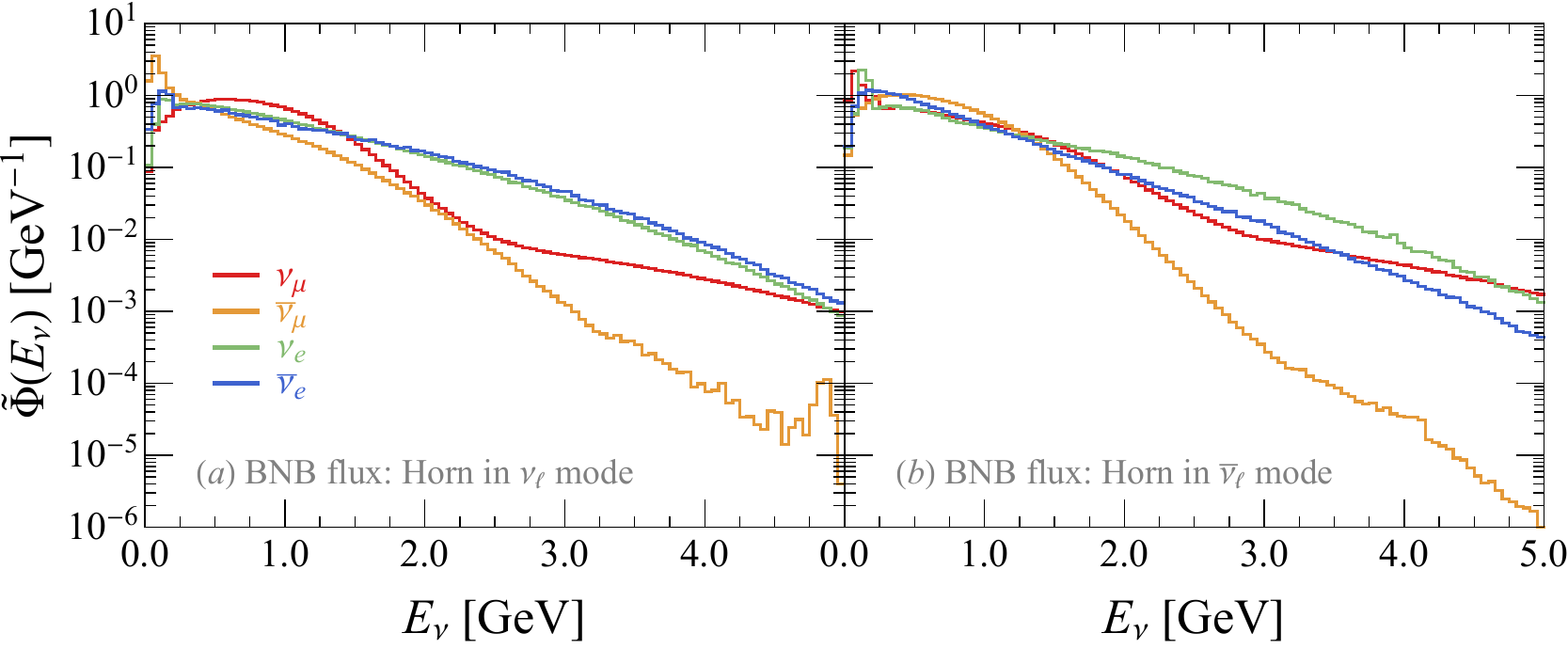}}
	\caption{Same as figure~\ref{fig:BNBfluxRaw} but for the normalized flux distributions $\tilde\Phi(E_\nu)$.}
	\label{fig:BNBflux}
\end{figure}

In figure~\ref{fig:BNBfluxRaw}, we present the total flux distribution $\Phi (E_\nu)$ of the muon- and electron-flavor neutrino and antineutrino at the MiniBooNE detector in BNB (labeled by ``BNB flux'') at Fermilab with magnetic focusing horn in neutrino (left panel) and antineutrino (right panel) modes based on the figures 27 and 28 of~\cite{MiniBooNE:2008hfu}, where POT stands for protons on target. After proper normalization for each (anti)neutrino species in figure~\ref{fig:BNBfluxRaw} with $E_\text{L}=0.025~\text{GeV}$ and $E_\text{U}=5~\text{GeV}$, we finally present in figure~\ref{fig:BNBflux} the corresponding normalized flux distribution $\tilde \Phi(E_\nu)$ used for the flux-averaged differential cross section~(\ref{eq:DiffSgm-fluxAve}).

\section{Nucleon spin structure}
\label{app:GPDs}

The vector $G_{E,M,S}^{q}(Q^2)$ and axial-vector $G_{A,P,T}^{q}(Q^2)$ quark FFs are directly related to unpolarized $X_\text{unpol}^q(x,\xi,t)$ and polarized $X_\text{pol}^q(x,\xi,t)$ quark generalized parton distributions (GPDs) for $X_\text{unpol}^q \in\{ H^q, E^q, T^q\}$ and $X_\text{pol}^q \in \{ \tilde H^q, \tilde E^q, \tilde T^q \}$, respectively, in exclusive processes at electron scattering facilities such as HERA~\cite{Barber:1994ew}, CERN SPS~\cite{Adams:2018pwt}, CEBAF~\cite{Arrington:2021alx,Accardi:2023chb}, EIC~\cite{Accardi:2012qut,Burkert:2022hjz}, and EicC~\cite{Anderle:2021wcy} through the first-moment sum rules~\cite{Ji:1996ek,Diehl:2003ny,Guidal:2013rya,Lorce:2025aqp,Chen:2024ksq,Chen:2025gmg}:
\begin{equation}
	\begin{aligned}\label{relations_to_GPDs}
		\begin{bmatrix}
			F_1^q(t)\\
			F_2^q(t)\\
			G_S^q(t)
		\end{bmatrix}
		= \int_{-1}^{1} \ud x
		\begin{bmatrix}
			H^q(x,\xi,t)\\
			E^q(x,\xi,t)\\
			T^q(x,\xi,t)\\
		\end{bmatrix},\qquad
        \begin{bmatrix}
			G_A^q(t)\\
			G_P^q(t)\\
			G_T^q(t)
		\end{bmatrix}
		= \int_{-1}^{1} \ud x
		\begin{bmatrix}
			\tilde H^q(x,\xi,t)\\
			\tilde E^q(x,\xi,t)\\
			\tilde T^q(x,\xi,t)\\
		\end{bmatrix},
	\end{aligned}
\end{equation}
with~\cite{Yennie:1957rmp,Ernst:1960zza}
\begin{equation}
	\begin{aligned}\label{GEM-F12-relations}
		G_E^q(t) &= F_1^q(t) + \frac{t}{4M^2} F_2^q(t),\\
        G_M^q(t) &= F_1^q(t) + F_2^q(t),
	\end{aligned}
\end{equation}
where $X^q(t) = X^q(Q^2)$ for $X\in \{ F_1, F_2, G_E, G_M,G_S,G_A,G_P,G_T \}$, $t\equiv \Delta^2=-Q^2$, $x \equiv (n \cdot k_q)/(n \cdot P) = k_q^+/P^+ \in [-1,1]$ is the momentum fraction carried by the initial quark with a four-momentum $k_q^\mu$ along the light-front\footnote{A generic instant-form four-vector $a^\mu=(a^0, a^1, a^2,a^3)$ is written in light-cone coordinates as $a^\mu=[a^+, a^-, a^1, a^2] = [a^+, a^-, \vec a_\perp ]$~\cite{Brodsky:1997de,Diehl:2003ny,Lorce:2025aqp}, with $a^{\pm} \equiv (a^0 \pm a^3)/\sqrt{2}$ and $\vec a_\perp \equiv (a^1, a^2)$.} plus (``$+$'') direction~\cite{Diehl:2003ny,Lorce:2025aqp}, and $\xi \equiv -(p'^+ -p^+)/(p'^+ + p^+ ) = - (n \cdot \Delta)/ (2 n \cdot P) = -\Delta^+/(2P^+) \in [-1,1]$ is the skewness parameter~\cite{Diehl:2003ny}, with $n_\mu=(1,0,0,1)/\sqrt{2}$ a light-like four-vector.

The unpolarized $X_\text{unpol}^q(x,\xi,t)$ and polarized $X_\text{pol}^q(x,\xi,t)$ quark GPDs are defined via the tensor decomposition of the integrals over the following correlation functions~\cite{Ji:1996ek,Diehl:2003ny,Guidal:2013rya,Lorce:2025aqp,Chen:2025gmg} of quark currents in coordinate space
\begin{equation}
	\begin{aligned}\label{eq:vector-GPDs}
		&\int \frac{\ud z^- }{4\pi} e^{i x P^+ z^- } \langle p',s'| \hat{\overline{q}}\left(- \tfrac{z}{2} \right) \gamma^\mu \hat{\mathcal W}\left[-\tfrac{z}{2}; \tfrac{z}{2} \right] \hat{q}\left( \tfrac{z}{2} \right)|p,s\rangle \Big|_{ \uvec z^+=|\vec z_\perp|=0 } \\
		&= \frac{1}{2P^+} \bar u(p',s')\! \left[ \gamma^\mu H^q(x,\xi,t) + \frac{ i\sigma^{\mu\nu} \Delta_\nu }{2M} E^q(x,\xi,t) + \frac{ i\Delta^\mu }{2M} T^q(x,\xi,t) + \cdots \right]\! u(p,s),
	\end{aligned}
\end{equation}
and
\begin{equation}
	\begin{aligned}\label{eq:axial-vector-GPDs}
		&\int \frac{\ud z^- }{4\pi} e^{i x P^+ z^- } \langle p',s'| \hat{\overline{q}}\left(- \tfrac{z}{2} \right) \gamma^\mu \gamma^5 \hat{\mathcal W}\left[-\tfrac{z}{2}; \tfrac{z}{2} \right] \hat{q}\left( \tfrac{z}{2} \right)|p,s\rangle \Big|_{ \uvec z^+=|\vec z_\perp|=0 } \\
		&= \frac{1}{2P^+} \bar u(p',s') \! \left[ \gamma^\mu \gamma^5 \tilde H^q(x,\xi,t) + \frac{ \Delta^\mu }{2M} \gamma^5 \tilde E^q(x,\xi,t) + \frac{ -\sigma^{\mu\nu} \Delta_\nu }{2M} \gamma^5 \tilde T^q(x,\xi,t) + \cdots \right]\!\gamma^5 u(p,s),
	\end{aligned}
\end{equation}
with
\begin{equation}	
    \begin{aligned}\label{eq:Wilson-line}		
        \hat{\mathcal W}\left[y_1; y_2 \right]		
        &= \mathcal P \exp\left( i g_s \int_{y_1}^{y_2} \ud y^\mu \hat A_\mu^a(y) T^a \right),
    \end{aligned}
\end{equation}
where $k_q \cdot z = k_q^+ z^- = x P^+ z^-$, $\hat{\mathcal W}\big[-\tfrac{z}{2}; \tfrac{z}{2} \big]$ denotes the Wilson line which connects the positions of quark fields by three straight lines along a certain light-like path from $-z/2$ to $z/2$ and ensures the color gauge invariance of the correlation function~\cite{Diehl:2003ny,Meissner:2009ww}, $\mathcal P$ indicates the path ordering, $g_s$ denotes the running strong coupling constant, $\hat A_\mu^a(y)$ denote the gauge (gluon) field operators, $T^a$ are generators of the $\text{SU}(N_c)$ group in the fundamental representation with $a \in \{1,\cdots,N_c^2-1 \}$. For simplicity, we have neglected the high-twist contributions that are denoted by the ellipsis.

Experimental results for $G_A^{(Z/W)}(Q^2)$ and $G_T^{(Z/W)}(Q^2)$, extracted from (anti)neutrino elastic~\cite{Horstkotte:1981ne,Ahrens:1986xe,MiniBooNE:2010xqw,MiniBooNE:2013dds,Ren:2022qop,Ren:2022qut} or quasielastic~\cite{Mann:1973pr,Baker:1981su,Belikov:1983kg,Ahrens:1988rr,MiniBooNE:2010bsu,MINERvA:2013bcy,MicroBooNE:2020fxd,MINERvA:2023avz} scattering data, will provide additional useful constraints at each $Q^2$ value on the first moments of quark GPDs $\tilde H^q(x,\xi,t)$ and $\tilde T^q(x,\xi,t)$, extracted at charged-lepton (e.g. electron, muon) or proton scattering facilities~\cite{Barber:1994ew,Adams:2018pwt,Arrington:2021alx,Accardi:2023chb,Accardi:2012qut,Burkert:2022hjz,Anderle:2021wcy}, via the sum rules (\ref{relations_to_GPDs}), namely~\cite{Chen:2024ksq,Chen:2025gmg}
\begin{equation}
	\begin{aligned}
		G_A^Z(t) &= \sum_q g_A^q \int_{-1}^{1} \ud x\, \tilde H^q(x,\xi,t),\\
        G_T^Z(t) &= \sum_q g_A^q \int_{-1}^{1} \ud x\, \tilde T^q(x,\xi,t),\\
        G_A^{W}(t) &= \int_{-1}^{1} \ud x \left[ \tilde H^u(x,\xi,t) - \tilde H^d(x,\xi,t) \right],\\
        G_T^{W}(t) &= \int_{-1}^{1} \ud x \left[ \tilde T^u(x,\xi,t) - \tilde T^d(x,\xi,t) \right].
	\end{aligned}
\end{equation}

\bibliographystyle{JHEP}
\bibliography{Refs}
\end{document}